\documentclass[preprint,12pt,authoryear]{elsarticle}

\usepackage{amssymb}
\usepackage{amsmath}

\usepackage{lineno}

\usepackage{changepage}
\usepackage{caption}
\usepackage{multirow}
\usepackage{physics}
\usepackage{stackengine} 
\usepackage{xcolor}
\usepackage[normalem]{ulem}

\definecolor{darkgreen}{rgb}{0.06, 0.6, 0.3}
\definecolor{dblue}{RGB}{6,69,173}
\definecolor{burgundy}{rgb}{0.5, 0.0, 0.13}
\definecolor{mulberry}{rgb}{0.7831, 0.0, 0.6549}

\newcommand{\of}[1]{\left(#1\right)}
\newcommand{\af}[1]{\left[#1\right]}
\newcommand{\uf}[1]{\left\{#1\right\}}

\newcommand{\skp}{\hspace{2pt}}
\newcommand{\skpp}{\hspace{1pt}}

\newcommand \stacksymbol[3]{\mathrel{\stackunder[2pt]{\stackon[4pt]{$#1$}{$\scriptscriptstyle#2$}}{$\scriptscriptstyle#3$}}}

\newcommand \stacksymbolL[3]{\mathrel{\stackunder[2pt]{\stackon[4pt]{$#1$}{$#2$}}{$#3$}}}

\journal{Journal of Theoretical Biology}

\begin{document}

\begin{frontmatter}



\title{Understanding how chromatin folding and enzyme competition affect rugged epigenetic landscapes} 


\author[lab_CRM]{Daria Stepanova\corref{cor_daria}}
\ead{dstepanova@crm.cat}
\cortext[cor_daria]{Corresponding author.}
\author[lab_edinburgh]{Meritxell Brunet Guasch}
\author[lab_WCMB,lab_Ludwig]{Helen M. Byrne}
\author[lab_ICREA,lab_CRM,lab_UAB,lab_collab]{Tom\'{a}s Alarc\'{o}n}

\affiliation[lab_CRM]{organization={Centre de Recerca Matem\`{a}tica},
            addressline={Campus de Bellaterra, Edifici C}, 
            city={Bellaterra},
            postcode={08193}, 
            state={Barcelona},
            country={Spain}}

\affiliation[lab_edinburgh]{organization={School of Mathematics and Maxwell Institute for Mathematical Sciences, University of Edinburgh},
            addressline={James Clerk Maxwell Building, Mayfield Rd}, 
            city={Edinburgh},
            postcode={EH9 3FD}, 
            state={Scotland},
            country={United Kingdom}}

\affiliation[lab_WCMB]{organization={Wolfson Centre for Mathematical Biology, Mathematical Institute, University of Oxford},
            addressline={Andrew Wiles Building, Radcliffe, Observatory Quarter, Woodstock Road}, 
            city={Oxford},
            postcode={OX2 6GG}, 
            state={England},
            country={United Kingdom}}
   
\affiliation[lab_Ludwig]{organization={Ludwig Institute for Cancer Research, Nuffield Department of Medicine, University of Oxford},
            addressline={Old Road Campus Research Building, Roosevelt Drive}, 
            city={Oxford},
            postcode={OX3 7DQ}, 
            state={England},
            country={United Kingdom}}

\affiliation[lab_ICREA]{organization={Instituci\'{o} Catalana de Recerca i Estudis Avan\c{c}ats},
            addressline={Passeig de Llu\'{i}s Companys, 23}, 
            city={Barcelona},
            postcode={08010}, 
            state={Barcelona},
            country={Spain}}

 \affiliation[lab_UAB]{organization={Departament de Matem\`{a}tiques, Universitat Aut\`{o}noma de Barcelona},
            addressline={Campus de Bellaterra, Edifici C}, 
            city={Bellaterra},
            postcode={08193}, 
            state={Barcelona},
            country={Spain}}

\affiliation[lab_collab]{organization={Barcelona Collaboratorium for Predictive and Theoretical Biology},
            addressline={Wellington, 30}, 
            city={Barcelona},
            postcode={08005}, 
            state={Barcelona},
            country={Spain}}
                        

\begin{abstract}
Epigenetics plays a key role in cellular differentiation and maintaining cell identity, enabling cells to regulate their genetic activity without altering the DNA sequence. Epigenetic regulation occurs within the context of hierarchically folded chromatin, yet the interplay between the dynamics of epigenetic modifications and chromatin architecture remains poorly understood. In addition, it remains unclear what mechanisms drive the formation of rugged epigenetic patterns, characterised by alternating genomic regions enriched in activating and repressive marks. In this study, we focus on post-translational modifications of histone H3 tails, particularly H3K27me3, H3K4me3, and H3K27ac. We introduce a mesoscopic stochastic model that incorporates chromatin architecture and competition of histone-modifying enzymes into the dynamics of epigenetic modifications in small genomic loci comprising several nucleosomes. Our approach enables us to investigate the mechanisms by which epigenetic patterns form on larger scales of chromatin organisation, such as loops and domains. Through bifurcation analysis and stochastic simulations, we demonstrate that the model can reproduce uniform chromatin states (open, closed, and bivalent) and generate previously unexplored rugged profiles. Our results suggest that enzyme competition and chromatin conformations with high-frequency interactions between distant genomic loci can drive the emergence of rugged epigenetic landscapes. Additionally, we hypothesise that bivalent chromatin can act as an intermediate state, facilitating transitions between uniform and rugged landscapes. This work offers a powerful mathematical framework for understanding the dynamic interactions between chromatin architecture and epigenetic regulation, providing new insights into the formation of complex epigenetic patterns.
\end{abstract}

\begin{graphicalabstract}
\vspace{1.5cm}
\begin{adjustwidth}{-1cm}{-1cm}
\includegraphics[width=1.2\textwidth]{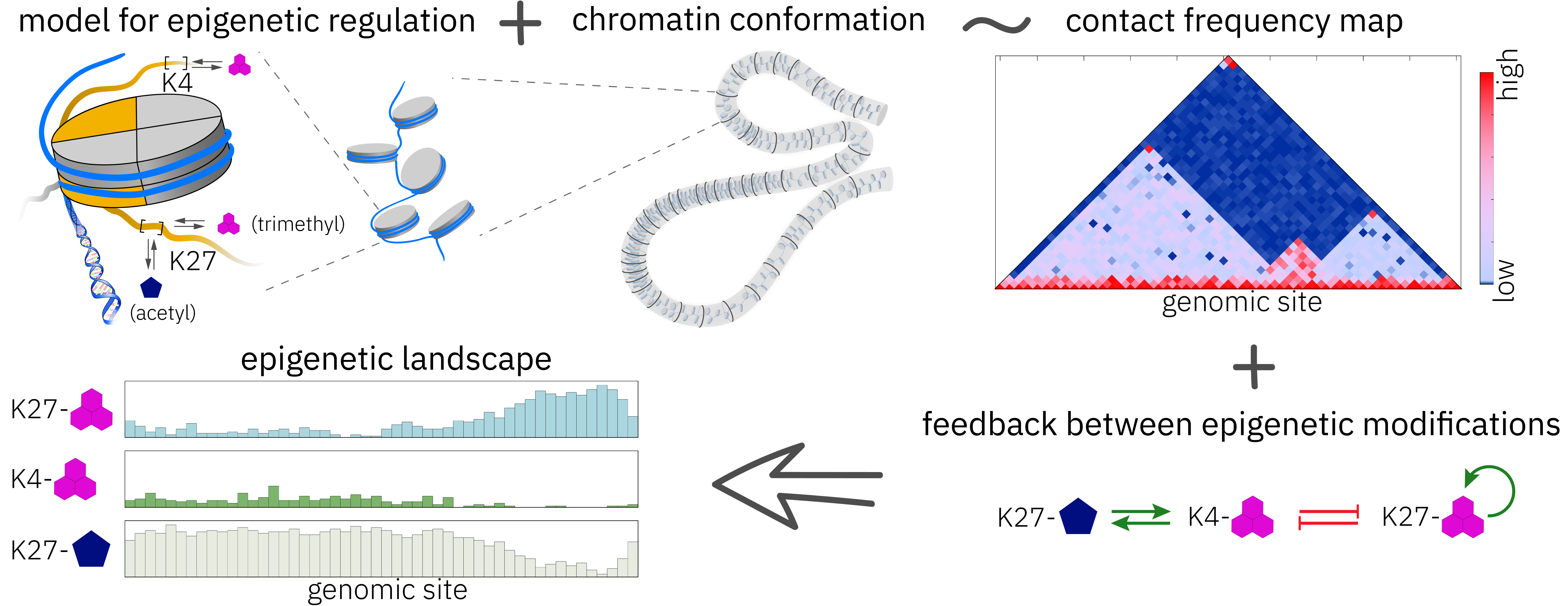}
\end{adjustwidth}
\end{graphicalabstract}

\begin{highlights}
\item Mesoscale model for histone modifications in the context of folded chromatin
\item Bifurcation analysis identifies parameter regimes for epigenetic pattern formation
\item Enzyme competition and chromatin conformation drive the emergence of rugged patterns
\item Bivalent chromatin may aid transitions between uniform and rugged epigenetic profiles
\end{highlights}


\end{frontmatter}


\emergencystretch 3em



\section{Introduction}
\label{sec1}

\begin{figure}[!t]
\begin{center}
\vspace{-0.3cm}
\includegraphics[width=\textwidth]{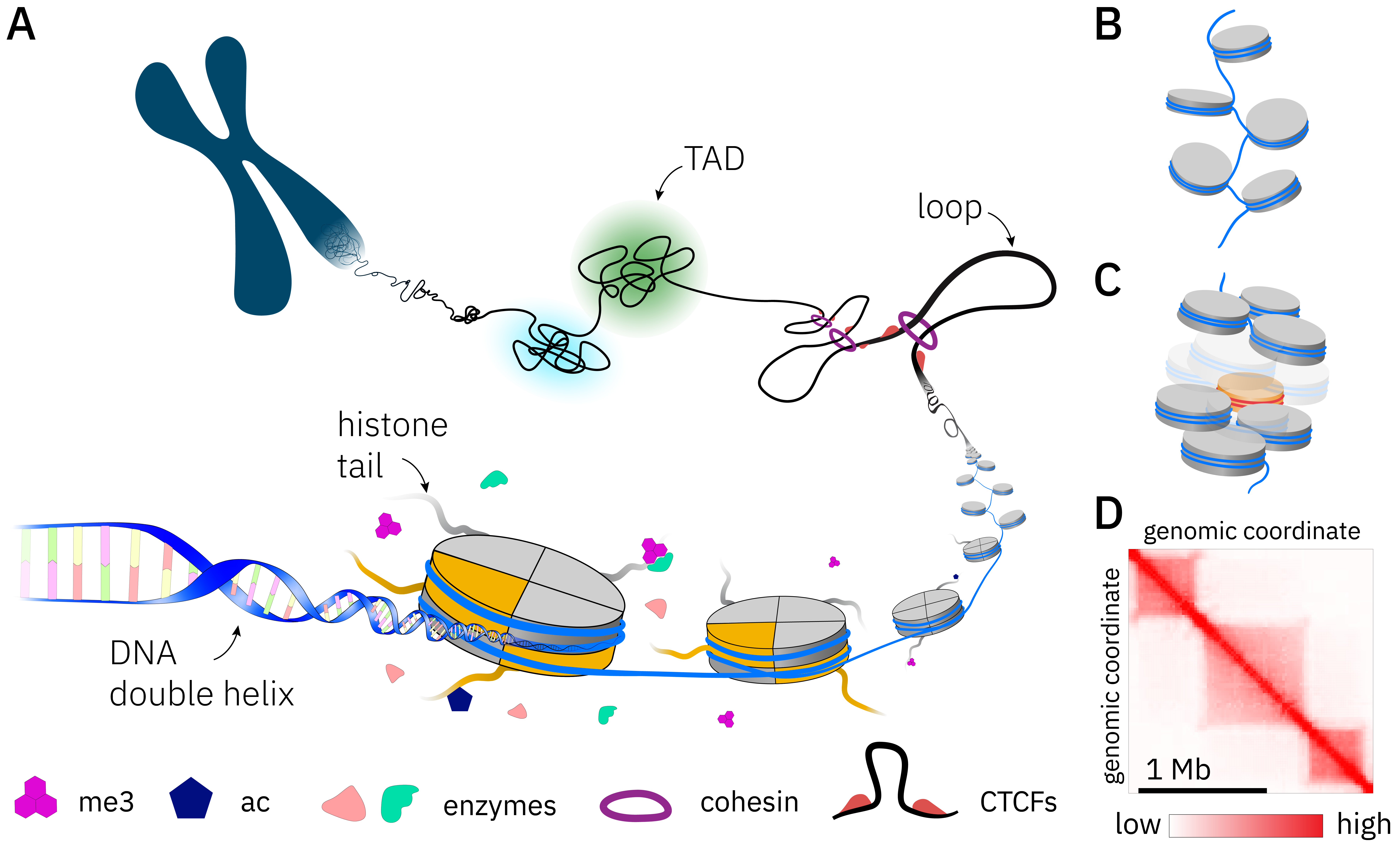}
\caption{\textbf{(A)} A cartoon illustrating the hierarchy of DNA packaging. Organisation of the DNA double helix (in blue) begins with it wrapping around nucleosomes, octamers consisting of eight histone proteins (the H3 core histones are highlighted in orange). The DNA fibre in each structural subunit comprises 147 base pairs (bp) wrapped around the nucleosome core and linker DNA, which can vary from a few nucleotides to 80 bp \citep{gross2015chromatin}. The complex of DNA and its associated proteins is called chromatin. Further DNA compaction involves chromatin coiling to form a denser 30-nm fibre (illustrated in black). Higher-order structures, such as loop domains, can form with the help of additional protein complexes, such as CTCF binding sites and cohesin \citep{gross2015chromatin}. CTCF insulators, oriented in a convergent manner, function as barriers for loop extrusion mediated by ring-shaped cohesin complexes that cannot slide past the CTCF sites. Subsequent compartmentalisation of chromatin involves its separation into Topological Associating Domains (TADs), which are characterised by frequent interactions between genomic loci within the TAD, whereas interactions between different TADs are limited (see also \textbf{(D)}). This schematic also depicts the addition of epigenetic marks, such as trimethyl (me3) and acetyl (ac), to histone terminal tails mediated by specific enzymes. \textbf{(B), (C)} Two schematics showing open and compacted states of the chromatin fibre, respectively. One nucleosome whose DNA cannot be accessed for transcription is coloured in orange (panel \textbf{C}). \textbf{(D)} A characteristic map showing interaction frequencies between distinct genomic loci. High-frequency diagonal interactions correspond to contacts between neighbouring genomic loci, while non-local square domains of increased interaction frequencies correspond to higher-order chromatin organisational units such as TADs \citep{szabo2019principles}.}
\label{Fig1}
\end{center}
\end{figure}

In eukaryotes, packaging genetic information within cells relies on the efficient organisation of DNA. Structured as long polymers, chromosomal DNA must fit within the confines of the cell nucleus while simultaneously ensuring accessibility for transcription of genomic regions containing genes essential for a particular cell phenotype and function. The primary level of packaging involves wrapping the DNA double helix around nucleosomes, protein complexes consisting of eight histones, two of each type --  H2A, H2B, H3, H4 (Figure~\ref{Fig1}A) \citep{gross2015chromatin}. The resulting structure of DNA and its associated proteins resembles `beads on a string' and is called chromatin. Chromatin is compacted by coiling and folding to form a condensed 30-nm fibre. Higher-order packaging includes the formation of loops, facilitated by DNA-binding proteins, such as the CCCTC-binding factor (CTCF) and the cohesin complex \citep{grubert2020landscape} (Figure~\ref{Fig1}A). Loop domains, with sizes ranging from a few kilobases (kb) to hundreds of kb \citep{rao20143d,grubert2020landscape}, help to regulate gene expression by enabling enhancer-promoter interactions and insulating regulatory elements from neighbouring regions to prevent abnormal activities \citep{grubert2020landscape}. At the megabase (Mb) scale, chromatin is organised into Topological Associating Domains (TADs), which are characterised by frequent interactions between genomic loci within the same TAD and limited inter-domain interactions (see Figure~\ref{Fig1}A, D) \citep{szabo2019principles}. Chromatin may undergo further condensation to form chromatids, and two identical sister chromatids together constitute a chromosome.

The packaging of DNA filaments is tightly coupled with the DNA's epigenetic state. Epigenetics regulate which genomic loci are accessible for transcription, without modifying the chromosomal DNA sequence. It involves the addition of specific epigenetic marks to the DNA itself (e.g. DNA methylation) or to the tails of core histones that induce conformational changes in the three-dimensional folding of the chromatin fibre, thus changing its accessibility for binding of transcription factors. One of the most common epigenetic modifications involves covalent modifications of residues of histone tails. Histone tails are long terminal sequences of amino acids which protrude from each core histone and at which post-translational changes can occur (Figures~\ref{Fig1}A) \citep{gross2015chromatin}. As the balance between different epigenetic marks changes, the chromatin architecture also changes to facilitate or inhibit access to genetic information. For example, the addition of acetyl groups to lysine residues at histone tails (acetylation) generally facilitates chromatin accessibility \citep{strahl2000language}. The cartoons in Figures~\ref{Fig1}B and C show open and compacted genomic loci, respectively. While open chromatin conformations allow access to the DNA (panel B), genetic information encoded in loci that are tightly coiled cannot be accessed (panel C). Chromatin architecture, in turn, also influences the epigenetic profiles via feedback between different epigenetic marks \citep{sneppen2019theoretical}. In this way, if the configuration of chromatin folding allows distant genomic loci to interact in space, feedback between existing epigenetic modifications at these loci can alter their state \citep{sneppen2019theoretical}. 

Epigenetics plays a key role in the development and maintenance of organism homeostasis. Without changing the underlying DNA sequence, it enables cells to adjust their gene expression profile according to their specialised function and environmental cues. It is directly implicated in cell differentiation, lineage commitment, maintenance of cell identity and adaptation to environmental changes. Specifically, a conceptual model proposed by \citep{waddington1957strategy} illustrates cell differentiation as a trajectory of a ball rolling down a hilly terrain. Hills and valleys in this metaphor define the epigenetic landscape, with each valley representing a distinct cell phenotype characterised by a specific pattern of open and silenced gene loci. This type of `rugged' epigenetic profile, characterised by alternating patterns of activating and repressive epigenetic modifications, plays a crucial role in determining cellular identity. Dysregulation of epigenetic mechanisms can lead to cancer, neurological disorders, metabolic dysregulation and autoimmune diseases \citep{zoghbi2016epigenetics,tollefsbol2018epigenetics,zhang2020epigenetics}. Despite extensive research \citep{felsenfeld2014brief}, many aspects of epigenetic regulation remain poorly understood. In particular, data are typically obtained as snapshots at specific time points, and little is known about the interplay between the dynamics of epigenetic modifications and chromatin architecture. Theoretical modelling offers a complimentary approach that can be used to gain insight into these interactions and the formation of epigenetic signatures.

Mathematical and computational modelling has been used to understand the mechanisms underlying key features of epigenetic regulation at the scale of small chromatin regions (10-60 nucleosomes, or equivalently $2-12$ kb). These locus-specific models focus on either covalent modifications of histone tails \citep{dodd2007theoretical,sneppen2008ultrasensitive,david2009inheritance,sneppen2012simple,sneppen2019theoretical,zhang2019quantifying}, DNA methylation \citep{sneppen2016nucleosome,chen2021mathematical}, or both \citep{thalheim2017regulatory,thalheim2018cooperation}. A crucial requirement of these models is their ability to exhibit bistability, allowing the formation of epigenetic landscapes corresponding to open and closed chromatin states. These states must remain stable under mitotic perturbations, where approximately half of the nucleosomes are replaced by newly synthesised ones that lack epigenetic modifications. Robustness in transmitting epigenetic information through cell division is referred to as epigenetic memory. One class of theoretical models focuses specifically on the formation of bivalent chromatin (e.g. \citep{zhao2021mathematical,sneppen2019theoretical,alarcon2021}), characterised by a mixture of activating and repressive marks. Bivalent chromatin is believed to preserve the developmental potential of stem cells by keeping key regulatory genes in a poised, undecided state. This poised state allows for rapid and flexible gene expression during development, enabling cells to quickly activate or silence genes in response to signals that determine their specific lineage \citep{macrae2023regulation}. \cite{sneppen2019theoretical} conducted an extensive literature review to summarise known feedback mechanisms between different epigenetic modifications. They then formulated a computational model to investigate whether these interactions could lead to the formation of bivalent chromatin. Their simulation results suggest that the experimentally observed mixture of activating and repressive marks is due to the rapidly switching bistable dynamics rather than the existence of a chromatin state with intermediate levels of both types of epigenetic marks. Another class of models investigates the interplay between epigenetic regulation and gene expression \citep{folguera2019multiscale,chen2021mathematical,alarcon2021}. These models investigate how epigenetic modifications influence gene activity and contribute to the complex regulatory networks governing cellular function and identity.

In recent work, \cite{newar2022dynamical} introduced a stochastic model to study the mechanisms driving the formation of epigenetic patterns of covalent modifications of H3K27 (histone H3 at lysine residue 27) around transcription start sites. By exploring experimental data on enzyme occupancy around target sites, they defined the binding profiles of enzymes that participate in the epigenetic modification of H3K27. The authors also introduced a matrix for contact frequency of interactions between different genomic loci. Nonetheless, for simplicity, they neglected any locus-specific chromatin architecture and assumed that the probability of contact between two genomic coordinates is inversely proportional to the distance along the chromatin polymer that separates them. Their simulation results confirmed that binding profiles of histone-modifying enzymes allow for the emergence of methylation and acetylation patterns in chromatin regions around transcription sites similar to those observed experimentally \citep{newar2022dynamical}. 

Another type of computational model focuses exclusively on the three-dimensional genome architecture. These studies employ polymer-based models to elucidate the specific features of the chromatin polymer required to replicate experimental data on chromatin conformation (e.g. \citep{buckle2018polymer}). More extensive reviews of existing models of epigenetic regulation can be found in \citep{cortini2016physics,ringrose2017dissecting}.

Most existing models \citep{dodd2007theoretical,sneppen2008ultrasensitive,david2009inheritance,sneppen2012simple,sneppen2016nucleosome,thalheim2017regulatory,thalheim2018cooperation,sneppen2019theoretical,zhang2019quantifying} focus on the stability of a uniform epigenetic state in an isolated locus, whereas the mechanisms underlying the formation of rugged epigenetic landscapes, which consist of patterns of open and closed chromatin loci, remain largely unexplored. Additionally, existing models frequently neglect the influence of chromatin architecture on the formation of epigenetic profiles. Interactions between nucleosomes in these models are simplified to either nearest-neighbour or all-to-all interactions. Several models also account for distance-dependent contact between genomic sites, assuming that the frequency at which these contacts occur decreases with separation distance \citep{dodd2007theoretical,sneppen2016nucleosome,newar2022dynamical,nickels2023confinement}. In practice, however, chromatin folding is more complex and can significantly influence the deposition of epigenetic marks through the activity of various `reader-writer' (`reader-eraser') enzymes \citep{sneppen2019theoretical}. These enzymes can recognise existing modifications and catalyse the deposition of new marks (or the removal of existing marks) at contact sites. Thus, chromatin folding can cause non-local feedback mechanisms that affect the spatial distribution of epigenetic modifications.

In this work, we propose a novel mesoscopic approach that considers epigenetic regulation between genomic loci of fixed length (on the scale of kb), each corresponding to DNA wrapped around several nucleosomes. We focus on covalent modifications of histone H3, accounting for covalent modifications of H3 tails at lysine residues 4 and 27 (H3K4 and H3K27, respectively). Trimethylation of H3K27 (H3K27me3) is recognised as a repressive mark, whereas acetylation of this residue (H3K27ac) and trimethylation of H3K4 (H3K4me3) represent activating epigenetic modifications. Despite our focus on these three modifications, we generalise their behaviour to represent broader classes of epigenetic marks, which result in closed and open chromatin states, respectively. Following previous work (e.g. \citep{dodd2007theoretical,david2009inheritance,sneppen2019theoretical}), our model integrates feedback mechanisms governing the addition and removal of these marks, influenced by the availability of catalysing enzymes and also by chromatin folding. We define chromatin architecture, a key input to our model, via a matrix of contact strengths reflecting interactions in 3D space between different genomic sites.

Experimental data for generating interaction frequency heatmaps among genomic loci (Figure~\ref{Fig1}D) can be acquired through chromatin conformation capture techniques. Variants of this method offer insights into chromatin organisation at different scales and resolutions \citep{sati2017chromosome}. For instance, the Hi-C method combines DNA proximity ligation with high-throughput sequencing on a genome-wide scale to capture pairwise interactions between loci \citep{lieberman2009comprehensive}, typically at the resolution of several kb (in the original work, the resolution was 1 Mb \citep{lieberman2009comprehensive}). 
 
The main goal of our work is to understand the mechanisms that drive the formation of rugged (inhomogeneous) epigenetic landscapes, with a particular focus on the roles played by chromatin conformation and enzyme competition. The structure of the paper is as follows. In Section~\ref{sec:model}, we present our modelling framework, formulated in a stochastic setting, to describe the dynamics of epigenetic modifications for a given chromatin conformation. We also derive the mean-field deterministic equations associated with the stochastic model. We use the mean-field equations to perform stability analysis for two interacting genomic sites. This analysis, presented in Section~\ref{sec:results_two_sites}, enables us to identify parameter regimes corresponding to different epigenetic landscapes: open, closed, bivalent, and rugged epigenetic landscapes. Section~\ref{sec:results_scaling} discusses the scalings which must be introduced to extend our model to multiple interacting sites. The multi-site model accounts for chromatin architecture, availability of enzymes, and their competition for binding to target sites. We then perform stability analysis and stochastic simulations for the particular case of two interacting regions spanning several genomic sites (Section~\ref{sec:results_twosubregion}). We conclude in Section~\ref{sec:discussion} with a summary of our findings and directions for future work.

\section{Materials and methods} \label{sec:model}

\begin{figure}[!t]
\begin{center}
\includegraphics[width=\textwidth]{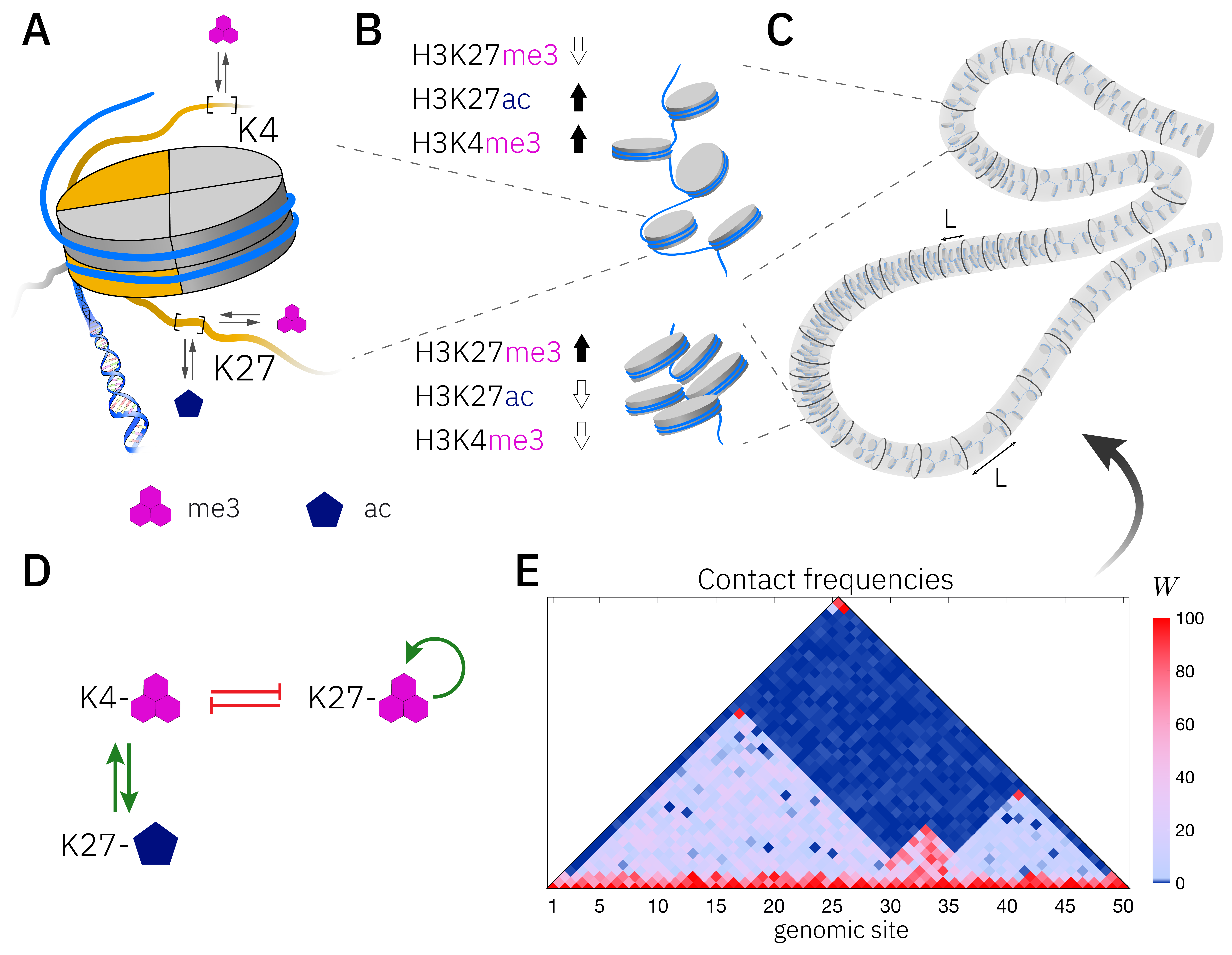}
\caption{A graphical summary of the modelling components. \textbf{(A)} A schematic highlighting the epigenetic modifications of lysine residues of the H3 histone (in orange) considered in our model. \textbf{(B)} We employ a mesoscopic approach in which the chromatin polymer is divided into genomic regions of fixed length, $L$ (kb), which may contain multiple nucleosomes. Each genomic region can be characterised by its epigenetic landscape. In particular, compacted chromatin regions are characterised by high levels of H3K27me3 repressive marks, whereas open chromatin contains high levels of activating marks, H3K4me3 and H3K27ac. \textbf{(C)} A cartoon illustrating the division of the chromatin fibre into subregions of constant length, $L$. \textbf{(D)} Epigenetic marks are known to affect the dynamics of histone modifications via multiple reinforcing and antagonistic feedback mechanisms that are mediated by reader-writer (-eraser) enzymes \citep{sneppen2019theoretical}. This diagram depicts the crosstalk between different epigenetic marks accounted for in this work. Green arrows indicate reinforcing interactions, while red, flat-end arrows indicate inhibitory interactions. \textbf{(E)} An example of a contact frequency map defining chromatin folding (see \textbf{(C)}), which serves as an input parameter in our model. Only the upper-diagonal part of the matrix, $W$, is shown due to its symmetry.}
\label{Fig2}
\end{center}
\end{figure}

In this section, we describe our model for epigenetic regulation (see Figure~\ref{Fig2}). As mentioned above, we focus on the epigenetic regulation of two lysine residues of histone H3, namely, H3K4 and H3K27. The dynamics of epigenetic marks at these two sites attract particular attention due to their role in regulating gene expression \citep{miller2012role}. Figure~\ref{Fig2}A illustrates those post-translational modifications of H3 residues that are included in the model -- trimethylation of H3K4 and H3K27 (H3K4me3 and H3K27me3, respectively) and acetylation of H3K27 (H3K27ac). Although we consider only three epigenetic modifications, we view them as representative of generic classes of marks with similar behaviour. For simplicity, we do not account for the intermediate stages of adding and removing individual methyl groups.

We also take into account the spatial folding of the chromatin fibre. Specifically, we decompose the chromatin polymer into fixed-length genomic regions that can include multiple nucleosomes (see Figure~\ref{Fig2}B-C). In this study, we assume that each genomic site spans a DNA chain of length $L$ kb (Figure~\ref{Fig2}C). This mesoscopic approach enables us to account for non-local interactions between distinct epigenetic marks. The pattern of these non-local interactions is given by the shape of chromatin folding, which can allow two genomic regions to be in close physical proximity without being neighbours in the DNA chain. Figure~\ref{Fig2}C depicts a simple domain with loops which bring into close 3D proximity genomic loci that are separated by many nucleotides along the genome. In general, interactions between different genomic sites are specified by contact frequency maps similar to those obtained through chromatin conformation capture techniques. Figure~\ref{Fig2}E shows an example of such a map. Since the contact frequency matrix is symmetric, it suffices to plot only its upper triangular part, which is commonly rotated by $45^{\circ}$ to align the main diagonal horizontally. This approach is widely used in experimental studies and will be employed in our work (Figure~\ref{Fig2}E). Higher frequency contacts at the diagonal entries (along the $x$-axis) indicate more frequent interactions between neighbouring loci, while loops and TADs are characterised by high contact frequencies off the main diagonal. In our work, the contact frequency map, $W$, is assumed to be fixed. While epigenetic marks are known to affect chromatin folding, here, we consider only the forward problem of determining the epigenetic landscape for a given (fixed) chromatin architecture.

Each genomic site is characterised by its pattern of epigenetic marks. In particular, closed chromatin regions are known to carry H3K27me3 repressive marks (Figure~\ref{Fig2}B, bottom panel), while open chromatin is typically characterised by acetylation of H3K27 and trimethylation of H3K4 (see Figure~\ref{Fig2}B, upper panel). Regions with excess of H3K27ac and H3K4me3 marks are associated with activation of the gene expression within those regions. We will also distinguish bivalent chromatin, which is characterised by a mixture of repressive and activating marks. 

In what follows, $i = 1,\ldots, N$ indicates the index of a particular genomic region, where $N$ is the total number of regions considered. Table~\ref{Table1} lists the epigenetic modifications considered in our model along with the enzymes responsible for adding and removing these marks \citep{sneppen2019theoretical}. Specifically, the addition of trimethyl groups to H3K4 is catalysed by the Mixed-Lineage Leukemia (MLL) complex of the Trithorax group, while removal of this mark is mediated by the histone lysine demethylase (KDM) family of enzymes. For the lysine residue H3K27, we include its trimethylation by the methyltransferase EZH2 (Enhancer of Zeste 2), which is an enzymatically-active component of the Polycomb Repressor Complex 2 (PRC2). Removal of this mark is mediated by the demethylase UTX (Ubiquitously Transcribed Tetratricopeptide Repeat on chromosome X). The acetyltransferase CBP (CREB-binding protein) catalyses addition of the acetyl group to H3K27, whereas deacetylation of this residue is mediated by the NuRD (Nucleosome Remodeling and Deacetylase) complex and the sirtuin family (SIRT) of proteins \citep{park2020short}. A comprehensive review of histone-modifying enzymes and the feedback mechanisms between them for Drosophila and vertebrates is provided in \citep{sneppen2019theoretical}.

\begin{table}[h!]
\centering
\begin{tabular}{| p{2.3cm} | p{2.8cm} | p{3.1cm} | p{3.7cm}|}
\hline
\textbf{Unmodified residues}                                                                 & \textbf{Modifications}  & \textbf{Writer enzyme (unbound)}                           & \textbf{Eraser enzyme (unbound)} \\ \hline
H3K4, \textcolor{dblue}{$S_4$} & 
H3K4me3, \textcolor{dblue}{$P_A$}  & 
MLL, \textcolor{dblue}{$E_M$}  
& KDM, \textcolor{dblue}{$E_K$} \\ \hline
\multicolumn{1}{| l |}{\multirow{2}{*}{H3K27, \textcolor{dblue}{$S_{27}$}}} & H3K27me3, \textcolor{dblue}{$P_I$} & EZH2, \textcolor{dblue}{$E_Z$} & UTX, \textcolor{dblue}{$E_U$}  \\ \cline{2-4}
\multicolumn{1}{|c|}{}                                                                      & H3K27ac, \textcolor{dblue}{$Q$}    & CBP, \textcolor{dblue}{$E_C$}  & NuRD and SIRT, \textcolor{dblue}{$E_S$} \\ \hline
\end{tabular}%
\caption{Histone modifications and histone-modifying enzymes considered in the model. The corresponding model variables are indicated in blue. We detail enzymes known to add and remove epigenetic marks in vertebrates \citep{sneppen2019theoretical}.}\label{Table1}
\end{table}

We model (de-)methylation and (de-)acetylation of lysine residues as enzymatic reactions catalysed by (eraser) writer enzymes (see Table~\ref{Table1}). Each reaction leads to the formation of an intermediate lysine residue-enzyme complex, $C_{J_i}$, which later yields the product reactant and the enzyme that initiated that reaction. If we introduce the subscript, $J \in \uf{M,Z,C,K,U,S}$ (see Table~\ref{Table1}) to indicate the enzyme that catalyses a particular reaction, then the following reactions represent the addition of epigenetic marks

\begin{align} \label{eq_add_marks_27me}
& S_{27_i} + E_Z \stacksymbol{\rightleftarrows}{\textcolor{burgundy}{\bar{k}_{1i}}}{\textcolor{burgundy}{\bar{k}_{2i}}} C_{Z_i} \stacksymbol{\rightarrow}{\textcolor{burgundy}{\bar{k}_{3}}}{} P_{I_i} + E_Z, \\ \label{eq_add_marks_4me}
& S_{4_i} + E_M \stacksymbol{\rightleftarrows}{\textcolor{burgundy}{\bar{k}_{4i}}}{\textcolor{burgundy}{\bar{k}_{5i}}} C_{M_i} \stacksymbol{\rightarrow}{\textcolor{burgundy}{\bar{k}_{6}}}{} P_{A_i} + E_M, \\ \label{eq_add_marks_27ac}
& S_{27_i} + E_C \stacksymbol{\rightleftarrows}{\textcolor{burgundy}{\bar{k}_{7i}}}{\textcolor{burgundy}{\bar{k}_{8}}} C_{C_i} \stacksymbol{\rightarrow}{\textcolor{burgundy}{\bar{k}_{9}}}{} Q_{{}_i} + E_C,
\end{align}
\noindent and the following reactions their removal:

\begin{align} \label{eq_remove_marks_27me}
& P_{I_i} + E_U \stacksymbol{\rightleftarrows}{\textcolor{burgundy}{\bar{k}_{10}}}{\textcolor{burgundy}{\bar{k}_{11}}} C_{U_i} \stacksymbol{\rightarrow}{\textcolor{burgundy}{\bar{k}_{12}}}{} S_{27_i} + E_U, \\ \label{eq_remove_marks_4me}
&  P_{A_i} + E_K \stacksymbol{\rightleftarrows}{\textcolor{burgundy}{\bar{k}_{13}}}{\textcolor{burgundy}{\bar{k}_{14}}} C_{\bar{k}_i} \stacksymbol{\rightarrow}{\textcolor{burgundy}{\bar{k}_{15}}}{} S_{4_i} + E_K, \\ \label{eq_remove_marks_27ac}
& Q_{{}_i} + E_S \stacksymbol{\rightleftarrows}{\textcolor{burgundy}{\bar{k}_{16}}}{\textcolor{burgundy}{\bar{k}_{17}}} C_{S_i} \stacksymbol{\rightarrow}{\textcolor{burgundy}{\bar{k}_{18}}}{} S_{27_i} + E_S. 
\end{align}
 
The reaction rates in Eqs~\eqref{eq_add_marks_27me}-\eqref{eq_add_marks_27ac} are stated in a way that emphasises reinforcing and antagonistic interactions of unmodified lysine residues with existing histone modifications \citep{sneppen2019theoretical}. Figure~\ref{Fig2}D illustrates the feedback mechanisms incorporated in our model. The crosstalk between different histone modifications is mediated by specific reader-writer and reader-eraser enzyme complexes. These complexes can recognise existing epigenetic marks and catalyse new modifications at neighbouring genomic loci. Chromatin folding can enhance these interactions by bringing distant genomic sites into close spatial proximity (see Figure~\ref{Fig2}E). 

Specifically, we assume that trimethylation of H3K27 is self-reinforcing \citep{sneppen2019theoretical}. Consequently, reaction rate, $\bar{k}_{1i}$, in Eq~\eqref{eq_add_marks_27me}, is upregulated if the same mark, H3K27me3, is present in the vicinity of genomic site $i$ (see Eq~\eqref{eq_kinetic_constants}). The proximity in 3D space of two genomic loci is defined by their contact frequency (see Figure~\ref{Fig2}E). Modifications H3K27me3 and H3K4me3 are known to inhibit each other \citep{sneppen2019theoretical} (Figure~\ref{Fig2}D). Therefore, we assume that the dissociation rates of the lysine residue-enzyme complex, $\bar{k}_{2i}$ and $\bar{k}_{5i}$, (from Eqs~\eqref{eq_add_marks_27me}-\eqref{eq_add_marks_4me}) are increasing functions of their antagonising epigenetic mark (H3K4me3 and H3K27me3, respectively; see Eq~\eqref{eq_kinetic_constants}). Additionally, the activating marks, H3K4me3 and H3K27ac, have been reported to reinforce each other's production \citep{sneppen2019theoretical} (Figure~\ref{Fig2}D). Therefore, we assume that the rates, $\bar{k}_{4i}$ and $\bar{k}_{7i}$, for the formation of complexes to catalyse the deposition of trimethyl in H3K4 and acetyl in H3K27, respectively, (Eqs~\eqref{eq_add_marks_4me}-\eqref{eq_add_marks_27ac}) are functions of the corresponding reinforcing mark. The above feedback mechanisms are incorporated into our model as follows:

\begin{align}
\bar{k}_{1i} & = \bar{k}_1 \of{\bar{k}_{1_0} + \displaystyle \sum_{l=1}^{N} {\overline{w}_{il}} P_{I_l}}, \qquad &&
\bar{k}_{2i}  = \bar{k}_2 \of{\bar{k}_{2_0} + \displaystyle \sum_{l=1}^{N} {\overline{w}_{il}} P_{A_l}}, \nonumber \\
\bar{k}_{4i} & = \bar{k}_4 \of{\bar{k}_{4_0} + \displaystyle \sum_{l=1}^{N} {\overline{w}_{il}} Q_{{}_l}}, &&
\bar{k}_{5i}  = \bar{k}_5 \of{\bar{k}_{5_0} + \displaystyle \sum_{l=1}^{N} {\overline{w}_{il}} P_{I_l}}, \label{eq_kinetic_constants} \\
\bar{k}_{7i} & = \bar{k}_7 \of{\bar{k}_{7_0} + \displaystyle \sum_{l=1}^{N} {\overline{w}_{il}} P_{A_l}}. \nonumber
\end{align}
\noindent The remaining reaction rates in Eqs~\eqref{eq_add_marks_27me}-\eqref{eq_remove_marks_27ac}, $\bar{k}_{r}$ for $r = \uf{3,6,8,9,\ldots,18}$, are assumed to be constant. In Eq~\eqref{eq_kinetic_constants}, $(\bar{k}_r \bar{k}_{r_0})$ is the baseline rate of the reaction $r = 1,2,4,5,7$. The matrix $\overline{W} = \uf{\overline{w}_{il}}_{il}$ accounts for the spatial embedding of the chromatin fibre in 3D space and describes the pattern of interactions among genomic regions, both proximal and distant along the fibre. In this work, we will consider two specific choices for $\overline{W}$. In the first scenario, we assume that the interaction strength between genomic sites $i$ and $l$ is determined by their contact frequency, $w_{il}$, and the level of H3K27me3 at those sites. The entries in contact frequency matrix, $W= \uf{w_{il}}_{il}$, illustrated in Figure~\ref{Fig2}E, are assumed to be fixed. These data can be obtained experimentally using chromatin conformation capture techniques. The formation of loops and non-local interactions between chromatin loci has been reported to be mediated by Polycomb Repressor Complex 1 (PRC1) recruited by trimethylation of H3K27 \citep{loubiere2019cell}. This motivated us to assume that the contact strength between sites $i$ and $l$ is proportional to the level of this histone modification, i.e. $\overline{w}_{il} = w_{il} P_{I_i} P_{I_l}$. For the second scenario, we assume that the interaction strength depends solely on the contact frequency between the relevant genomic loci, i.e. $\overline{w}_{il} = w_{il}$. Nonetheless, for most of our results, we use the H3K27me3-dependent interaction between genomic sites, unless otherwise stated.

The reactions, given by Eqs~\eqref{eq_add_marks_27me}-\eqref{eq_remove_marks_27ac} are complemented by the conservation laws for the total number of lysine residues (unmodified and modified) at each genomic region $i$,

\begin{align} \label{conservation_law_lysine27}
S_{27_0} & = S_{27_i} + P_{I_i} + Q_{{}_i} + C_{Z_i} + C_{U_i} + C_{C_i} + C_{S_i}, \\ \label{conservation_law_lysine4}
S_{4_0} & = S_{4_i} + P_{A_i} + C_{M_i} + C_{K_i},
\end{align}
\noindent and the total numbers of the writer (eraser) enzymes,

\begin{align} \label{conservation_law_enzyme}
E_J + \displaystyle \sum_{l=1}^N C_{J_l} = E_{J_0}, \qquad J \in \uf{M,Z,C,K,U,S},
\end{align}
\noindent where the summation is taken over all the genomic regions, $l = 1,\ldots,N$. Here, $E_{J_0}$, $ J \in \uf{M,Z,C,K,U,S}$, represents the total amount of enzyme $J$ in the system. We note that our approach explicitly accounts for the competition of enzyme binding to the available lysine residues. 

To simplify the model and reduce its dimensionality, we follow the Briggs-Haldane approach and assume the following scalings for the model variables \citep{keener2009mathematical}. Lysine residues (modified and unmodified) scale with $\Omega_S$ representing the characteristic number of lysine residues in a genomic site, i.e. $\of{S_{27_i}, S_{4_i}, P_{I_i}, P_{A_i}, Q_{{}_i} } = \Omega_S \of{s_{27_i}, s_{4_i}, p_{I_i}, p_{A_i}, q_{{}_i}}$ for all $i$. By contrast, enzymes and the complexes they form scale with the characteristic enzyme level, $\Omega_E$, which leads to $\of{E_J, E_{J_0}, C_{J_i}} = \Omega_E \of{e_J, e_{J_0}, c_{J_i}}$ for all enzyme types, $J \in \uf{M,Z,C,K,U,S}$, and all genomic sites, $i$. We assume $\Omega_S \gg \Omega_E$. Under this scaling, the dynamics of intermediate enzymatic complexes are fast compared to those associated with product formation. As such, we can assume that the enzyme distributions are at a quasi-equilibrium on the slow timescale of epigenetic modifications. In the stochastic version of the quasi-steady state (QSS) approximation \citep{ball2006asymptotic,folguera2019multiscale,alarcon2021}, the fast variables quickly settle on to their QSS distribution, which is conditioned to the (instantaneous) values of the slow variables. The evolution of the slow variables takes place on a much slower timescale than that required for the fast variables to reach the quasi-equilibrium. Their evolution takes a simplified form where the values of the fast variables are sampled from their QSS distribution. It can be shown (see Section~I of Supplementary Material; \ref{app1}) that this results in multinomial QSS distributions for the probabilities of finding an enzyme $J$ in a complex at one of the genomic sites or in a free state. The corresponding QSS probability density function (PDF), $\phi_{J}$, reads:
\begin{adjustwidth}{-1cm}{-0.0cm}
\begin{align} \label{dist_multinomial_enzymeJ}
\phi_{J} (E_J, C_{J_1},\ldots,C_{J_N} \mid \mathbf{p_I},\mathbf{p_A},\mathbf{q}) = \frac{\of{E_{J_0}}!}{\of{E_J}! ~ \displaystyle \prod_{l = 1}^N \of{C_{J_l}}!} ~ \of{p_{E_J}}^{E_J} ~ \displaystyle \prod_{l = 1}^N \of{p_{C_{J_l}}}^{C_{J_l}},
\end{align}
\end{adjustwidth}
\noindent where $E_{J_0}$ is the total amount of enzyme $J$ (free and bound in complexes) and $p_{E_J}$ and $p_{C_{J_l}}$ are the probabilities that an enzyme of this type is in a free or bound state at genomic site $l$, $l = 1,\ldots, N$. In particular, we have that $p_{E_J} + \sum_{l=1}^N p_{C_{J_l}} = 1$. Furthermore, in Eq~\eqref{dist_multinomial_enzymeJ}, we note the dependency of the QSS PDFs on the levels of existing histone modifications, $(\mathbf{p_I},\mathbf{p_A},\mathbf{q})$, where $\mathbf{p_I} = \uf{p_{I_1}, \ldots, p_{I_N}}$ denotes the H3K27me3 landscape, $\mathbf{p_A} = \uf{p_{A_1}, \ldots, p_{A_N}}$ denotes the levels of H3K4me3 marks and $\mathbf{q} = \uf{q_{{}_1}, \ldots, q_{{}_N}}$ denotes the H3K27ac profile. A detailed derivation of the QSS approximation is included in Section~I of Supplementary Material (see \ref{app1}). For instance, the QSS probabilities for free and bound EZH2 methyltransferase, $p_{E_Z}$ and $p_{C_{Z_i}}$, are given by the following expressions:

\begin{align} \label{dist_multinomial_prob_enzymeZ}
p_{E_Z} (\mathbf{p_I},\mathbf{p_A},\mathbf{q}) &= \frac{1}{1+\displaystyle \sum_{j=1}^N \frac{\kappa_{1j}}{\kappa_{2j} + \kappa_3}(s_{27_0} - p_{I_j} - q_{{}_j})}, \\  \label{dist_multinomial_prob_enzymeZ2}
 p_{C_{Z_i}} (\mathbf{p_I},\mathbf{p_A},\mathbf{q}) &= \frac{\frac{\kappa_{1i}}{\kappa_{2i}+\kappa_3}(s_{27_0} - p_{I_i} - q_{{}_i})}{1+\displaystyle \sum_{j=1}^N \frac{\kappa_{1j}}{\kappa_{2j} + \kappa_3}(s_{27_0} - p_{I_j} - q_{{}_j})},
\end{align}
\noindent where $\kappa_r$, $r = \uf{1i, 2i, 3}$, are appropriately rescaled reaction rates. Here, the number of unmodified H3K27 residues available for enzyme binding at genomic region $i$ is approximated by $(s_{27_0} - p_{I_i} - q_{{}_i})$.

The reactions for addition and removal of epigenetic modifications of the reduced model under the assumption of QSS distribution of enzyme complexes are then simplified as follows: 

\begin{align} \label{marks_reduced}
& S_{27_i} \stacksymbolL{\rightleftarrows}{\textcolor{burgundy}{{\small{\kappa_3 c_{Z_i}}}}}{\textcolor{burgundy}{{\small{\kappa_{12} c_{U_i}}}}} P_{I_i} , \qquad \qquad
S_{4_i}  \stacksymbolL{\rightleftarrows}{\textcolor{burgundy}{{\small{\kappa_{6} c_{M_i}}}}}{\textcolor{burgundy}{{\small{\kappa_{15} c_{K_i}}}}} P_{A_i}, \qquad \qquad
S_{27_i} \stacksymbolL{\rightleftarrows}{\textcolor{burgundy}{{\small{\kappa_{9} c_{C_i}}}}}{\textcolor{burgundy}{{\small{\kappa_{18} c_{S_i}}}}} Q_{{}_i}.
\end{align}
\noindent Here, $\kappa_r$, $r = \uf{3,6,9,12,15,18}$, are appropriately rescaled reaction rates. These reactions can be simulated using any algorithm for stochastic simulations (e.g. Gillespie algorithm \citep{gillespie1976general}). At each time step of these simulations, the levels of enzymatic complexes, $c_{J_i}$, of type $J$ are sampled from the QSS multinomial distribution (see \ref{app1}). Thus, our stochastic model consists of reactions for the dynamics of epigenetic marks given by Eq~\eqref{marks_reduced} complemented by appropriate initial conditions for each genomic site $i$, $P_{I_i}(t=0)$, $P_{A_i}(t=0)$ and $Q_{{}_i}(t=0)$.

We also derive the associated mean-field equations for the dynamics of epigenetic modifications which we use to perform the stability analysis of our system. The resulting system of coupled ordinary differential equations is given by: 

\begin{align}
\dv{p_{I_i}}{\tau} &=  \kappa_3 e_{Z_0} p_{C_{Z_i}} (\mathbf{p_I},\mathbf{p_A},\mathbf{q}) - \kappa_{12} e_{U_0} p_{C_{U_i}} (\mathbf{p_I}),  \nonumber \\
\dv{p_{A_i}}{\tau} &= \kappa_6 e_{M_0} p_{C_{M_i}} (\mathbf{p_I},\mathbf{p_A},\mathbf{q}) - \kappa_{15} e_{K_0} p_{C_{K_i}} (\mathbf{p_A}), \label{eq_marks_meanfield} \\
\dv{q_{{}_i}}{\tau} &= \kappa_9 e_{C_0} p_{C_{C_i}} (\mathbf{p_I},\mathbf{p_A},\mathbf{q}) - \kappa_{18} e_{S_0} p_{C_{S_i}} (\mathbf{q}). \nonumber
\end{align}
\noindent Here, $i = 1, \ldots, N$ and we emphasise the dependency of the QSS probabilities, $p_{C_{J_i}}$, on the levels of epigenetic marks along chromatin, $(\mathbf{p_I},\mathbf{p_A},\mathbf{q})$, leading to nonlocal coupling between genomic sites. These equations are supplemented with initial conditions $\mathbf{p_I}(t=0)$, $\mathbf{p_A}(t=0)$ and $\mathbf{q}(t=0)$. In our stochastic simulations, unless otherwise stated, we start with random initial conditions, where the initial levels of epigenetic marks are uniformly distributed on $[0,1]$.

\section{Results} \label{sec:results}
\subsection{Emergence of rugged epigenetic landscapes: Bifurcation analysis of the two-site model}  \label{sec:results_two_sites}

As can be seen from Section~\ref{sec:model}, our model contains a large number of parameters. Although estimates exist for the (de-) methylation and (de-) acetylation rates of individual lysine residues (e.g. see \citep{newar2022dynamical} and references therein) for several cell types, little is known about the feedback between epigenetic modifications and the effects of chromatin folding. Epigenetic landscapes and chromatin organisation can vary significantly between organisms and cell differentiation stages. For these reasons, we focus on understanding the qualitative behaviour of our model without attempting to calibrate it against experimental data. To ensure that our predictions are robust, we performed an extensive parameter sweep to discern different modes of behaviour as enzyme levels, $e_{J_0}$, and contact frequencies, $w_{il}$, vary. We note that all Supplementary Figures and Tables can be found in Section~V of Supplementary Material (see \ref{app1}).

\begin{figure}[!h]
\begin{center}
\includegraphics[width=0.67\textwidth]{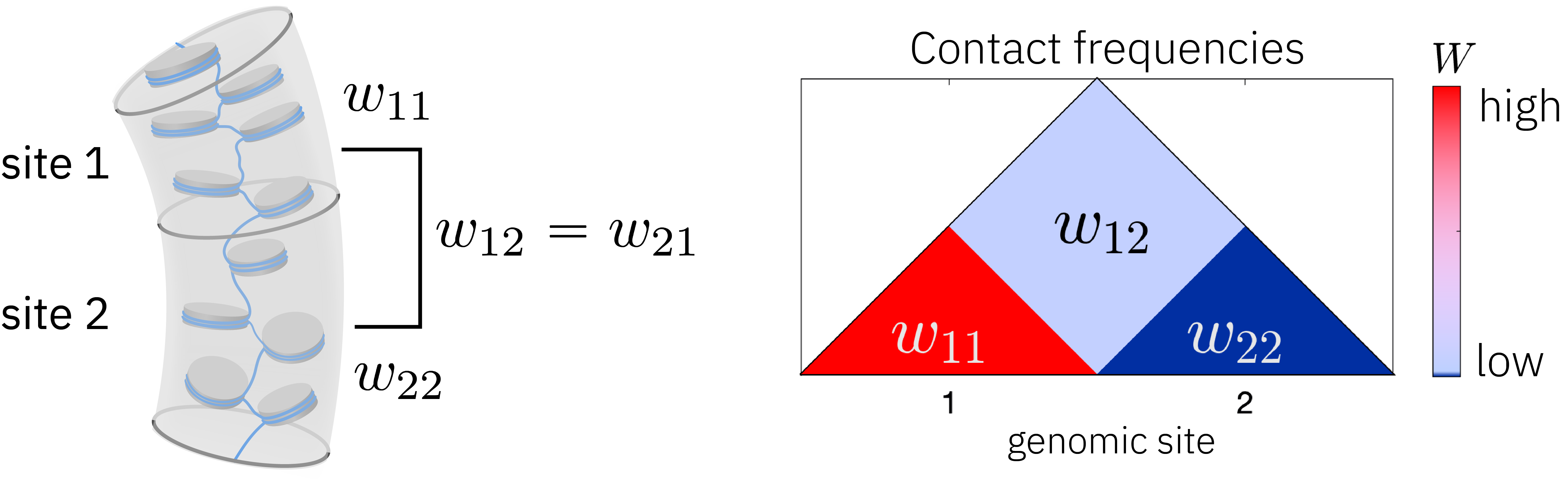}
\caption{A cartoon illustrating the two-site system with the frequency of interactions within site $1$ (site $2$), $w_{11}$ ($w_{22}$), and the cross-interaction between the sites given by $w_{12}$. The cross-interactions are symmetric, i.e. $w_{21} = w_{12}$.}
\label{Fig3}
\end{center}
\end{figure}

We start by performing a model analysis for the simplified case when the entire chromatin domain under consideration is split into two interacting genomic loci, as shown in Figure~\ref{Fig3}. Specifically, we used the mean-field, deterministic equations associated with our stochastic model (see Section~II of Supplementary Material, \ref{app1}). Within this simplified scenario, we can perform an exhaustive bifurcation analysis of the system regarding the stable epigenetic landscapes that our model can produce. We are particularly interested in proving that our model produces both uniform and rugged epigenetic landscapes and in determining the parameter regimes that lead to either of them. Rugged epigenetic profiles are characterised by an alternating pattern of domains enriched in activating marks and regions exhibiting higher levels of repressive marks. We introduce a ruggedness measure, $\mathcal{R}$, which quantifies the heterogeneity of the epigenetic profiles for the two genomic sites: 

\begin{align} \label{rug_measure}
\mathcal{R} = \displaystyle \max_{s.s.s.} \of{\abs{p_{I_1} - p_{I_2}}}.
\end{align}
\noindent Here, the maximum is taken over all stable steady states (s.s.s.) of the system for a given parameter set. Although this measure quantifies differences in H3K27me3 levels between sites, similar metrics can be considered for activating epigenetic marks, H3K4me3 and H3K27ac, due to their inverse relationship with H3K27me3 (see Figure~\ref{Fig2}B).

\begin{figure}[htbp]
\vspace{-2.5cm}
\begin{adjustwidth}{-1cm}{0cm}
\includegraphics[width=1.2\textwidth]{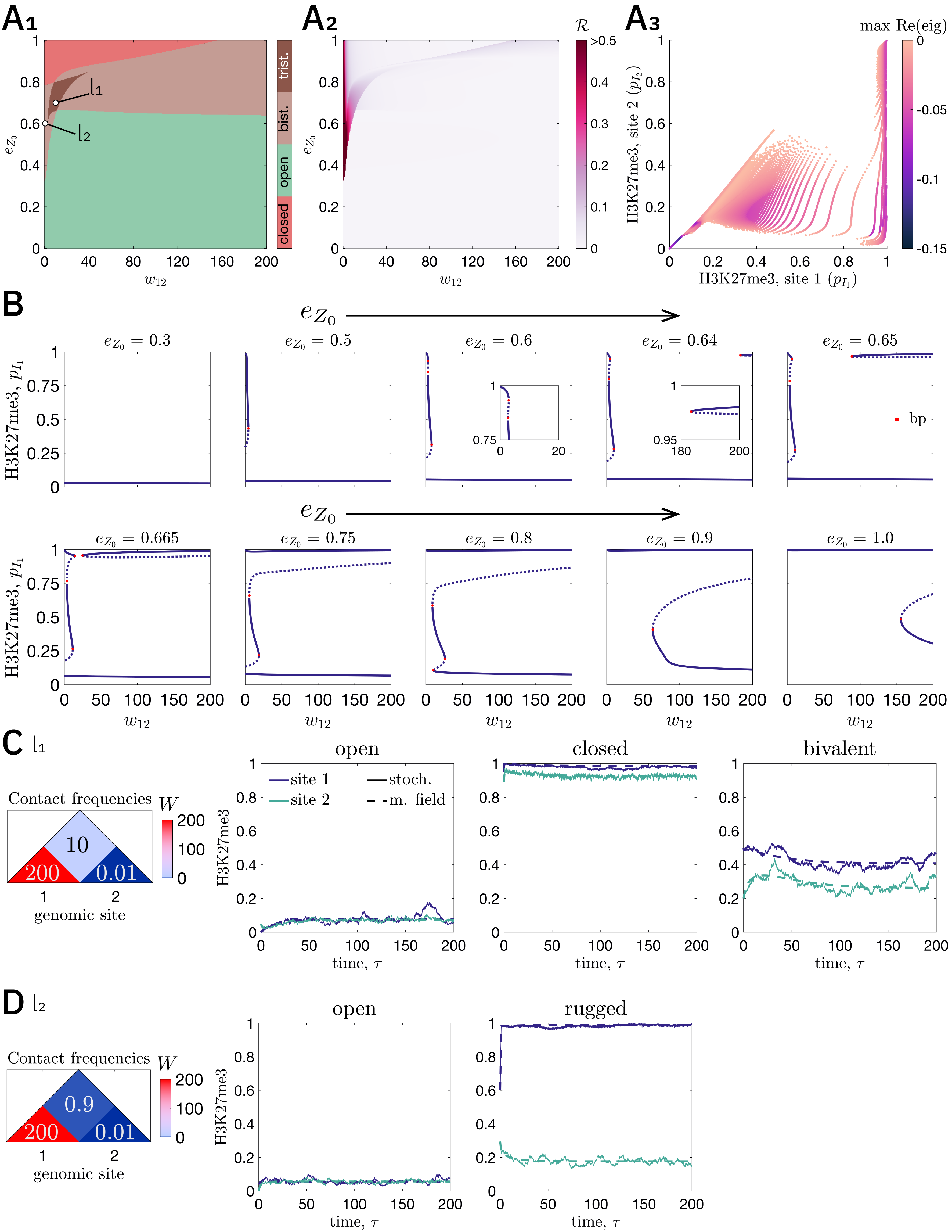}
\end{adjustwidth}
\captionsetup{singlelinecheck=off,justification=raggedright}
\caption{  \textit{(Caption on the next page.)}}
\end{figure}
\addtocounter{figure}{-1}
\makeatletter
\setlength{\@fptop}{0pt}
\makeatother
\begin{figure} [t!]
\caption{\textbf{(A)} Stability analysis of the two-site system in response to changes in levels of the enzyme EZH2, $e_{Z_0}$, and the interaction strength between sites 1 and 2, $w_{12}$. \textbf{(A\textsubscript{1})} A state space diagram where the monostable region corresponding to open (closed) chromatin is shown in green (red). A state is classified as open or closed depending on the equilibrium values of the epigenetic marks at site 1; an open (closed) state corresponds to $p_{I_1}<0.3$ ($p_{I_1}>0.7$). Bistable (tristable) parameter regions are indicated in (dark) brown. \textbf{(A\textsubscript{2})} A heatmap of the ruggedness metric, $\mathcal{R}$, defined by Eq~\eqref{rug_measure}. \textbf{(A\textsubscript{3})} A scatter plot for the values of H3K27me3 at sites 1 and 2 for all stable equilibria obtained by varying $e_{Z_0}$ and $w_{12}$. The colour bar indicates the maximum value of the real part of the eigenvalues of the linearised system. \textbf{(B)} A series of bifurcation plots showing how, for fixed values of $e_{Z_0}$, levels of H3K27me3 at site 1 change as $w_{12}$ varies. Saddle-node bifurcation points are indicated by red dots (bp). \textbf{(C)}, \textbf{(D)} Temporal evolution of the two-site model for two choices of parameter values: $e_{Z_0} = 0.75$, $w_{12} = 10$ (labelled $l_1$ in \textbf{(A\textsubscript{1})}) and $e_{Z_0} = 0.6$, $w_{12} = 0.9$ ($l_2$ label in \textbf{(A\textsubscript{1})}). In each case, the stable equilibrium to which the system converges depends on the initial distribution of marks. We include trajectories obtained from stochastic simulations (solid lines) and the associated mean-field equations (dashed lines) for site 1 (deep blue) and site 2 (green). Except for $e_{Z_0}$ and $w_{il}$, all parameters are fixed at the baseline levels listed in Supplementary Table~1.} 
\label{Fig4}
\end{figure}

We begin by imposing heterogeneous behaviour between the two sites, assuming that one site has a high frequency of interactions while the other is weakly connected. Thus, we fix $w_{11} = 200$ and $w_{22} = 0.01$ and vary the strength of the cross-interaction, $w_{12}$, and enzyme levels. In Supplementary Table~1, we list the baseline values of parameters used in our model, except where we explicitly note the variations in certain parameters. Figure~\ref{Fig4} illustrates the characteristic behaviour of our system as we vary levels of EZH2 enzyme, $e_{Z_0}$, and the interaction strength between the sites, $w_{12}$. The state space diagram shown in Figure~\ref{Fig4}A\textsubscript{1} delineates different stability regimes as these parameters change. 

We determine that low levels of the methyltransferase, $e_{Z_0}$, are insufficient to induce trimethylation of H3K27, resulting instead in mostly hypomethylated chromatin configurations. At higher values of $e_{Z_0}$, the system switches to a stable steady state corresponding to hypermethylated chromatin. A rugged pattern, where one site is open, and the other closed, appears when $e_{Z_0} \gtrsim 0.3$ and cross-interactions, $w_{12}$, are low (see Figure~\ref{Fig4}A\textsubscript{2}). Additionally, given the values of $w_{11}$ and $w_{22}$, site $1$ is more connected and tends to be closed (high H3K27me3), while site 2 remains open (low H3K27me3). This pattern is shown in Figure~\ref{Fig4}A\textsubscript{3}, which reports the steady-state levels of this mark at both sites, $p_{I_1}$ and $p_{I_2}$. The colouring in this plot indicates the maximum value of the real part of the eigenvalues of the linearised system for the corresponding steady state, which describes its local stability.

The transition to bi- and tristability can be better observed in the bifurcation plots shown in Figure~\ref{Fig4}B. Here, we demonstrate the bifurcation curve for H3K27me3 levels at site 1, $p_{I_1}$, as a function of the interaction strength between the sites, $w_{12}$. The level of EZH2 in each panel is fixed to the value indicated in its title. The system exhibits several saddle-node bifurcations, which facilitate the appearance of the closed state for site 1 for higher values of EZH2 methyltransferase, $e_{Z_0}$. 

It is noteworthy that our model also reproduces the emergence of the so-called bivalent state, which is characterised by intermediate levels of negative (H3K27me3) and positive marks (H3K4me3 and H3K27ac). The bivalent state appears less robust than the open and closed configurations (see Figure~\ref{Fig4}A\textsubscript{3}). We further confirmed this by plotting the distributions of the maximum value of the real part of the eigenvalues of the linearised system in a breakdown for open, closed, and bivalent chromatin. Supplementary Figure~2A shows these distributions for the data presented in Figure~\ref{Fig4}A. It can be seen that for all stable equilibria obtained by varying $e_{Z_0}$ and $w_{12}$, intermediate bivalent states are characterised by the highest values of their maximum real part of the eigenvalues as compared to open and closed chromatin. Similar results are obtained for other parameter values (see Supplementary Figures~2B-F, 3C-D and 4).

We illustrate the (stochastic) temporal evolution of the system for two particular combinations of parameters, one within the tristable region (Figure~\ref{Fig4}C) and another one within the bistable region (Figure~\ref{Fig4}D). These parameter combinations are referred to as $l_1$ and $l_2$, respectively (see Figure~\ref{Fig4}A\textsubscript{1}). For $l_1$, the system possesses three stable steady states corresponding to open, closed, and bivalent chromatin. The convergence of its time evolution to a specific equilibrium depends on the initial conditions. We observe similar tristable regions in the parameter space for variations of all other enzymes incorporated in the model and other kinetic parameters (Supplementary Figures~1 and 3A-B). For all these cases, the stable equilibria in tristable regions correspond to open, closed, and bivalent chromatin. No rugged pattern is observed for the tristable regime (see Figures~\ref{Fig4}A\textsubscript{1}-A\textsubscript{2} and Supplementary Figures~1 and 3A-B; \ref{app1}). For asymmetric self-interactions of $w_{11}=200$ and $w_{22}=0.01$, a rugged pattern of epigenetic profiles at sites 1 and 2 is observed in the mono- and bistable regimes. Figure~\ref{Fig4}D illustrates this behaviour for the bistable case (parameter combination, $l_2$). 

\begin{figure}[htbp]
\begin{center}
\includegraphics[width=\textwidth]{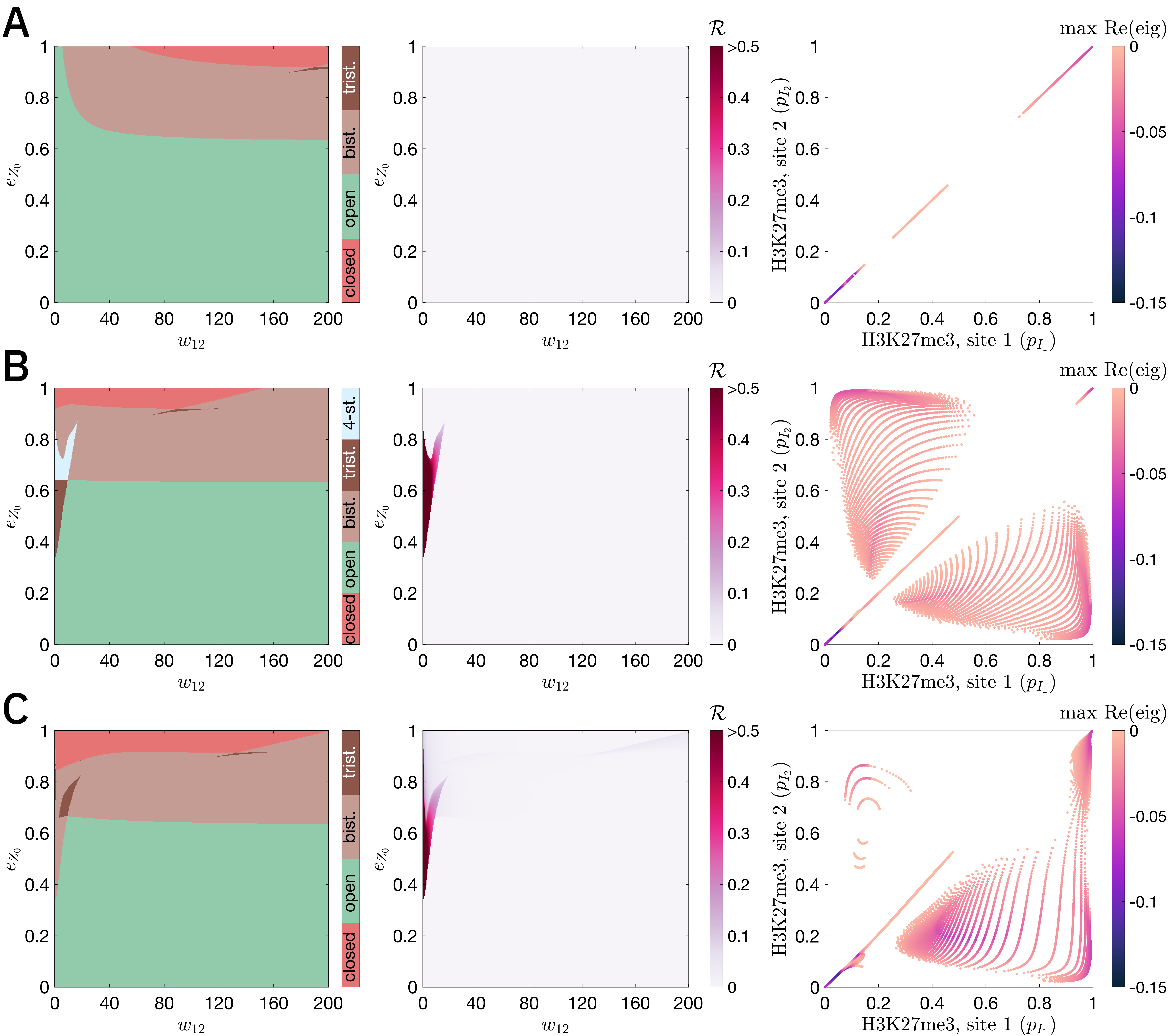}
\caption{Stability analysis of the two-site system for varying cross-interaction, $w_{12}$, and the EZH2 levels, $e_{Z_0}$ for \textbf{(A)} $w_{11}=w_{22}=1$, \textbf{(B)} $w_{11}=w_{22}=100$, and \textbf{(C)} $w_{11}=100$, $w_{22}=10$. The case of $w_{11}=200$, $w_{22}=0.01$ is shown in Figure~\ref{Fig4}A. The leftmost plots present state space diagrams with monostable regions corresponding to the open (closed) state are shown in green (red). The classification of the state as open or closed is determined by the equilibrium value of epigenetic marks at site 1; an open (closed) state corresponds to $p_{I_1}<0.3$ ($p_{I_1}>0.7$). The bistable and tristable parameter regions are indicated in brown and dark brown, respectively. In panel \textbf{(B)}, we observe the emergence of regions where there exist four stable equilibria (uniform open, uniform closed and two opposite rugged states). These regions are coloured in light blue. The middle plots show heatmaps for the value of the ruggedness metric, $\mathcal{R}$, defined by Eq~\eqref{rug_measure}. The rightmost plots present scatter plots of the H3K27me3 levels at sites 1 and 2 for all stable equilibria included in the analysis of the corresponding panel. The colour bar indicates the maximum value of the real part of the eigenvalues of the linearised system. The remaining parameters are set to their baseline levels listed in Supplementary Table~1.}
\label{Fig5}
\end{center}
\end{figure}

We now revise the assumption of asymmetric strengths in connectivity within sites 1 and 2 and investigate the behaviour of the system when interaction frequencies in both sites have the same value, i.e. $w_{11}=w_{22}$. This scenario represents the worst-case setting for observing rugged patterns in epigenetic profiles. Figure~\ref{Fig5}A-C shows these results for increasing frequency of self-interactions. We observe that low frequencies of self-interaction, i.e. $w_{ii}=1$, fail to generate heterogeneous patterns of epigenetic marks in sites 1 and 2 (see Figure~\ref{Fig5}A). Here, low levels of EZH2 methyltransferase result in a uniform open configuration at both sites, while high levels lead to a uniformly closed configuration. Between these extremes lies a bistable region where both (homogeneous) stable equilibria coexist. The bivalent state can also be observed in a small region of tristability. We note the symmetry in the distributions of H3K27me3 marks for the two sites (all equilibria lie on the diagonal in Figure~\ref{Fig5}A, right plot).

For larger values of $w_{11}=w_{22}$, a rugged landscape emerges when the cross-interaction, $w_{12}$, is small and levels of EZH2, $e_{Z_0}$, are high (Figure~\ref{Fig5}B). By symmetry, two opposite rugged states emerge simultaneously, corresponding to equilibria falling outside the main diagonal in Figure~\ref{Fig5}B (right plot). One rugged state corresponds to site 1 being in an open and site 2 in a closed configuration. The other rugged state is characterised by site 1 being closed and site 2 open. In this case, the rugged states lie within a tristable regime (coexisting with a uniform open state) and in regions with four stable equilibria (uniform open, uniform closed, and two rugged states). The two rugged states are also observed when the contact patterns are asymmetric, provided the self-interaction frequencies are sufficiently high. For example, in Figure~\ref{Fig5}C, we fix $w_{11}=100$ and $w_{22} = 10$, so interactions within site 2 are less frequent than for site 1. Nonetheless, Figure~\ref{Fig5}C (right plot) confirms the existence of the second rugged state (high H3K27me3 for site 2 and low H3K27me3 for site 1). For completeness, we report the distributions of the maximum value of the real part of the eigenvalues for the data presented in Figure~\ref{Fig5} (see Supplementary Figure~4).

We observe similar trends when all kinetic parameters and enzyme levels are fixed, and the contact frequencies, $w_{il}$, vary. Specifically, Supplementary Figure~5 presents a stability analysis of the two-site system as $w_{12}$ and $w_{22}$ are varied for increasing interaction within site 1, $w_{11}$. It can be seen that larger values of cross-interaction, $w_{12}$, induce the formation of uniform epigenetic landscapes. In contrast, rugged patterns emerge for low $w_{12}$ and when self-interaction in at least one site is sufficiently strong. Higher self-interaction values at both genomic sites support the formation of rugged landscapes for a larger range of $w_{12}$ values.
 
Summarising, our model exhibits bistability where uniform open and closed states coexist for a wide range of parameter values. These regions can be seen in our stability maps in brown colour (Figure~\ref{Fig4}A\textsubscript{1}, leftmost plots in Figure~\ref{Fig5} and Supplementary Figures~1 and 3A-B) whenever the corresponding values of the ruggedness metric (middle panels) are low. Rugged landscapes can be identified by higher values (greater than $0.5$) of the ruggedness metric, $\mathcal{R}$. Typically, rugged equilibria appear in regions of multistability, accompanied by a uniform open and/or closed state(s), for relatively low values of cross-interaction between genomic sites. When the contact frequency between sites 1 and 2 increases, the system tends to have uniform epigenetic profiles (i.e. no ruggedness). Bivalent chromatin emerges in regions of bi- and tristability and is more sensitive to external perturbations. This feature is consistent with the function bivalent chromatin is normally assumed to perform, namely, facilitating rapid transitions to either uniform open or uniform closed equilibria that coexist with it. We did not observe coexistence of bivalent and rugged states. Instead, bivalent states appear to act as intermediaries, facilitating transitions between uniform and rugged landscapes (and vice versa). This behaviour can be observed, for instance, in the bifurcation plot shown in Figure~\ref{Fig4}B for $e_{Z_0}=0.6$. As we trace the bifurcation curve, starting at high values of the cross-interaction parameter $w_{12}$, the system initially exhibits monostability, characterised by an open chromatin state. As $w_{12}$ decreases, a second stable steady state, representing a bivalent state (intermediate values of $p_{I_1}$), emerges via a saddle-node bifurcation. With further reduction in $w_{12}$, site 1 in this bivalent state shifts towards a closed state (indicated by higher values of $p_{I_1}$). Finally, when $w_{12}$ reaches lower values (e.g., $w_{12}=0.9$ as in Figure~\ref{Fig4}D), the system transitions to a regime in which open and rugged states coexist.

\subsection{Emergence of rugged epigenetic landscapes in multi-site systems: A scaling approach} \label{sec:results_scaling}

Under general conditions, an exhaustive bifurcation and parameter sensitivity analysis like the one carried out in Section~\ref{sec:results_two_sites} for a two-site system is unfeasible when we move on to the more realistic multi-site cases where $N\gg 1$. Rather than attempting to tackle this situation directly, our strategy is to leverage our understanding of the system's behaviour in response to parameter variations obtained from the stability analysis of the two-site system. To do so, we apply a scaling argument. This scaling must ensure the invariance of the system's behaviour as we change the number of genomic sites onto which the entire chromatin domain is decomposed. In particular, the rescaling of the system should enable us to increase the resolution of our system by subdividing the total chromatin domain into smaller bins. Thus, we now assume that the number of genomic sites is greater than two, i.e. $N>2$.

The scaling argument goes as follows. In our model, since the total concentration of enzymes (free and chromatin-bound) is conserved, the number of genomic sites influences the competition for enzyme binding, and this is reflected in the QSS enzyme distributions of the type given by Eqs~\eqref{dist_multinomial_prob_enzymeZ}-\eqref{dist_multinomial_prob_enzymeZ2} (see Supplementary Material for the complete list of QSS probabilities; \ref{app1}). They can be rewritten in the following general way:

\begin{align} \label{dist_multinomial_prob_generalised}
p_{E_J} = \frac{1}{1+\displaystyle \sum_{j=1}^N \widetilde{\kappa}_{j} \skpp s_{{}_j}}, \qquad \qquad p_{C_{J_i}} = \frac{\widetilde{\kappa}_{i} \skpp s_{{}_i}}{1+\displaystyle \sum_{j=1}^N \widetilde{\kappa}_{j} \skpp s_{{}_j}}.
\end{align}
\noindent Here, $\widetilde{\kappa}_{i}$ denotes the inverse of a generalised Michaelis constant \citep{ingalls2013mathematical} for the enzymatic reaction catalysed by enzyme $J$ and $s_{{}_i} \in \af{0,1}$ is the normalised substrate level at binding position $i$. Inspection of the expression for $p_{E_J}$ reveals that adding more binding sites will reduce the probability of finding enzyme in an unbound state to zero, i.e. $p_{E_J} \rightarrow 0$ as $N \rightarrow \infty$.

To avoid this dilution effect, we employ a scaling for which $p_{E_J}$ remains fixed as the resolution is successively refined by dividing the chromatin domain into larger numbers of genomic sites. If then, for simplicity, we assume that all binding sites have the same affinity (i.e. $\widetilde{\kappa}_{i} = \widetilde{\kappa}$), this is equivalent to the following:
\begin{align} \label{dist_multinomial_prob_S}
& \displaystyle \lim_{N\rightarrow \infty} \frac{1}{N/2} \skp \displaystyle \sum_{j=1}^N s_{{}_j} = \text{const}.
\end{align}
\noindent Here, the factor $N/2$ is due to the fact that the initial analysis was performed for two sites. 

\begin{figure}[!t]
\captionsetup{singlelinecheck=off,justification=raggedright}
\begin{center}
\includegraphics[width=\textwidth]{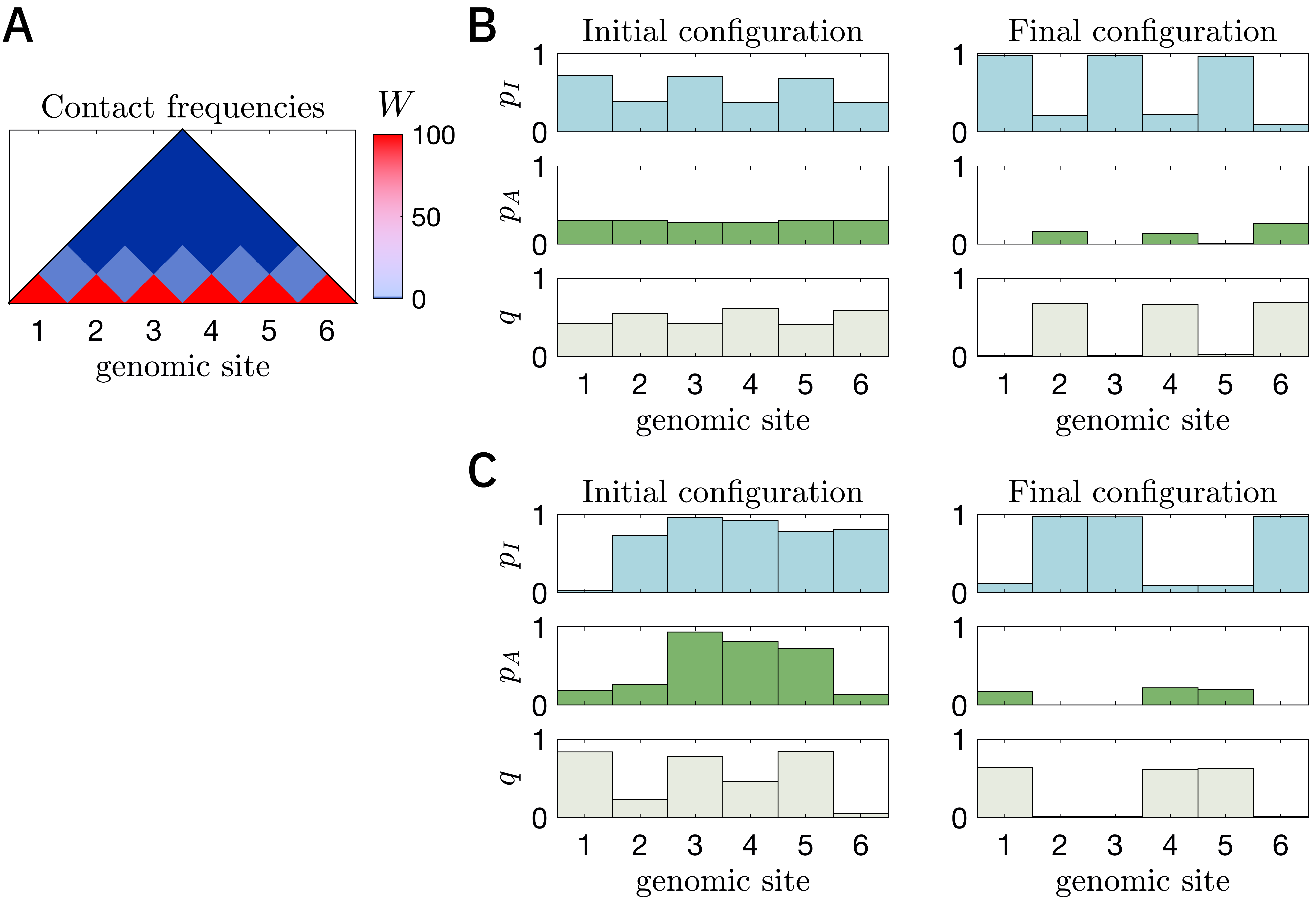}
\end{center}
\caption{Two stochastic simulations of the multi-site system for $N=6$. The upper-diagonal part of the connectivity matrix, $W$, is shown in \textbf{(A)}. Here, $w_{ii}=100$, $w_{il}=0.9$ for nearest-neighbour interaction and $w_{il}=0$ otherwise. \textbf{(B)} Initial and final configurations of a simulation initialised with alternating expressions of repressive and activating marks. \textbf{(C)} Initial and final configurations of a simulation with a random initial condition. Here, $p_{I}$ denotes levels of H3K27me3, $p_{A}$ -- H3K4me3, and $q_{}$ -- H3K27ac. In these simulations, we used $e_{Z_0}=0.6$ (EZH2 methyltransferase) and the final simulation time, $\tau_{final}=2000$ (dimensionless). The remaining parameters are set to their baseline levels listed in Supplementary Table~1.} 
\label{Fig6}
\end{figure}

The above scaling is used to rescale the amount of available substrate for all QSS probabilities and leads to the following transformation of Eq~\eqref{dist_multinomial_prob_generalised}:

\begin{align} \label{dist_multinomial_prob_generalised_rescaled}
p_{E_J} = \frac{1}{1+{\color{mulberry}{\frac{2}{N}}} \displaystyle \sum_{j=1}^N \widetilde{\kappa}_{j} s_{{}_j}}, \qquad p_{C_{J_i}} = \frac{\widetilde{\kappa}_{i} {\color{mulberry}{\frac{2}{N}}} \skpp s_{{}_i}}{1+{\color{mulberry}{\frac{2}{N}}} \displaystyle \sum_{j=1}^N \widetilde{\kappa}_{j} s_{{}_j}}.
\end{align}
\noindent A full list of the rescaled QSS probabilities is included in Section~III of Supplementary Material, \ref{app1}.

We illustrate the behaviour of the rescaled system by performing stochastic simulations of our model for $N=6$, using the parameter set corresponding to tristability in Figure~\ref{Fig5}B (with $e_{Z_0}=0.6$ and $w_{12}=0.9$). We recall that this parameter combination is characterised by the coexistence of a uniformly open and two rugged states. Figure~\ref{Fig6} shows stochastic simulations for two different initial conditions. For a suitable choice of the initial conditions, we obtain a regular rugged pattern of alternating open and closed regions (Figure~\ref{Fig6}B). For random initial conditions, the rugged pattern persists, but the distribution of open and closed loci varies (Figure~\ref{Fig6}C). Nonetheless, a general trend establishes so that half of the loci are open (and half closed), forming a rugged pattern as predicted by the two-site analysis.

\subsection{Stability-analysis for two-region system} \label{sec:results_twosubregion}

The scaling argument introduced in Section~\ref{sec:results_scaling} enables us to extend our model to an arbitrary number of genomic sites. However, instead of conducting an extensive exploration of characteristic chromatin conformations, we focus on a specific chromatin architecture involving two interacting regions, which allows for simulation via our stochastic model and stability analysis using the associated mean-field equations. We consider a chromatin domain comprising $N$ genomic loci divided into two distinct regions. In region 1, genomic sites interact only within the same site, so that $w_{ii} = w_{diag}$, for $i = 1, \ldots R$, where $R$ is the size of region 1, and  $w_{il}=0$ for $i = 1, \ldots R$, $l = 1, \ldots R$ and $i \neq l$. Region 2 behaves like a TAD-like domain, with all-to-all interactions between genomic sites at a uniform frequency, $w_{TAD}$, i.e. $w_{il} = w_{TAD}$ for $i = R+1, \ldots N$ and $l = R+1, \ldots N$. We assume further that regions 1 and 2 do not interact. A schematic of this chromatin architecture is provided in Section~IV of Supplementary Material (\ref{app1}). We note that, even though not all genomic sites interact in this setup, their dynamics are coupled due to competition for enzyme binding.

The mean-field equations associated with the stochastic model of this two-region scenario are provided in Section~IV of Supplementary Material (\ref{app1}). Specifically, the uniform connectivity of all genomic sites within each region means that their evolution can be described by identical equations. Further, if we assume uniform initial conditions within each region, then the system's dynamics can be reduced to a system of two ordinary differential equations, each representing a distinct region. Consequently, we can perform stability analysis as in Section~\ref{sec:results_two_sites}.

\begin{figure}[htbp]
\begin{center}
\includegraphics[width=\textwidth]{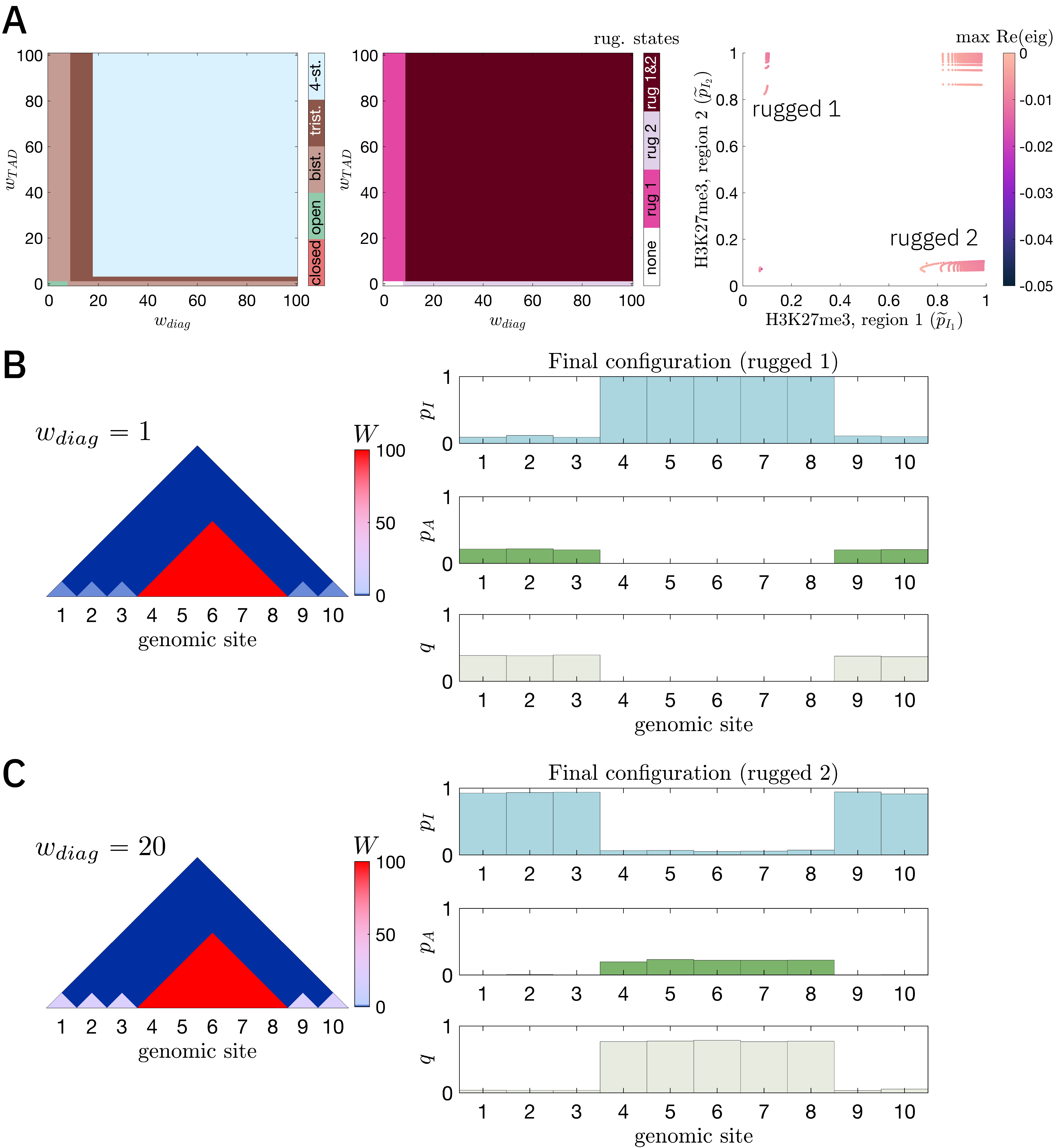}
\end{center}
\captionsetup{singlelinecheck=off,justification=raggedright}
\caption{  \textit{(Caption on the next page.)}}
\end{figure}
\addtocounter{figure}{-1}
\makeatletter
\setlength{\@fptop}{0pt}
\makeatother
\begin{figure} [t!]
\caption{Analysis and simulations of the two-region system, with region 1 exhibiting only connections within the same site with frequency, $w_{diag}$, while TAD-like region 2 is characterised by all-to-all interactions with the uniform strength of $w_{TAD}$. \textbf{(A)} Stability analysis of the system in response to variations of $w_{diag}$ and $w_{TAD}$. The full modelling equations are presented in Supplementary Material (\ref{app1}). We note that this analysis is valid only for a particular case when the initial conditions within each region are the same. The colour schemes of the leftmost and rightmost panels are as in Figure~\ref{Fig5}B. The middle panel shows a heatmap describing which rugged equilibrium patterns exist for the corresponding values of parameters. Here, rugged pattern 1 (2) shown in magenta (pastel purple) is characterised by region 1 enriched with activating (repressive) marks, whereas TAD-like region 2 displays high levels of repressive (activating) modifications. The parameter space where both rugged patterns coexist is shown in dark red. \textbf{(B)}, \textbf{(C)} Stochastic simulations of the multi-site system for the two-region scenario with a fixed value of $w_{TAD}=100$. The value of $w_{diag}= 1$ in \textbf{(B)} and $w_{diag}= 20$ in \textbf{(C)}. Final configurations are marked as rugged pattern 1 and rugged pattern 2 as described for panel \textbf{(A)}. Here, we used the total number of sites, $N=10$, the size of the region 1, $R=5$, and the final simulation time, $\tau_{final}=5000$ (dimensionless). All other parameters are set to their baseline values listed in Supplementary Table~1.}
\label{Fig7}
\end{figure}

Figure~\ref{Fig7}A shows the results of the stability analysis as we vary $w_{diag}$ and $w_{TAD}$, the components of the contact frequency matrix in regions 1 and 2, respectively. Given that TAD-like regions are typically characterised by a high frequency of contacts among genomic loci, we focus on the dynamics for intermediate-to-high values of $w_{TAD}$. Figure~\ref{Fig7}A shows that, for any fixed value of $w_{TAD}>2$, the system undergoes a transition from one to two coexisting rugged states when $w_{diag} \approx 10$. Specifically, for low values of $w_{diag}$, the system exhibits bistability between uniformly open chromatin and rugged pattern 1. For rugged pattern 1, region 1, characterised by diagonal connections, is enriched with activating marks (H3K4me3 and H3K27ac), while the TAD-like region 2 shows high levels of repressive modifications (H3K27me3). As $w_{diag}$ increases, the system transitions to rugged pattern 2 (high repressive marks in region 1, high activating marks in region 2), which coexists with rugged pattern 1.

We performed stochastic simulations for this chromatin geometry, with a fixed value of $w_{TAD}$ and two values of $w_{diag}$. The simulation results are presented in Figures~\ref{Fig7}B and C. For the first scenario with the low value of $w_{diag}=1$, only rugged pattern 1 exists. In the second scenario, we employed a higher value of $w_{diag}=20$, representing the parameter regime where two opposite rugged patterns coexist. We note that for the chromatin geometries considered in these figures, region 1 consists of genomic sites $\uf{1, 2, 3, 9, 10}$, while region 2 comprises genomic loci 4 to 8. For low values of $w_{diag}$ (Figure~\ref{Fig7}B), we consistently observed an epigenetic landscape corresponding to rugged pattern 1, with the TAD region 2 being closed (high H3K27me3 levels). However, for large values of $w_{diag}$ (e.g. $w_{diag}=20$ in Figure~\ref{Fig7}C), the only equilibrium pattern observed in our stochastic simulations is rugged pattern 2 (TAD region open with acetylated H3K27). The emergence of this epigenetic landscape was robust regardless of the choice of initial conditions used in the simulations. This outcome arises because the stability analysis of the two-region system was conducted for a simplified case (ODE system of 6 equations), which assumed identical initial conditions within each region-- homogeneous initial conditions for all sites in the TAD-like region and identical initial conditions for loci in the region with diagonal self-interactions. The complete set of equations associated with the stochastic simulations shown in Figures~\ref{Fig7}B and C consists of 30 coupled ODEs, representing the evolution of 3 epigenetic marks across $10$ genomic sites. Interactions among $10$ genomic sites and competition for enzyme binding in the full system increase the size of the basin of attraction of rugged pattern 2.

Similar results are obtained when the interaction matrix is constant, $\overline{W}=W$ (see Section~\ref{sec:model}). Supplementary Figure~6 presents the stability analysis and simulations that give rise to the two rugged patterns in this scenario. It is important to note that the results shown in Figure~\ref{Fig7} were obtained using an H3K27me3-dependent interaction matrix, $\overline{w}_{il} = w_{il} p_{I_l} p_{I_l}$. This non-linearity necessitates larger values of $W = \uf{w_{il}}_{i,l}$ for transitions between distinct behaviours, compared to the constant $\overline{W}=W$ matrix used in Supplementary Figure~6.

In conclusion, the results presented in Figure~\ref{Fig7} and Supplementary Figure~6 confirm that, regardless of the functional form of the interaction matrix, $\overline{W}$, our model predicts that the epigenetic profile within a TAD-like region is characterised by activating marks if the diagonal interactions in the rest of the domain are sufficiently strong. Experimentally obtained contact frequency maps (see Figure~\ref{Fig1}D) indicate that diagonal interactions within individual genomic loci are more frequent than interactions with distant loci (away from the main diagonal), suggesting higher values of $w_{diag}$ in our theoretical framework. Consequently, our model indicates that rugged pattern 2, with an acetylation peak at the TAD-like region, is more likely to be observed than rugged pattern 1. Additionally, since these results also hold for a constant interaction matrix, $\overline{W}$, we hypothesise that this behaviour is due to a combination of chromatin architecture and enzyme competition for binding at different genomic loci.

\section{Discussion} \label{sec:discussion}

In this study, we have presented a stochastic model for epigenetic regulation within the context of folded chromatin polymers and enzyme competition. These two effects, frequently overlooked in existing epigenetic models \citep{dodd2007theoretical,sneppen2008ultrasensitive,david2009inheritance,sneppen2012simple,sneppen2016nucleosome,thalheim2017regulatory,thalheim2018cooperation,sneppen2019theoretical,zhang2019quantifying,nickels2023confinement}, are essential for understanding the dynamics of epigenetic modifications. Our approach specifically focuses on the post-translational covalent modifications of histone H3 tails, particularly H3K27me3, H3K4me3, and H3K27ac. By employing a mesoscopic framework, we were able to examine the dynamics of epigenetic marks at genomic loci comprising multiple nucleosomes and to explore the formation of epigenetic patterns on a larger scale of DNA compaction, such as loops and TADs.

Our primary focus has been to analyse the model in order to identify parameter regimes that drive the formation of rugged epigenetic profiles. The mathematical formulation of our model offers a significant advantage compared to purely computational approaches \citep{dodd2007theoretical,sneppen2008ultrasensitive,sneppen2012simple,sneppen2019theoretical,zhang2019quantifying,nickels2023confinement}, allowing us to efficiently explore parameter space and characterise the qualitative behaviour it generates. One of our key findings involves the identification of conditions that lead to either uniform or rugged epigenetic patterns. Specifically, our analysis of a two-site system suggests that rugged patterns tend to form for intermediate enzyme concentrations. In contrast, low or high enzyme levels tend to produce more uniform epigenetic landscapes, with the specific outcome depending on the enzymes involved. Our analysis of the model's behaviour for different choices of the interaction map between genomic sites suggests that increasing cross-interaction generally leads to more uniform epigenetic profiles and that a higher frequency of self-interaction within genomic regions increases the robustness of rugged patterns.

\begin{figure}[!t]
\begin{center}
\includegraphics[width=\textwidth]{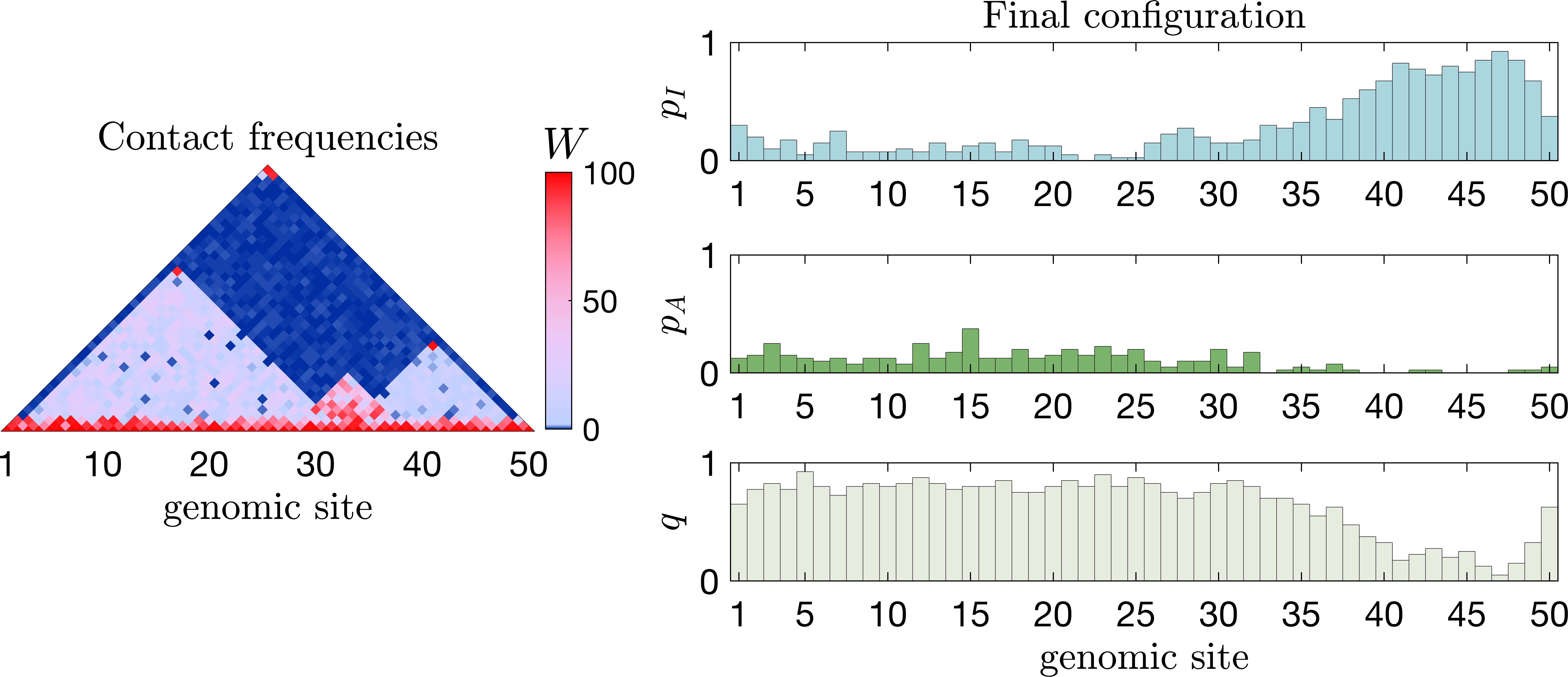}
\caption{A stochastic simulation of the multi-site system with $N = 50$ conducted for the chromatin architecture shown in Figure~\ref{Fig2}E (and Graphical Abstract). Left panel: the upper-diagonal part of the contact frequency matrix, $W$. Right panel: final configuration of the epigenetic profile for this chromatin geometry. Here, $p_I$ denotes levels of H3K27me3, $p_A$ – H3K4me3, and $q$ – H3K27ac. Different realisations yield similar epigenetic profiles regardless of the initial conditions. The final simulation time was set to $\tau_{final} = 30000$ (dimensionless). The remaining parameters are set to their baseline levels listed in Supplementary Table~1. For a movie of this numerical simulation, see Supplementary Movie~1 (\ref{app1}).} 
\label{Fig8}
\end{center}
\end{figure}

Our model successfully captures the emergence of bivalent chromatin states characterised by a mixture of activating and repressive epigenetic marks. These bivalent states arise in bi- and tristable parameter regions, where they coexist with, and are less robust than, uniformly open and/or closed chromatin states. Notably, we observe that variations in parameter values (especially the strength of cross-interactions between genomic sites) can transform bivalent states into rugged states. Given the dynamic nature of chromatin and its folding \citep{dekker20163d}, we hypothesise that the bivalent state may facilitate transitions between distinct epigenetic states in response to subtle changes in chromatin's local environment. We also emphasise that our mesoscopic modelling framework is consistent with the hypothesis that mixed levels of activating and repressive epigenetic marks observed in bivalent chromatin may arise from rapid bistable switching dynamics \citep{sneppen2019theoretical}. 

Our approach enables us to shift the focus from the dynamics of specific genomic loci to larger-scale epigenetic patterns, as our model can be extended to account for an arbitrary number of genomic sites by suitable rescaling of the QSS probabilities for enzyme distributions. The limit on the number of sites that can be simulated with our modelling approach depends on the resolution of each site (its length in base pairs) and enzyme availability. Provided the assumption of enzyme competition holds, our model remains valid. In the simulations presented in this work, we assume that the total simulated chromatin region spans approximately $50$ kb. This corresponds to a length of $1$ kb (around $5$ nucleosomes) per genomic site in the simulation shown in Figure~\ref{Fig8}, or $5$ kb ($25$ nucleosomes) per genomic site in Figure~\ref{Fig7}. Most gene domains in the human genome typically range between $10$ and $100$ kb, while the length of TADs can extend from $100$ kb to several megabases \cite{bickmore2013genome}. Therefore, our modelling framework is well-suited for simulating chromatin regions on biologically relevant scales, including gene domains and smaller TADs. We also note that our model can be extended to account for differential enzyme availability, which simplifies certain aspects of its formulation.

Epigenetic landscapes are influenced by many factors, including chromatin folding, which can significantly vary along the genome and across cell types and states. Although exploring characteristic chromatin geometries is beyond the current scope of our work, we highlight the potential of our modelling framework with a simulation based on a chromatin conformation shown in Figure~\ref{Fig2}E (and also in Graphical Abstract). Figure~\ref{Fig8} illustrates the resulting epigenetic landscape for this simulation, demonstrating the model's applicability to more realistic chromatin conformations.

A limitation of our current model is the assumption that the matrix of contact frequencies remains fixed. In biological systems, epigenetic modifications can influence chromatin folding, thus changing the map of contact frequencies.

Looking ahead, we plan to explore the formation of epigenetic landscapes for characteristic chromatin architectures such as loops, TADs, and architectural stripes. Additionally, we aim to extend our model to account for the dynamic coupling between the epigenetic landscape and interaction frequencies between genomic sites. By addressing these aspects, we hope to provide a more comprehensive understanding of epigenetic regulation in the context of 3D chromatin organisation.

\section*{Declaration of competing interest}
The authors declare that they have no known competing financial interests or personal relationships that could have appeared to influence the work reported in this paper.

\section*{Data availability statement}
The detailed description of our model provided in this manuscript and/or in the supplementary material allows for the reproducibility of all the presented results. The code for model simulations used in this work is available upon request.

\section*{Acknowledgements}
We are grateful to James Davies from the MRC Molecular Haematology Unit, MRC Weatherall Institute of Molecular Medicine, Radcliffe Department of Medicine, University of Oxford, for insightful discussions on the biological interpretation of our modelling results. This work has been funded by the Spanish Research Agency (AEI), through the Severo Ochoa and Maria de Maeztu Program for Centers and Units of Excellence in R\&D (CEX2020-001084-M). D.S. and T.A. thank CERCA Program/Generalitat de Catalunya for institutional support. D.S. and T.A. have been funded by grant PID2021-127896OB-I00 funded by MCIN/AEI/10.13039/501100011033 `ERDF A way of making Europe'.

\appendix
\section{Supplementary material} \label{app1}

\textbf{Supplementary Material.pdf} Supplementary material containing a more detailed description of our model and additional figures and tables.

\textbf{Supplementary Movie 1.mp4} A movie of a stochastic simulation of the multi-site system with $N = 50$ conducted for the chromatin architecture shown in Figure~\ref{Fig2}E (and Graphical Abstract). The final configuration of this simulation is shown in Figure~\ref{Fig8}. Left panel: the upper-diagonal part of the contact frequency matrix, $W$. Right panel: final configuration of the epigenetic profile for this chromatin geometry. Here, $p_I$ denotes levels of H3K27me3, $p_A$ – H3K4me3, and $q$ – H3K27ac. Different realisations yield similar epigenetic profiles regardless of the initial conditions. The final simulation time was set to $\tau_{final} = 30000$ (dimensionless). The remaining parameters are set to their baseline levels listed in Supplementary Table~1.






\end{document}





\title{Understanding how chromatin folding and enzyme competition affect rugged epigenetic landscapes -- Supplementary material} 


\author[lab_CRM]{Daria Stepanova\corref{cor_daria}}
\ead{dstepanova@crm.cat}
\cortext[cor_daria]{Corresponding author.}
\author[lab_edinburgh]{Meritxell Brunet Guasch}
\author[lab_WCMB,lab_Ludwig]{Helen M. Byrne}
\author[lab_ICREA,lab_CRM,lab_UAB,lab_collab]{Tom\'{a}s Alarc\'{o}n}

\affiliation[lab_CRM]{organization={Centre de Recerca Matem\`{a}tica},
            addressline={Campus de Bellaterra, Edifici C}, 
            city={Bellaterra},
            postcode={08193}, 
            state={Barcelona},
            country={Spain}}

\affiliation[lab_edinburgh]{organization={School of Mathematics and Maxwell Institute for Mathematical Sciences, University of Edinburgh},
            addressline={James Clerk Maxwell Building, Mayfield Rd}, 
            city={Edinburgh},
            postcode={EH9 3FD}, 
            state={Scotland},
            country={United Kingdom}}

\affiliation[lab_WCMB]{organization={Wolfson Centre for Mathematical Biology, Mathematical Institute, University of Oxford},
            addressline={Andrew Wiles Building, Radcliffe, Observatory Quarter, Woodstock Road}, 
            city={Oxford},
            postcode={OX2 6GG}, 
            state={England},
            country={United Kingdom}}
   
\affiliation[lab_Ludwig]{organization={Ludwig Institute for Cancer Research, Nuffield Department of Medicine, University of Oxford},
            addressline={Old Road Campus Research Building, Roosevelt Drive}, 
            city={Oxford},
            postcode={OX3 7DQ}, 
            state={England},
            country={United Kingdom}}

\affiliation[lab_ICREA]{organization={Instituci\'{o} Catalana de Recerca i Estudis Avan\c{c}ats},
            addressline={Passeig de Llu\'{i}s Companys, 23}, 
            city={Barcelona},
            postcode={08010}, 
            state={Barcelona},
            country={Spain}}

 \affiliation[lab_UAB]{organization={Departament de Matem\`{a}tiques, Universitat Aut\`{o}noma de Barcelona},
            addressline={Campus de Bellaterra, Edifici C}, 
            city={Bellaterra},
            postcode={08193}, 
            state={Barcelona},
            country={Spain}}

\affiliation[lab_collab]{organization={Barcelona Collaboratorium for Predictive and Theoretical Biology},
            addressline={Wellington, 30}, 
            city={Barcelona},
            postcode={08005}, 
            state={Barcelona},
            country={Spain}}
                       
\MaketitleBox
\small{
\tableofcontents}
\vfill
\printFirstPageNotes
\thispagestyle{pprintTitle}%
  \gdef\thefootnote{\arabic{footnote}}%


\pagebreak

\emergencystretch 3em

\section{Quasi-steady state approximation for the model of epigenetic modifications}
\label{secI}

In this section, we describe in more detail the derivation of the quasi-steady-state distributions for the histone-modifying enzymes in our stochastic model and the associated mean-field equations. We recall that the enzymatic reactions mediating histone modifications included in our model are given by Eqs~\eqref{eq_add_marks_27me}-\eqref{eq_remove_marks_27ac} of the main text. They are complemented by the reaction rate functions, Eq~\eqref{eq_kinetic_constants}, which reflect the reinforcing and antagonistic relations between the considered epigenetic marks in the context of a folded chromatin polymer. Finally, conservation laws, given by Eqs~\eqref{conservation_law_lysine27}-\eqref{conservation_law_enzyme}, ensure that total numbers of (unmodified and modified) lysine residues within each genomic region and overall enzyme levels are conserved. The model is closed by imposing appropriate initial conditions for the epigenetic marks at all genomic loci.

Enzymatic reactions are typically characterised by small enzyme-to-substrate ratios. This leads to fast dynamics for the formation of substrate-enzyme complexes and slower product formation. In our model, the unmodified and modified lysine residues play the role of substrates for the binding of writer- and eraser-type enzymes, respectively. This enables us to employ the Briggs-Haldane approximation \citep{keener2009mathematical} to justify a separation of timescales between the fast dynamics of enzyme complexes and the slower release of product reactants (epigenetic modifications). To do this, we introduced the characteristic scalings, $\Omega_E$ and $\Omega_S$, for the enzyme levels and epigenetic marks (i.e. substrate), respectively \citep{ball2006asymptotic,folguera2019multiscale,alarcon2021}:
\begin{adjustwidth}{-0.7cm}{-0.7cm}
\begin{align}
&S_{27_i} = \Omega_S s_{27_i}, \quad S_{4_i} = \Omega_S s_{4_i}, \quad P_{I_i} = \Omega_S p_{I_i}, \quad P_{A_i} = \Omega_S p_{A_i}, \quad Q_{{}_i} = \Omega_S q_{{}_i},  \label{eq_scaling}  \\ 
&E_J = \Omega_E e_J, \quad C_{J_i} = \Omega_E c_{J_i},\qquad J \in \uf{M,Z,C,K,U,S}. \nonumber
\end{align}
\end{adjustwidth}
\noindent We recall that the model variables are listed in Table~\ref{Table1} of the main text. We assume that the substrate is more abundant than enzymes, i.e. $\Omega_S \gg \Omega_E$. We now utilise Watanabe’s characterisation of Poisson processes \citep{anderson2015stochastic} and the aforementioned scaling to state the dynamics of the modified lysine residues corresponding to Eqs~\eqref{eq_add_marks_27me}-\eqref{eq_kinetic_constants} 

\begin{adjustwidth}{-0.7cm}{-0.7cm}
\begin{align}
p_{I_i} (t) &= p_{I_i} (0) + \frac{1}{\Omega_S} \mathcal{Y}\of{\Omega_S \displaystyle \int_0^t \bar{k}_3 \frac{\Omega_E}{\Omega_S} c_{Z_i} \diff t } + \frac{1}{\Omega_S} \mathcal{Y}\of{\Omega_S \displaystyle \int_0^t \bar{k}_{11} \frac{\Omega_E}{\Omega_S} c_{U_i} \diff t } \\ & - \frac{1}{\Omega_S} \mathcal{Y}\of{\Omega_S \displaystyle \int_0^t \bar{k}_{10} \Omega_E p_{I_i} e_U \diff t }, \nonumber \displaybreak \\
p_{A_i} (t) &= p_{A_i} (0) + \frac{1}{\Omega_S} \mathcal{Y}\of{\Omega_S \displaystyle \int_0^t \bar{k}_6 \frac{\Omega_E}{\Omega_S} c_{M_i} \diff t } + \frac{1}{\Omega_S} \mathcal{Y}\of{\Omega_S \displaystyle \int_0^t \bar{k}_{14} \frac{\Omega_E}{\Omega_S} c_{K_i} \diff t } \\ &- \frac{1}{\Omega_S} \mathcal{Y}\of{\Omega_S \displaystyle \int_0^t \bar{k}_{13} \Omega_E p_{A_i} e_K \diff t }, \nonumber \\
q_{{}_i} (t) &= q_{{}_i} (0) + \frac{1}{\Omega_S} \mathcal{Y}\of{\Omega_S \displaystyle \int_0^t \bar{k}_9 \frac{\Omega_E}{\Omega_S} c_{C_i} \diff t } + \frac{1}{\Omega_S} \mathcal{Y}\of{\Omega_S \displaystyle \int_0^t \bar{k}_{17} \frac{\Omega_E}{\Omega_S} c_{S_i} \diff t } \\ &- \frac{1}{\Omega_S} \mathcal{Y}\of{\Omega_S \displaystyle \int_0^t \bar{k}_{16} \Omega_E q_{{}_i} e_S \diff t }, \nonumber
\end{align}
\end{adjustwidth}

\noindent and the enzymatic complexes:
\begin{adjustwidth}{-0.7cm}{-0.7cm}
\begin{align}
\allowdisplaybreaks
c_{Z_i} (t) &= c_{Z_i} (0) + \frac{1}{\Omega_E} \mathcal{Y}\of{\Omega_E \displaystyle \int_0^t \Omega_S^2 \bar{k}_1 \of{\frac{\bar{k}_{1_0}}{\Omega_S} + \displaystyle \sum_{l=1}^N \overline{w}_{il} p_{I_l}} s_{27_i} e_Z \diff t } \nonumber \\ &- \frac{1}{\Omega_E} \mathcal{Y}\of{\Omega_E \displaystyle \int_0^t \af{ \Omega_S \bar{k}_2 \of{\frac{\bar{k}_{2_0}}{\Omega_S} + \displaystyle \sum_{l=1}^N \overline{w}_{il} p_{A_l}} + \bar{k}_3 } c_{Z_i} \diff t }, \nonumber \\
c_{M_i} (t) &= c_{M_i} (0) + \frac{1}{\Omega_E} \mathcal{Y}\of{\Omega_E \displaystyle \int_0^t \Omega_S^2 \bar{k}_4 \of{\frac{\bar{k}_{4_0}}{\Omega_S} + \displaystyle \sum_{l=1}^N \overline{w}_{il} q_{{}_l}} s_{4_i} e_M \diff t } \nonumber \\ &- \frac{1}{\Omega_E} \mathcal{Y}\of{\Omega_E \displaystyle \int_0^t \af{ \Omega_S \bar{k}_5 \of{\frac{\bar{k}_{5_0}}{\Omega_S} + \displaystyle \sum_{l=1}^N \overline{w}_{il} p_{I_l}} + \bar{k}_6 } c_{M_i} \diff t }, \nonumber  \\
c_{C_i} (t) &= c_{C_i} (0) + \frac{1}{\Omega_E} \mathcal{Y}\of{\Omega_E \displaystyle \int_0^t \Omega_S^2 \bar{k}_7 \of{\frac{\bar{k}_{7_0}}{\Omega_S} + \displaystyle \sum_{l=1}^N \overline{w}_{il} p_{A_l}} s_{27_i} e_C \diff t } \label{eq_watanabe_complexes} \\ & - \frac{1}{\Omega_E} \mathcal{Y}\of{\Omega_E \displaystyle \int_0^t \of{\bar{k}_8 + \bar{k}_9 } c_{C_i} \diff t }, \nonumber  \\
c_{U_i} (t) &= c_{U_i} (0) + \frac{1}{\Omega_E} \mathcal{Y}\of{\Omega_E \displaystyle \int_0^t \bar{k}_{10} \Omega_S p_{I_i} e_U \diff t } - \frac{1}{\Omega_E} \mathcal{Y}\of{\Omega_E \displaystyle \int_0^t \of{\bar{k}_{11} + \bar{k}_{12} } c_{U_i} \diff t }, \nonumber \\
c_{K_i} (t) &= c_{K_i} (0) + \frac{1}{\Omega_E} \mathcal{Y}\of{\Omega_E \displaystyle \int_0^t \bar{k}_{13} \Omega_S p_{A_i} e_K \diff t } - \frac{1}{\Omega_E} \mathcal{Y}\of{\Omega_E \displaystyle \int_0^t \of{\bar{k}_{14} + \bar{k}_{15} } c_{K_i} \diff t }, \nonumber \\
c_{S_i} (t) &= c_{S_i} (0) + \frac{1}{\Omega_E} \mathcal{Y}\of{\Omega_E \displaystyle \int_0^t \bar{k}_{16} \Omega_S q_{{}_i} e_S \diff t } - \frac{1}{\Omega_E} \mathcal{Y}\of{\Omega_E \displaystyle \int_0^t \of{\bar{k}_{17} + \bar{k}_{18} } c_{S_i} \diff t }. \nonumber
\end{align}
\end{adjustwidth}
\noindent Here, $\mathcal{Y}(\lambda)$ is a Poisson process with intensity $\lambda$. We denote by $p_{I_i} (0)$, $p_{A_i} (0)$ and $q_{{}_i} (0)$ the initial levels of epigenetic marks H3K27me3, H3K4me3 and H3K27ac at the $i$-th genomic site, respectively. The initial levels of enzyme complexes are represented by $c_{J_i}(0)$ for $J \in \uf{M,Z,C,K,U,S}$. Kinetic constants, $\bar{k}_r$, are as defined in Eqs~\eqref{eq_add_marks_27me}-\eqref{eq_kinetic_constants}. The temporal evolution of the unmodified lysine residues and free (unbound) enzymes can be obtained by exploiting the conservation laws given by Eq~\eqref{conservation_law_lysine27}-\eqref{conservation_law_enzyme} (and applying the same scaling, Eq~\eqref{eq_scaling}). 

We now rescale the time variable, $\tau = \bar{k}_7 \Omega_S \Omega_E t$, and the kinetic constants to obtain the characteristic timescales for slow (epigenetic marks) and fast (enzymatic complexes) variables \citep{folguera2019multiscale}. The equations for the epigenetic marks become:

\begin{adjustwidth}{-0.7cm}{-0.7cm}
\begin{align} \label{eq_watanabe_marks_tau1} 
p_{I_i} (\tau) &= p_{I_i} (0) + \frac{1}{\Omega_S} \mathcal{Y}\of{\Omega_S \displaystyle \int_0^{\tau} \kappa_3 c_{Z_i} \diff \tau } + \frac{1}{\Omega_S} \mathcal{Y}\of{\Omega_S \displaystyle \int_0^{\tau} \kappa_{11} c_{U_i} \diff \tau } \\ &- \frac{1}{\Omega_S} \mathcal{Y}\of{\Omega_S \displaystyle \int_0^{\tau} \kappa_{10} p_{I_i} e_U \diff \tau }, \nonumber \\ \label{eq_watanabe_marks_tau2}
p_{A_i} (\tau) &= p_{A_i} (0) + \frac{1}{\Omega_S} \mathcal{Y}\of{\Omega_S \displaystyle \int_0^{\tau} \kappa_6 c_{M_i} \diff \tau } + \frac{1}{\Omega_S} \mathcal{Y}\of{\Omega_S \displaystyle \int_0^{\tau} \kappa_{14} c_{K_i} \diff \tau } \\ &- \frac{1}{\Omega_S} \mathcal{Y}\of{\Omega_S \displaystyle \int_0^{\tau} \kappa_{13} p_{A_i} e_K \diff \tau }, \nonumber \\ \label{eq_watanabe_marks_tau3}
q_{{}_i} (\tau) &= q_{{}_i} (0) + \frac{1}{\Omega_S} \mathcal{Y}\of{\Omega_S \displaystyle \int_0^{\tau} \kappa_9 c_{C_i} \diff \tau } + \frac{1}{\Omega_S} \mathcal{Y}\of{\Omega_S \displaystyle \int_0^{\tau} \kappa_{17} c_{S_i} \diff \tau } \\ & - \frac{1}{\Omega_S} \mathcal{Y}\of{\Omega_S \displaystyle \int_0^{\tau} \kappa_{16} q_{{}_i} e_S \diff \tau }. \nonumber
\end{align}
\end{adjustwidth}
\noindent The fast dynamics of the enzyme complexes are given as follows: 

\begin{align}
\allowdisplaybreaks
c_{Z_i} (\tau) &= c_{Z_i} (0) + \frac{1}{\Omega_E} \mathcal{Y}\of{\Omega_E \frac{1}{\epsilon} \displaystyle \int_0^{\tau} \kappa_1 \of{\kappa_{1_0} + \displaystyle \sum_{l=1}^N \overline{w}_{il} p_{I_l}} s_{27_i} e_Z \diff \tau } \nonumber \\ &- \frac{1}{\Omega_E} \mathcal{Y}\of{\Omega_E \frac{1}{\epsilon} \displaystyle \int_0^{\tau} \af{ \kappa_2 \of{\kappa_{2_0} + \displaystyle \sum_{l=1}^N \overline{w}_{il} p_{A_l}} + \kappa_3 } c_{Z_i} \diff \tau }, \nonumber \\
c_{M_i} (\tau) &= c_{M_i} (0) + \frac{1}{\Omega_E} \mathcal{Y}\of{\Omega_E \frac{1}{\epsilon} \displaystyle \int_0^{\tau} \kappa_4 \of{\kappa_{4_0} + \displaystyle \sum_{l=1}^N \overline{w}_{il} q_{{}_l}} s_{4_i} e_M \diff \tau } \nonumber \\ &- \frac{1}{\Omega_E} \mathcal{Y}\of{\Omega_E \frac{1}{\epsilon} \displaystyle \int_0^{\tau} \af{ \kappa_5 \of{\kappa_{5_0} + \displaystyle \sum_{l=1}^N \overline{w}_{il} p_{I_l}} + \kappa_6 } c_{M_i} \diff \tau }, \nonumber \\
c_{C_i} (\tau) &= c_{C_i} (0) + \frac{1}{\Omega_E} \mathcal{Y}\of{\Omega_E \frac{1}{\epsilon} \displaystyle \int_0^{\tau} \of{\kappa_{7_0} + \displaystyle \sum_{l=1}^N \overline{w}_{il} p_{A_l}} s_{27_i} e_C \diff \tau } \label{eq_watanabe_complexes_tau}  \\ &- \frac{1}{\Omega_E} \mathcal{Y}\of{\Omega_E \frac{1}{\epsilon} \displaystyle \int_0^{\tau} \of{\kappa_8 + \kappa_9 } c_{C_i} \diff \tau }, \nonumber \displaybreak \\
c_{U_i} (\tau) &= c_{U_i} (0) + \frac{1}{\Omega_E} \mathcal{Y}\of{\Omega_E \frac{1}{\epsilon} \displaystyle \int_0^{\tau} \kappa_{10} p_{I_i} e_U \diff \tau } \nonumber \\ & - \frac{1}{\Omega_E} \mathcal{Y}\of{\Omega_E \frac{1}{\epsilon} \displaystyle \int_0^{\tau} \of{\kappa_{11} + \kappa_{12} } c_{U_i} \diff \tau }, \nonumber\\
c_{K_i} (\tau) &= c_{K_i} (0) + \frac{1}{\Omega_E} \mathcal{Y}\of{\Omega_E \frac{1}{\epsilon} \displaystyle \int_0^{\tau} \kappa_{13} p_{A_i} e_K \diff \tau } \nonumber \\ & - \frac{1}{\Omega_E} \mathcal{Y}\of{\Omega_E \frac{1}{\epsilon} \displaystyle \int_0^{\tau} \of{\kappa_{14} + \kappa_{15} } c_{K_i} \diff \tau }, \nonumber \\
c_{S_i} (\tau) &= c_{S_i} (0) + \frac{1}{\Omega_E} \mathcal{Y}\of{\Omega_E \frac{1}{\epsilon} \displaystyle \int_0^{\tau} \kappa_{16} q_{{}_i} e_S \diff \tau } \nonumber \\ & - \frac{1}{\Omega_E} \mathcal{Y}\of{\Omega_E \frac{1}{\epsilon} \displaystyle \int_0^{\tau} \of{\kappa_{17} + \kappa_{18} } c_{S_i} \diff \tau }. \nonumber
\end{align}
\noindent Here, we have introduced a small parameter, $\epsilon =\Omega_E/{\Omega_S} \ll 1$, and the rescaled kinetic constants are defined in Table~\ref{TableRescaledRates}. 

\begin{table}[h!]
\centering
\renewcommand{\arraystretch}{1.5}
\setlength{\arrayrulewidth}{0.1mm} 
\begin{tabular}{| p{2cm} p{2cm} p{2cm} p{2cm} p{2cm}|}
\hline
$\kappa_1 = \frac{\bar{k}_1}{\bar{k}_7}$ & $\kappa_{1_0} = \frac{\bar{k}_{1_0}}{\Omega_S}$ & $\kappa_2 = \frac{\bar{k}_2}{\bar{k}_7 \Omega_S}$ & $\kappa_{2_0} = \frac{\bar{k}_{2_0}}{\Omega_S}$ &  $\kappa_3 = \frac{\bar{k}_3}{\bar{k}_7 \Omega_S^2}$ \\[3pt] \hline
$\kappa_4 = \frac{\bar{k}_4}{\bar{k}_7}$ & $\kappa_{4_0} = \frac{\bar{k}_{4_0}}{\Omega_S}$ & $\kappa_5 = \frac{\bar{k}_5}{\bar{k}_7 \Omega_S}$ & $\kappa_{5_0} = \frac{\bar{k}_{5_0}}{\Omega_S}$ &  $\kappa_6 = \frac{\bar{k}_6}{\bar{k}_7 \Omega_S^2}$ \\[3pt] \hline
$\kappa_{7_0} = \frac{\bar{k}_{7_0}}{\Omega_S}$ &  $\kappa_8 = \frac{\bar{k}_8}{\bar{k}_7 \Omega_S^2}$ &  $\kappa_9 = \frac{\bar{k}_9}{\bar{k}_7 \Omega_S^2}$  & & \\[3pt] \hline
$\kappa_{10} = \frac{\bar{k}_{10}}{\bar{k}_7\Omega_S}$ &  $\kappa_{11} = \frac{\bar{k}_{11}}{\bar{k}_7 \Omega_S^2}$ &  $\kappa_{12} = \frac{\bar{k}_{12}}{\bar{k}_7 \Omega_S^2}$  & & \\[3pt] \hline
$\kappa_{13} = \frac{\bar{k}_{13}}{\bar{k}_7\Omega_S}$ &  $\kappa_{14} = \frac{\bar{k}_{14}}{\bar{k}_7 \Omega_S^2}$ &  $\kappa_{15} = \frac{\bar{k}_{15}}{\bar{k}_7 \Omega_S^2}$  & & \\[3pt] \hline
$\kappa_{16} = \frac{\bar{k}_{16}}{\bar{k}_7\Omega_S}$ &  $\kappa_{17} = \frac{\bar{k}_{17}}{\bar{k}_7 \Omega_S^2}$ &  $\kappa_{18} = \frac{\bar{k}_{18}}{\bar{k}_7 \Omega_S^2}$  & & \\[3pt] \hline
$\kappa_{19} = \frac{\bar{k}_{19}}{\bar{k}_7\Omega_S^2}$ &  $\kappa_{20} = \frac{\bar{k}_{20}}{\bar{k}_7 \Omega_S^2}$ &  & & \\[3pt] \hline
\end{tabular}
\caption{Rescaled kinetic rate constants used to obtain Eqs~\eqref{eq_watanabe_marks_tau1}-\eqref{eq_watanabe_complexes_tau}.} \label{TableRescaledRates}
\end{table}

We note the difference in characteristic timescales for modified lysine residues (slow variables) given by Eqs~\eqref{eq_watanabe_marks_tau1}-\eqref{eq_watanabe_marks_tau3} and enzymatic complexes (fast variables), defined by Eq~\eqref{eq_watanabe_complexes_tau}. This allows us to consider separately the relaxation dynamics of the fast variables on short timescales (slow variables remain unchanged) and the dynamics of slow variables (with quasi-equilibrium for fast variables) on longer timescales. The former defines the \textit{inner} solution of the system, while the latter defines the \textit{outer} solution. We first illustrate the derivation of the inner solution for a simple case involving one enzymatic reaction and a single binding site, $N=1$. We then extend this approach to account for the six enzymatic reactions and multiple binding sites ($N>1$) involved in our epigenetic model. 

\subsection*{Inner solution: single enzymatic reaction at one binding site}

To avoid confusion with the notation used for our model for epigenetic modifications, we repeat the first steps of the derivation for the following stochastic process: 
\begin{align} \label{app_reaction_enzyme}
S + E \stacksymbol{\rightleftarrows}{\textcolor{burgundy}{k_1}}{\textcolor{burgundy}{k_2}} C \stacksymbol{\rightarrow}{\textcolor{burgundy}{k_3}}{} P + E.
\end{align} 
\noindent Here, substrate, $S$, binds reversibly to an enzyme, $E$, to form an intermediate complex, $C$, that is transformed into the product reactant, $P$, and free enzyme at rate $k_3$. The reaction rates, $k_r$, ($r = 1,2,3$) are assumed to be positive constants. These reactions are complemented by a conservation law for the total number of enzyme molecules, $E_0$: 
\begin{align} \label{app_conservation_law}
C+E = E_0.
\end{align} 

We employ the Briggs-Haldane assumption to perform a separation of timescale \citep{keener2009mathematical} and introduce the characteristic scalings, $\Omega_S$ and $\Omega_E$, for the substrate and enzyme concentrations, respectively \citep{ball2006asymptotic,folguera2019multiscale,alarcon2021}: 
\begin{align} \label{app_scaling}
S = \Omega_S s, \quad E = \Omega_E e, \quad C = \Omega_E c, \quad \epsilon = \frac{\Omega_E}{\Omega_S} \ll 1.
\end{align} 
Using the Watanabe representation of Poisson processes \citep{anderson2015stochastic}, the stochastic dynamics of the substrate (slow variable) and the enzymatic complex (fast variable) can be written as follows
\begin{adjustwidth}{-0.7cm}{-0.7cm}
\begin{align}
s (t) &= s (0) + \frac{1}{\Omega_S} \mathcal{Y}\of{ \Omega_S \displaystyle \int_0^t \frac{\Omega_E}{\Omega_S} \of{k_2 + k_3} c \diff t } - \frac{1}{\Omega_S} \mathcal{Y}\of{\Omega_S \displaystyle \int_0^t \Omega_E k_1 s e \diff t }, \\
c (t) &= c (0) + \frac{1}{\Omega_E} \mathcal{Y}\of{\Omega_E \displaystyle \int_0^t \Omega_S k_1 s e \diff t } - \frac{1}{\Omega_E} \mathcal{Y}\of{\Omega_E \displaystyle \int_0^t \of{k_2 + k_3} c \diff t },
\end{align}
\end{adjustwidth}
\noindent where $s (0)$ and $c (0)$ are the initial levels of substrate and enzymatic complex, respectively. We also introduce the rescaled time variable $\tau = k_1 \Omega_E t$ \citep{folguera2019multiscale}: 
\begin{adjustwidth}{-0.7cm}{-0.7cm}
\begin{align} \label{single_reaction_s}
s (\tau) &= s (0) + \frac{1}{\Omega_S} \mathcal{Y}\of{ \Omega_S \displaystyle \int_0^{\tau} \of{\kappa_2 + \kappa_3} c \diff \tau } - \frac{1}{\Omega_S} \mathcal{Y}\of{\Omega_S \displaystyle \int_0^{\tau}  s e \diff \tau }, \\
c (\tau) &= c (0) + \frac{1}{\Omega_E} \mathcal{Y}\of{\Omega_E \frac{1}{\epsilon}  \displaystyle \int_0^{\tau} s e \diff \tau } - \frac{1}{\Omega_E} \mathcal{Y}\of{\Omega_E \frac{1}{\epsilon} \displaystyle \int_0^{\tau} \of{\kappa_2 + \kappa_3} c \diff \tau }, \label{single_reaction_c}
\end{align}
\end{adjustwidth}
\noindent where $\kappa_2 = \frac{k_2}{\Omega_S k_1}$ and $\kappa_3 = \frac{k_3}{\Omega_S k_1}$.

\vspace{0.2cm}
The inner solution focuses on the rapid relaxation of the enzymatic complex, $c$, onto its quasi-equilibrium. On this fast timescale, variation in the slow variables (substrate, $s$) is negligible. In order to study the dynamics on the longer timescale, we introduce $T = \tau/\epsilon$. Then Eqs~\eqref{single_reaction_s}-\eqref{single_reaction_c} reduce to:

\begin{align}
s (T) &= s (0) + \mathcal{O}(\epsilon), \label{app_eq_inner_substrate}\\
c (T) &= c (0) + \frac{1}{\Omega_E} \mathcal{Y}\of{\Omega_E  \displaystyle \int_0^{T} s (e_0 - c) \diff T } \label{app_eq_inner_complex} \\
& - \frac{1}{\Omega_E} \mathcal{Y}\of{\Omega_E \displaystyle \int_0^{T} \of{\kappa_2 + \kappa_3} c \diff T }, \nonumber
\end{align}
\noindent where we have used the (rescaled) conservation law for the enzyme to eliminate $e = e_0 - c$. From Eq~\eqref{app_eq_inner_substrate}, we can assume that, on this timescale, the substrate concentration, $s$, is approximately constant. On the other hand, Eq~\eqref{app_eq_inner_complex} can be viewed as a birth-and-death process for the dynamics of the enzyme complex with the following transition rates: 

\begin{align} \label{app_eq_reaction_wplus}
\textbf{birth: }\quad &w_{+}\of{C,C+1}= s (E_0 - C), \\ \label{app_eq_reaction_wminus}
\textbf{death: }\quad &w_{-}\of{C,C-1} = (\kappa_2+\kappa_3) C.
\end{align}
\noindent At this point, we return to unscaled variables for the fast variables (i.e number of substrate-enzyme complexes, $C$, and the total number of enzyme molecules, $E_0$), and formulate the Master Equation for the time evolution of the probability function, $\mathbf{P}(C,T)$, for the birth-and-death process corresponding to the transition rates given by Eqs~\eqref{app_eq_reaction_wplus}-\eqref{app_eq_reaction_wminus} (the corresponding stoichiometric vectors are $r_{+} = +1$ and $r_{-} = -1$, respectively).
\begin{adjustwidth}{-0.7cm}{-0.7cm}
\begin{align} 
& \pdv{\mathbf{P}(C,T)}{T} = \mathbf{P}(C-1,T) w_{+}(C-1,C) + \mathbf{P}(C+1,T) w_{-}(C+1,C) \\
& - \mathbf{P}(C,T) \af{w_{+}(C,C+1) + w_{-}(C,C-1}), \nonumber \\[5pt]
&\text{or, equivalently} \hspace{32pt}& \nonumber \\[5pt]
& \pdv{\mathbf{P}(C,T)}{T} = \mathbf{P}(C-1,T) s (E_0 - C+1) + \mathbf{P}(C+1,T) (\kappa_2+\kappa_3) (C+1) \label{app_inner_master_eq}\\ 
& \hspace{50pt} - \mathbf{P}(C,T) \af{s (E_0 - C) + (\kappa_2+\kappa_3)C}. \nonumber
\end{align}
\end{adjustwidth}
We note that the variable $s$ is frozen as we are solving the inner dynamics. We recall that the generating function is given by $\mathbf{G}(z,T) = \sum_C z^C \mathbf{P}(C,T) $. An evolution equation for $\mathbf{G}(z,T)$ is obtained by multiplying Eq~\eqref{app_inner_master_eq} by $z^C$ and summing over all possible values of $C$:
\begin{align} \nonumber
\pdv{\mathbf{G}(z,T)}{T} &= \displaystyle \sum_C z^C \big( \mathbf{P}(C-1,T) s (E_0 - C+1) \\ &+ \mathbf{P}(C+1,T) (\kappa_2+\kappa_3) (C+1)  \label{app_inner_G_eq} \\
&- \mathbf{P}(C,T) \af{s (E_0 - C) + (\kappa_2+\kappa_3)C} \big) . \nonumber
\end{align}
This equation can be simplified by using the following auxiliary results: 
\begin{adjustwidth}{-0.7cm}{-0.7cm}
\begin{subequations}  \label{app_aux}
\begin{align} 
& \displaystyle \sum_C  z^C  \mathbf{P}(C-1,T) s (E_0 - (C-1)) = z \displaystyle \sum_C  z^C \mathbf{P}(C,T) s (E_0 - C);  \label{app_aux1}  \\
& \displaystyle \sum_C  z^C  \mathbf{P}(C+1,T) (\kappa_2 + \kappa_3)(C+1) = z^{-1} \displaystyle \sum_C  z^C \mathbf{P}(C,T) (\kappa_2 + \kappa_3)C; \label{app_aux2} \\
& \displaystyle \sum_C  z^C C~ \mathbf{P}(C,T) = \displaystyle \sum_C z \pdv{z^C}{z}  \mathbf{P}(C,T)  = z \pdv{}{z}\of{\displaystyle \sum_C z^C \mathbf{P}(C,T) }  = z \pdv{\mathbf{G}(z,T)}{z}.  \label{app_aux3} 
\end{align}
\end{subequations}
\end{adjustwidth}
Exploiting Eq~\eqref{app_aux}, we finally obtain: 

\begin{align} \label{app_eq_gen_fun}
\pdv{\mathbf{G}(z,T)}{T} &= (1-z)\af{\of{sz + \kappa_2+\kappa_3}\pdv{\mathbf{G}(z,T)}{z} - s E_0 \mathbf{G}(z,T)}.
\end{align}

We emphasise that for the inner solution, the slow variables are assumed to be `frozen', i.e. $s = const$. We assume that at $T=0$, there are no enzymatic complexes ($C=0$) so that $\mathbf{P}(C,T=0) = \delta_{C,0}$, where we use the standard notation for the Dirac-delta function, $\delta_{x,x_0}$. Given the definition of the generating function, it is straightforward to show that this initial condition corresponds to $\mathbf{G}(z,T=0) = \mathbf{G_0}(z) = 1$.

We now sketch the solution of Eq~\eqref{app_eq_gen_fun}, using the method of characteristics. The equations for the characteristic curves are given by: 

\begin{align}
&\dot{T} = 1,\nonumber \\
&\dot{z} = - (1-z)(\kappa_2+\kappa_3 + sz),\\
&T(0) = 0,\nonumber \\
&z(0) = z_0, \nonumber
\end{align}
\noindent with solution: 

\begin{align}
z(T) = \frac{\kappa_2+\kappa_3 + sz_0 - (1-z_0)(\kappa_2+\kappa_3)\exp\af{\of{\kappa_2+\kappa_3 + s}T}}{\kappa_2+\kappa_3 + sz_0 + (1-z_0)s \exp\af{\of{\kappa_2+\kappa_3 + s}T}}.
\end{align}

If we define $\mathbf{\Phi}(T) = \mathbf{G}(z(T),T)$, then from Eq~\eqref{app_eq_gen_fun}, we have that: 
\begin{align}
\dv{\mathbf{\Phi}}{T} = - (1-z(T))s E_0 \mathbf{\Phi}. 
\end{align}
\noindent This ODE can be solved in a straightforward manner to determine the generating function $\mathbf{G}(z(T),T)$:
\begin{adjustwidth}{-0.7cm}{0cm}
\begin{align}
&\mathbf{G}(z,T) = \mathbf{G_0}\of{z_0} \af{\frac{\kappa_2+\kappa_3 + sz + (1-z)s\exp\of{-\of{\kappa_2+\kappa_3 + s}T}}{\kappa_2+\kappa_3 + s}}^{E_0},
\end{align}
\end{adjustwidth}
\noindent where
\begin{align}
& z_0 = \frac{\of{\kappa_2+\kappa_3 + sz} - (\kappa_2+\kappa_3)(1-z)\exp\af{-\of{\kappa_2+\kappa_3 + s}T}}{\of{\kappa_2+\kappa_3 + sz} + s(1-z)\exp\af{-\of{\kappa_2+\kappa_3 + s}T}}.
\end{align}

Imposing the initial condition for the generating function, $G(z_0,0) = \mathbf{G_0}(z) = 1$, the final solution becomes: 
\begin{align} \label{app_gen_fun_final_sol}
&\mathbf{G}(z,T) = \af{\frac{\kappa_2+\kappa_3 + sz + (1-z)s\exp\of{-\of{\kappa_2+\kappa_3 + s}T}}{\kappa_2+\kappa_3 + s}}^{E_0}.
\end{align}

We recall that the expression for the generating function of a binomial distribution with probability, $p_C(T)$, is given by $\mathbf{G}(z,T) = \of{1- p_C(T) + p_C(T)z}^n$, where $n$ is the number of trials. Then, it is straightforward to deduce from Eq~\eqref{app_gen_fun_final_sol} that $\mathbf{G}(z,T)$ corresponds to a binomial distribution with probability:

\begin{align}
p_C(T) =\frac{s}{\kappa_2+\kappa_3 + s}\af{1 - \exp\of{-\of{\kappa_2+\kappa_3 + s}T}}
\end{align}
\noindent In this context, $E_0$ is viewed as the `number of trials' since we sample from all $E_0$ enzyme molecules and these may be either bound to the substrate (with probability, $p_C(T)$) or unbound (with probability, $(1-p_C(T))$).

We obtain the stationary probabilities corresponding to the quasi-equilibrium of the fast variable of enzymatic complexes by taking the limit $T \rightarrow \infty$;

\begin{align}
&p_C^{st}=\frac{s}{\kappa_2+\kappa_3 + s} \qquad &&p_E^{st} =\frac{\kappa_2+\kappa_3}{\kappa_2+\kappa_3 + s},
\end{align}
\noindent or, equivalently,
\begin{align}
&p_C^{st}=\frac{\frac{s}{\kappa_2+\kappa_3}}{1 + \frac{s}{\kappa_2+\kappa_3}}, && p_E^{st}=\frac{1}{1 + \frac{s}{\kappa_2+\kappa_3}}.
\end{align}
\noindent The corresponding QSS probability density function (PDF) is given by a PDF of the binomial distribution with these stationary probabilities (given the `frozen' substrate level, $s$):

\begin{align}
\phi(E,C \mid s) = \frac{\of{E_0}!}{E!~ C!} \of{p_E^{st}}^E \of{p_C^{st}}^C.
\end{align}

For the extended scenario of multiple binding sites, $\mathbf{s} = \uf{s_1,\ldots,s_N}$, the generalisation of the current result is immediate. The probabilities and the QSS PDF of the resulting multinomial distribution read: 

\begin{align} \label{one_enzyme_reaction_pdf}
&\phi(E,C_1,\ldots,C_N \mid \mathbf{s}) = \frac{\of{E_0}!}{E!~ \of{C_1}! \cdots \of{C_N} !} \of{p_E}^E \of{p_{C_1}}^{C_1} \cdots \of{p_{C_N}}^{C_N}, \\
& \text{where} \quad p_E = \frac{1}{1 + \frac{\sum_j s_j}{\kappa_2+\kappa_3}}, \qquad p_{C_i} = \frac{\frac{s_i}{\kappa_2+\kappa_3}}{1 + \frac{\sum_j s_j}{\kappa_2+\kappa_3}}. \label{app_prob_multi}
\end{align}

We note that when there are multiple enzymatic reactions, depending on how the time variable is initially rescaled (here, we used $\tau = \textcolor{blue}{k_1} \Omega_E t$), a rescaled kinetic constant $\kappa_1$ may appear in front of terms of type $s_j$ or $s_i$ in Eq~\eqref{app_prob_multi}.

\subsection*{Inner solution: full model for epigenetic modifications}

We now generalise the derivation of the QSS distributions of enzyme complexes for our epigenetic model, with six enzymatic reactions and multiple binding sites. We rescale the time variable as $\tau = \epsilon T$. Recalling Eqs~\eqref{eq_watanabe_marks_tau1}-\eqref{eq_watanabe_complexes_tau}, which describe the dynamics of our stochastic model, this yields:

\begin{align} 
p_{I_i} (T) &= p_{I_i} (0) + \mathcal{O}(\epsilon), \nonumber \\ 
p_{A_i} (T) &= p_{A_i} (0) + \mathcal{O}(\epsilon), \label{eq_watanabe_marks_inner} \\
q_{{}_i} (T) &= q_{{}_i} (0) + \mathcal{O}(\epsilon). \nonumber
\end{align}

Similarly, for the enzymatic complexes, we have: 

\begin{align}
\allowdisplaybreaks
c_{Z_i} (T) &= c_{Z_i} (0) + \frac{1}{\Omega_E} \mathcal{Y}\of{\Omega_E  \displaystyle \int_0^{T} \kappa_1 \of{\kappa_{1_0} + \displaystyle \sum_{l=1}^N \overline{w}_{il} p_{I_l}} s_{27_i} e_Z \diff T } \nonumber \\&- \frac{1}{\Omega_E} \mathcal{Y}\of{\Omega_E  \displaystyle \int_0^{T} \af{ \kappa_2 \of{\kappa_{2_0} + \displaystyle \sum_{l=1}^N \overline{w}_{il} p_{A_l}} + \kappa_3 } c_{Z_i} \diff T }, \nonumber  \\
c_{M_i} (T) &= c_{M_i} (0) + \frac{1}{\Omega_E} \mathcal{Y}\of{\Omega_E  \displaystyle \int_0^{T} \kappa_4 \of{\kappa_{4_0} + \displaystyle \sum_{l=1}^N \overline{w}_{il} q_{{}_l}} s_{4_i} e_M \diff T } \nonumber \\&- \frac{1}{\Omega_E} \mathcal{Y}\of{\Omega_E  \displaystyle \int_0^{T} \af{ \kappa_5 \of{\kappa_{5_0} + \displaystyle \sum_{l=1}^N \overline{w}_{il} p_{I_l}} + \kappa_6 } c_{M_i} \diff T }, \nonumber \\
c_{C_i} (T) &= c_{C_i} (0) + \frac{1}{\Omega_E} \mathcal{Y}\of{\Omega_E  \displaystyle \int_0^{T} \of{\kappa_{7_0} + \displaystyle \sum_{l=1}^N \overline{w}_{il} p_{A_l}} s_{27_i} e_C \diff T }  \label{eq_watanabe_complexes_inner} \\&- \frac{1}{\Omega_E} \mathcal{Y}\of{\Omega_E  \displaystyle \int_0^{T} \of{\kappa_8 + \kappa_9 } c_{C_i} \diff T }, \nonumber  \\
c_{U_i} (T) &= c_{U_i} (0) + \frac{1}{\Omega_E} \mathcal{Y}\of{\Omega_E  \displaystyle \int_0^{T} \kappa_{10} p_{I_i} e_U \diff T } \nonumber \\& - \frac{1}{\Omega_E} \mathcal{Y}\of{\Omega_E  \displaystyle \int_0^{T} \of{\kappa_{11} + \kappa_{12} } c_{U_i} \diff T }, \nonumber  \\
c_{K_i} (T) &= c_{K_i} (0) + \frac{1}{\Omega_E} \mathcal{Y}\of{\Omega_E  \displaystyle \int_0^{T} \kappa_{13} p_{A_i} e_K \diff T } \nonumber \\& - \frac{1}{\Omega_E} \mathcal{Y}\of{\Omega_E  \displaystyle \int_0^{T} \of{\kappa_{14} + \kappa_{15} } c_{K_i} \diff T }, \nonumber \displaybreak \\
c_{S_i} (T) &= c_{S_i} (0) + \frac{1}{\Omega_E} \mathcal{Y}\of{\Omega_E  \displaystyle \int_0^{T} \kappa_{16} q_{{}_i} e_S \diff T } \nonumber \\& - \frac{1}{\Omega_E} \mathcal{Y}\of{\Omega_E  \displaystyle \int_0^{T} \of{\kappa_{17} + \kappa_{18} } c_{S_i} \diff T }. \nonumber
\end{align}

We note that terms contributing to the dynamics of the epigenetic marks are of order $\mathcal{O}(\epsilon)$. As such, we can assume that these variables remain constant on shorter timescales. As for the case of a single enzymatic reaction, we can also formulate QSS PDFs for enzymatic complexes conditioned on the given (`frozen') distribution of epigenetic marks. We define $\mathbf{p_I} = \uf{p_{I_1}, \ldots, p_{I_N}}$, $\mathbf{p_A} = \uf{p_{A_1}, \ldots, p_{A_N}}$, and $\mathbf{q} = \uf{q_{{}_1}, \ldots, q_{{}_N}}$. Generalising Eq~\eqref{one_enzyme_reaction_pdf} to an enzyme with index $J$ which can bind to $i = 1,\ldots,N$ sites, the QSS PDF conditioned on the static variables for epigenetic modifications reads
\begin{adjustwidth}{-1.5cm}{-0.0cm}
\begin{align} \label{dist_multinomial_enzymeJ_app}
\phi_{J} (E_J, C_{J_1},\ldots,C_{J_N} \mid \mathbf{p_I},\mathbf{p_A},\mathbf{q}) = \frac{\of{E_{J_0}}!}{\of{E_J}! ~ \displaystyle \prod_{i = 1}^N \of{C_{J_i}}!} ~ \of{p_{E_J}}^{E_J} ~ \displaystyle \prod_{i = 1}^N \of{p_{C_{J_i}}}^{C_{J_i}},
\end{align}
\end{adjustwidth}
\noindent where $E_{J_0}$ is the total amount of enzyme $J$ (free and bound) and the probabilities $p_{E_J}$ and $p_{C_{J_i}}$ ($i = 1,\ldots,N$) sum to $1$ and may depend on the slow variables $(\mathbf{p_I},\mathbf{p_A},\mathbf{q})$. This QSS PDF corresponds to a multinomial distribution for the free enzyme of type $J$, $E_J$, and its complexes at each genomic region, $C_{J_i}$. Building on the derivation for a single enzymatic reaction, we now state the QSS multinomial distributions for the enzymatic complexes corresponding to the dynamics described by Eq~\eqref{eq_watanabe_complexes_inner}:
\begin{adjustwidth}{-1.4cm}{-1.4cm}
\begin{subequations} \label{dist_multinomial_enzyme}
\begin{align}
\allowdisplaybreaks
& \phi_{Z} (E_Z, C_{Z_1},\ldots,C_{Z_N} \mid \mathbf{p_I},\mathbf{p_A},\mathbf{q}) = \frac{\of{E_{Z_0}}!}{\of{E_Z}! ~ \displaystyle \prod_{i = 1}^N \of{C_{Z_i}}!} ~ \of{p_{E_Z}}^{E_Z} ~ \displaystyle \prod_{i = 1}^N \of{p_{C_{Z_i}}}^{C_{Z_i}}, \\
&\text{where }\quad p_{E_Z} = \frac{1}{1+\displaystyle \sum_j \frac{\kappa_{1j}}{\kappa_{2j} + \kappa_3}(s_{27_0} - p_{I_j} - q_{{}_j})},  \\ 
& \hspace{43pt} p_{C_{Z_i}} = \frac{\frac{\kappa_{1i}}{\kappa_{2i}+\kappa_3}(s_{27_0} - p_{I_i} - q_{{}_i})}{1+\displaystyle \sum_j \frac{\kappa_{1j}}{\kappa_{2j} + \kappa_3}(s_{27_0} - p_{I_j} - q_{{}_j})}, \nonumber \displaybreak \\[10pt]   
& \phi_{M} (E_M, C_{M_1},\ldots,C_{M_N} \mid \mathbf{p_I},\mathbf{p_A},\mathbf{q}) = \frac{\of{E_{M_0}}!}{\of{E_M}! ~ \displaystyle \prod_{i = 1}^N \of{C_{M_i}}!} ~ \of{p_{E_M}}^{E_M} ~ \displaystyle \prod_{i = 1}^N \of{p_{C_{M_i}}}^{C_{M_i}},   \\
&\text{where }\quad p_{E_M} = \frac{1}{1+\displaystyle \sum_j \frac{\kappa_{4j}}{\kappa_{5j} + \kappa_6}(s_{4_0} - p_{A_j})}, \\
& \hspace{43pt} p_{C_{M_i}} = \frac{\frac{\kappa_{4i}}{\kappa_{5i}+\kappa_6}(s_{4_0} - p_{A_i})}{1+\displaystyle \sum_j \frac{\kappa_{4j}}{\kappa_{5j} + \kappa_6}(s_{4_0} - p_{A_j})}, \nonumber \\[10pt]
& \phi_{C} (E_C, C_{C_1},\ldots,C_{C_N} \mid \mathbf{p_I},\mathbf{p_A},\mathbf{q}) = \frac{\of{E_{C_0}}!}{\of{E_C}! ~ \displaystyle \prod_{i = 1}^N \of{C_{C_i}}!} ~ \of{p_{E_C}}^{E_C} ~ \displaystyle \prod_{i = 1}^N \of{p_{C_{C_i}}}^{C_{C_i}}, \\
&\text{where }\quad p_{E_C} = \frac{1}{1+\displaystyle \sum_j \frac{\kappa_{7j}}{\kappa_{8} + \kappa_9}(s_{27_0} - p_{I_j} - q_{{}_j})}, \\
& \hspace{43pt} p_{C_{C_i}} = \frac{\frac{\kappa_{7i}}{\kappa_{8}+\kappa_9}(s_{27_0} - p_{I_i} - q_{{}_i})}{1+\displaystyle \sum_j \frac{\kappa_{7j}}{\kappa_{8} + \kappa_9}(s_{27_0} - p_{I_j} - q_{{}_j})},  \nonumber \\[10pt]
& \phi_{U} (E_U, C_{U_1},\ldots,C_{U_N} \mid \mathbf{p_I}) = \frac{\of{E_{U_0}}!}{\of{E_U}! ~ \displaystyle \prod_{i = 1}^N \of{C_{U_i}}!} ~ \of{p_{E_U}}^{E_U} ~ \displaystyle \prod_{i = 1}^N \of{p_{C_{U_i}}}^{C_{U_i}}, \\
&\text{where }\quad p_{E_U} = \frac{1}{1+\displaystyle \sum_j \frac{\kappa_{10}}{\kappa_{11} + \kappa_{12}} p_{I_j}}, \qquad p_{C_{U_i}} = \frac{\frac{\kappa_{10}}{\kappa_{11}+\kappa_{12}} p_{I_i} }{1+\displaystyle \sum_j \frac{\kappa_{10}}{\kappa_{11} + \kappa_{12}} p_{I_j} }, \\[10pt]
& \phi_{K} (E_K, C_{K_1},\ldots,C_{K_N} \mid \mathbf{p_A}) = \frac{\of{E_{K_0}}!}{\of{E_K}! ~ \displaystyle \prod_{i = 1}^N \of{C_{K_i}}!}  ~ \of{p_{E_K}}^{E_K} ~ \displaystyle \prod_{i = 1}^N \of{p_{C_{K_i}}}^{C_{K_i}},  \\
&\text{where }\quad p_{E_K} = \frac{1}{1+\displaystyle \sum_j \frac{\kappa_{13}}{\kappa_{14} + \kappa_{15}} p_{A_j}}, \qquad p_{C_{K_i}} = \frac{\frac{\kappa_{13}}{\kappa_{14}+\kappa_{15}} p_{A_i} }{1+\displaystyle \sum_j \frac{\kappa_{13}}{\kappa_{14} + \kappa_{15}} p_{A_j} }, \displaybreak \\[10pt]
& \phi_{S} (E_S, C_{S_1},\ldots,C_{S_N} \mid \mathbf{q}) = \frac{\of{E_{S_0}}!}{\of{E_S}! ~ \displaystyle \prod_{i = 1}^N \of{C_{S_i}}!} ~ \of{p_{E_S}}^{E_S} ~ \displaystyle \prod_{i = 1}^N \of{p_{C_{S_i}}}^{C_{S_i}}, \\
&\text{where }\quad p_{E_S} = \frac{1}{1 +  \displaystyle \sum_j \frac{\kappa_{16}}{\kappa_{17} + \kappa_{18}} q_{{}_j}}, \qquad p_{C_{S_i}} = \frac{\frac{\kappa_{16}}{\kappa_{17}+\kappa_{18}} q_{{}_i} }{1+ + \displaystyle \sum_j \frac{\kappa_{16}}{\kappa_{17} + \kappa_{18}} q_{{}_j} }.
\end{align}
\end{subequations}
\end{adjustwidth}
\noindent Here, we used the conservation laws, Eqs~\eqref{conservation_law_lysine27}-\eqref{conservation_law_lysine4}, and the scaling given by Eq~\eqref{eq_scaling} to obtain $s_{27_i} = s_{27_0} - p_{I_i} - q_{{}_i} + \mathcal{O}(\epsilon) \approx s_{27_0} - p_{I_i} - q_{{}_i}$ and $s_{4_i} = s_{4_0} - p_{A_i} + \mathcal{O}(\epsilon) \approx s_{4_0} - p_{A_i}$, for any $i = 1,\ldots,N$. In the above probabilities, the summation is over all genomic regions ($j = 1,\ldots,N$). In addition, we have introduced the following kinetic constants: 

\begin{align} 
& \kappa_{1i} = \kappa_1 \of{\kappa_{1_0} + \displaystyle \sum_{l=1}^N \overline{w}_{il} p_{I_l}}, \quad &&  \kappa_{2i} = \kappa_2 \of{\kappa_{2_0} + \displaystyle \sum_{l=1}^N \overline{w}_{il} p_{A_l}}, \nonumber \\
&  \kappa_{4i} = \kappa_4 \of{\kappa_{4_0} + \displaystyle \sum_{l=1}^N \overline{w}_{il} q_{{}_l}},
&& \kappa_{5i} = \kappa_5 \of{\kappa_{5_0} + \displaystyle \sum_{l=1}^N \overline{w}_{il} p_{I_l}}, \label{eq_constants_kappai} \\
& \kappa_{7i} = \of{\kappa_{7_0} + \displaystyle \sum_{l=1}^N \overline{w}_{il} p_{A_l}}. \nonumber
\end{align}
\noindent We emphasise that the kinetic rate functions incorporate reinforcing and antagonistic feedback mechanisms between epigenetic marks and the chromatin architecture.

\subsection*{Outer solution} 
The outer solution, also known as the quasi-steady-state approximation (QSSA), describes the evolution of the slow variables (i.e. epigenetic marks) when the fast variables have relaxed onto their QSS PDFs. From Eq~\eqref{eq_watanabe_complexes_tau}, $\epsilon \of{ C_{J_i}(\tau) - C_{J_i}(0)} \approx 0$, and thus, we can simplify the equations for the dynamics of histone modifications as follows:

\begin{align} \label{eq_watanabe_marks_outer1}
p_{I_i} (\tau+ \Delta \tau) &= p_{I_i} (\tau) + \frac{1}{\Omega_S} \mathcal{Y}\of{\Omega_S \displaystyle \int_{\tau}^{\Delta \tau} \kappa_3 c_{Z_i} \diff \tau } \\
&- \frac{1}{\Omega_S} \mathcal{Y}\of{\Omega_S \displaystyle \int_{\tau}^{\Delta \tau} \kappa_{12} c_{U_i} \diff \tau }, \nonumber \displaybreak \\ \label{eq_watanabe_marks_outer2}
p_{A_i} (\tau + \Delta \tau) &= p_{A_i} (\tau) + \frac{1}{\Omega_S} \mathcal{Y}\of{\Omega_S \displaystyle \int_{\tau}^{\Delta \tau} \kappa_6 c_{M_i} \diff \tau } \\
& - \frac{1}{\Omega_S} \mathcal{Y}\of{\Omega_S \displaystyle \int_{\tau}^{\Delta \tau} \kappa_{15} c_{K_i} \diff \tau }, \nonumber \\ \label{eq_watanabe_marks_outer3}
q_{{}_i} (\tau + \Delta \tau) &= q_{{}_i} (\tau) + \frac{1}{\Omega_S} \mathcal{Y}\of{\Omega_S \displaystyle \int_{\tau}^{\Delta \tau} \kappa_9 c_{C_i} \diff \tau } \\
& - \frac{1}{\Omega_S} \mathcal{Y}\of{\Omega_S \displaystyle \int_{\tau}^{\Delta \tau} \kappa_{18} c_{S_i} \diff \tau }. \nonumber
\end{align}

These equations describe the stochastic dynamics for the addition and removal of epigenetic marks. The former process is described by the first Poisson counting process (second term) in Eqs~\eqref{eq_watanabe_marks_outer1}-\eqref{eq_watanabe_marks_outer3}, while the removal reactions are described by the last terms in these equations. The corresponding reactions of the reduced system (after the separation of timescales) can be found in Eq~\eqref{marks_reduced} of the main text. This stochastic system can be simulated numerically using stochastic simulation algorithms such as the Gillespie algorithm \citep{gillespie1976general}. At each time step, the number of enzyme complexes at each genomic site, $c_{J_i}$ for $J \in \uf{M,Z,C,K,U,S}$, is sampled from the corresponding QSS multinomial distribution with the PDF, $\phi_{J} (E_J, C_{J_1},\ldots,C_{J_N} \mid \mathbf{p_I},\mathbf{p_A},\mathbf{q})$, given by Eq~\eqref{dist_multinomial_enzyme}. For all stochastic simulations presented in this work, we use Eqs~\eqref{eq_watanabe_marks_outer1}-\eqref{eq_watanabe_marks_outer3}.

\subsection*{Mean-field limit} 

In order to perform the stability analysis of the model, it is useful to derive the mean-field equations associated with our stochastic model (Eqs~\eqref{eq_watanabe_marks_outer1}-\eqref{eq_watanabe_marks_outer3} or, equivalently, reactions Eq~\eqref{marks_reduced} in the main text). To do so, we note that, due to the separation of the timescales, the dynamics of the enzymatic complexes (fast variables) can be approximated by the corresponding QSS probabilities as follows \citep{cao2005slow}

\begin{align}
c_{J_i} \simeq \langle c_{J_i} \mid  \mathbf{p_I},\mathbf{p_A},\mathbf{q} \rangle = e_{J_0} p_{C_{J_i}} (\mathbf{p_I},\mathbf{p_A},\mathbf{q}),
\end{align} 
with $J \in \uf{M,Z,C,K,U,S}$ and the QSS probabilities given by Eq~\eqref{dist_multinomial_enzyme}. Under these assumptions, Eqs~\eqref{eq_watanabe_marks_outer1}-\eqref{eq_watanabe_marks_outer3} become:

\begin{align} \label{eq_watanabe_marks_final_outer1}
p_{I_i} (\tau+ \Delta \tau) &= p_{I_i} (\tau) + \frac{1}{\Omega_S} \mathcal{Y}\of{\Omega_S \displaystyle \int_{\tau}^{\Delta \tau} \kappa_3 e_{Z_0} p_{C_{Z_i}} (\mathbf{p_I},\mathbf{p_A},\mathbf{q}) \diff \tau } \\ &- \frac{1}{\Omega_S} \mathcal{Y}\of{\Omega_S \displaystyle \int_{\tau}^{\Delta \tau} \kappa_{12} e_{U_0} p_{C_{U_i}} (\mathbf{p_I}) \diff \tau }, \nonumber \displaybreak \\
 \label{eq_watanabe_marks_final_outer2}
p_{A_i} (\tau + \Delta \tau) &= p_{A_i} (\tau) + \frac{1}{\Omega_S} \mathcal{Y}\of{\Omega_S \displaystyle \int_{\tau}^{\Delta \tau} \kappa_6 e_{M_0} p_{C_{M_i}} (\mathbf{p_I},\mathbf{p_A},\mathbf{q}) \diff \tau } \\ &- \frac{1}{\Omega_S} \mathcal{Y}\of{\Omega_S \displaystyle \int_{\tau}^{\Delta \tau} \kappa_{15} e_{K_0} p_{C_{K_i}} (\mathbf{p_A}) \diff \tau }, \nonumber\\ 
\label{eq_watanabe_marks_final_outer3}
q_{{}_i} (\tau + \Delta \tau) &= q_{{}_i} (\tau) + \frac{1}{\Omega_S} \mathcal{Y}\of{\Omega_S \displaystyle \int_{\tau}^{\Delta \tau} \kappa_9 e_{C_0} p_{C_{C_i}} (\mathbf{p_I},\mathbf{p_A},\mathbf{q}) \diff \tau } \\ &- \frac{1}{\Omega_S} \mathcal{Y}\of{\Omega_S \displaystyle \int_{\tau}^{\Delta \tau} \kappa_{18} e_{S_0} p_{C_{S_i}} (\mathbf{q}) \diff \tau }. \nonumber
\end{align}

We can use Eqs~\eqref{eq_watanabe_marks_final_outer1}-\eqref{eq_watanabe_marks_final_outer1} to derive mean-field equations for the temporal evolution of epigenetic modifications \citep{anderson2015stochastic}. We formulate the corresponding ordinary differential equations (ODEs)

\begin{align}
\dv{p_{I_i}}{\tau} &=  \kappa_3 e_{Z_0} p_{C_{Z_i}} (\mathbf{p_I},\mathbf{p_A},\mathbf{q}) - \kappa_{12} e_{U_0} p_{C_{U_i}} (\mathbf{p_I}),  \nonumber \\
\dv{p_{A_i}}{\tau} &= \kappa_6 e_{M_0} p_{C_{M_i}} (\mathbf{p_I},\mathbf{p_A},\mathbf{q}) - \kappa_{15} e_{K_0} p_{C_{K_i}} (\mathbf{p_A}), \label{eq_marks_meanfield_app} \\
\dv{q_{{}_i}}{\tau} &= \kappa_9 e_{C_0} p_{C_{C_i}} (\mathbf{p_I},\mathbf{p_A},\mathbf{q}) - \kappa_{18} e_{S_0} p_{C_{S_i}} (\mathbf{q}), \nonumber
\end{align}
\noindent which describe the dynamics of H3K27me3, H3K4me3 and H3K27ac marks, respectively. We explicitly note the dependency of QSS probabilities for enzymatic complexes, $p_{C_{J_i}}$, on the levels of epigenetic modifications (e.g. $p_{C_{Z_i}}  = p_{C_{Z_i}} (\mathbf{p_I},\mathbf{p_A},\mathbf{q})$). Finally, by incorporating into these equations the expressions for $p_{C_{J_i}}$ given by Eq~\eqref{dist_multinomial_enzyme}, we obtain the mean-field system of equations for the dynamics of epigenetic marks:

\begin{adjustwidth}{-0.7cm}{-0.7cm}
\begin{align}
\text{H3K27me3:}~~ &\dv{p_{I_i}}{\tau} =  \frac{\kappa_3 e_{Z_0} k_{1_i}(\mathbf{p_I},\mathbf{p_A})\of{s_{27_0} - p_{I_i} - q_{{}_i}}}{1+\displaystyle \sum_j  k_{1_j}(\mathbf{p_I},\mathbf{p_A}) \of{s_{27_0} - p_{I_j} - q_{{}_j}}}  - \frac{\kappa_{12} e_{U_0} \alpha_{10} ~p_{I_i}}{1+ \alpha_{10} \displaystyle \sum_j p_{I_j}}, \label{eq_marks_meanfield_final1} \\
\text{H3K4me3:} ~~~ &\dv{p_{A_i}}{\tau} =  \frac{\kappa_6 e_{M_0} k_{4_i}(\mathbf{p_I},\mathbf{q}) \of{s_{4_0} - p_{A_i}}}{1+\displaystyle \sum_j  k_{4_j}(\mathbf{p_I},\mathbf{q}) \of{s_{4_0} - p_{A_j}}}  - \frac{\kappa_{15} e_{K_0} \alpha_{13} ~p_{A_i}}{1+ \alpha_{13} \displaystyle \sum_j p_{A_j}},  \label{eq_marks_meanfield_final2} \\
\text{H3K27ac:} ~~ ~&\dv{q_{{}_i}}{\tau} ~ =  \frac{\kappa_9 e_{C_0} k_{7_i}(\mathbf{p_I},\mathbf{p_A})\of{s_{27_0} - p_{I_i} - q_{{}_i}}}{1+\displaystyle \sum_j  k_{7_j}(\mathbf{p_I},\mathbf{p_A}) \of{s_{27_0} - p_{I_j} - q_{{}_j}}}  - \frac{\kappa_{18} e_{S_0} \alpha_{16} ~ q_{{}_i}}{1 + \alpha_{16} \displaystyle \sum_j q_{{}_j}}.  \label{eq_marks_meanfield_final3}
\end{align}
 \end{adjustwidth}
\noindent Here, the summation is performed over all genomic coordinates ($j = 1,\ldots,N$), and we introduced the following notation:
\begin{adjustwidth}{-1.2cm}{-0.7cm}
\begin{align}
& k_{1_i}(\mathbf{p_I},\mathbf{p_A}) = \frac{\alpha_{I_0} + \kappa_1 \displaystyle \sum_l \overline{w}_{il} \skpp p_{I_l}}{\overline{\kappa}_3 + \kappa_2 \displaystyle \sum_l \overline{w}_{il} \skpp p_{A_l}}, \hspace{5pt} && \overline{\kappa}_3 = \kappa_3 + \kappa_2 \kappa_{2_0}, \hspace{3pt} &&& \alpha_{I_0} = \kappa_1 \kappa_{1_0}, \hspace{22pt}  \nonumber\\
& k_{4_i}(\mathbf{p_I},\mathbf{q}) = \frac{\alpha_{A_0} + \kappa_4 \displaystyle \sum_l \overline{w}_{il} \skpp q_{{}_l}}{\overline{\kappa}_6 + \kappa_5 \displaystyle \sum_l \overline{w}_{il} \skpp p_{I_l} },  && \overline{\kappa}_6 = \kappa_6 + \kappa_5 \kappa_{5_0},  &&& \alpha_{A_0} = \kappa_4 \kappa_{4_0},  \hspace{22pt}  \label{eq_meanfield_constants}  \\
& k_{7_i}(\mathbf{p_I},\mathbf{p_A}) = \frac{1}{\alpha_8}\of{\alpha_{Q_0} +  \displaystyle \sum_l \overline{w}_{il} \skpp p_{A_l}},  && \alpha_8 = \kappa_8 + \kappa_9,  &&& \alpha_{Q_0} = \kappa_{7_0}, \hspace{33pt}   \nonumber \\
& \alpha_{10} = \frac{\kappa_{10}}{\kappa_{11}+\kappa_{12}}, &&  \alpha_{13} = \frac{\kappa_{13}}{\kappa_{14}+\kappa_{15}}, &&& \alpha_{16} = \frac{\kappa_{16}}{\kappa_{17}+\kappa_{18}}. \nonumber
\end{align}
 \end{adjustwidth}
 
 We use the mean-field Eqs~\eqref{eq_marks_meanfield_final1}-\eqref{eq_marks_meanfield_final3} for the steady state analysis. This analysis is performed for the simple cases of two interacting genomic sites (see Section~\ref{secII}) and two interacting regions for a specific case of $W$ matrix (see Section~\ref{secIV}).
 
\section{Mean-field equations for the two-site system}
\label{secII}

For completeness, we state below the equations for epigenetic regulation of the two-site model. We use them to perform stability analysis of our model. The governing equations are obtained from Eqs~\eqref{eq_marks_meanfield_final1}-\eqref{eq_marks_meanfield_final3} by fixing the number of genomic sites for enzyme binding, $N=2$, which leads to the following system of ODEs:
\begin{adjustwidth}{-0.7cm}{0.0cm}
\begin{align}
\dv{p_{I_i}}{\tau} &=  \frac{\kappa_3 e_{Z_0} k_{1_i}\of{s_{27_0} - p_{I_i} - q_{{}_i}}}{1+ k_{1_1}\of{s_{27_0} - p_{I_1} - q_{{}_1}} + k_{1_2}\of{s_{27_0} - p_{I_2} - q_{{}_2}}}  - \frac{\kappa_{12} e_{U_0} \alpha_{10} ~p_{I_i}}{1+ \alpha_{10} \of{ p_{I_1} + p_{I_2}}}, \label{eq_two_sites_full1} \\
\dv{p_{A_i}}{\tau} &=  \frac{\kappa_6 e_{M_0} k_{4_i} \of{s_{4_0} - p_{A_i}}}{1+ k_{4_1}\of{s_{4_0} - p_{A_1}} + k_{4_2}\of{s_{4_0} - p_{A_2}} }  - \frac{\kappa_{15} e_{K_0} \alpha_{13} ~p_{A_i}}{1+ \alpha_{13}\of{ p_{A_1} +  p_{A_2} }},  \label{eq_two_sites_full2}  \\
\dv{q_{{}_i}}{\tau} &=  \frac{\kappa_9 e_{C_0} k_{7_i}\of{s_{27_0} - p_{I_i} - q_{{}_i}}}{1+ k_{7_1}\of{s_{27_0} - p_{I_1} - q_{{}_1}} + k_{7_2}\of{s_{27_0} - p_{I_2} - q_{{}_2}}}  - \frac{\kappa_{18} e_{S_0} \alpha_{16} ~q_{{}_i}}{1 + \alpha_{16} \of{ q_{{}_1} + q_{{}_2}}}. \label{eq_two_sites_full3} 
\end{align}
\end{adjustwidth}
\noindent Here, $i = 1,2$ and all notation is as described in Section~\ref{secI}. The functional forms of $k_{1_i}$, $k_{4_i}$ and $k_{7_i}$ depend on our assumptions about the interactions between genomic sites (see the main text). For the case of H3K27me3-dependent interaction frequency, $\overline{w}_{il} = w_{il}\skpp p_{I_i} p_{I_l}$ and we have:

\begin{align}
& k_{1_1} = \frac{\alpha_{I_0} + \kappa_1\of{ w_{11} p_{I_1}^3 + w_{12} p_{I_1} p_{I_2}^2} }{\overline{\kappa}_3 + \kappa_2 \of{w_{11} p_{I_1}^2 p_{A_1} + w_{12} p_{I_1} p_{I_2} p_{A_2}}},  \nonumber \\
& k_{1_2} = \frac{\alpha_{I_0} + \kappa_1\of{ w_{21} p_{I_1}^2 p_{I_2} + w_{22} p_{I_2}^3} }{\overline{\kappa}_3 + \kappa_2 \of{w_{21} p_{I_1} p_{I_2} p_{A_1} + w_{22} p_{I_2}^2 p_{A_2}}},  \nonumber \\
& k_{4_1} = \frac{\alpha_{A_0} + \kappa_4\of{ w_{11} p_{I_1}^2 q_{{}_1} + w_{12} p_{I_1} p_{I_2} q_{{}_2}}}{\overline{\kappa}_6 + \kappa_5\of{ w_{11} p_{I_1}^3 + w_{12} p_{I_1} p_{I_2}^2}}, \label{eq_two_sites_full_k} \\
&  k_{4_2} = \frac{\alpha_{A_0} + \kappa_4\of{ w_{21} p_{I_1} p_{I_2} q_{{}_1} + w_{22} p_{I_2}^2 q_{{}_2}}}{\overline{\kappa}_6 + \kappa_5\of{ w_{21} p_{I_1}^2  p_{I_2} + w_{22} p_{I_2}^3}},  \nonumber \\
& k_{7_1} = \frac{1}{\alpha_8}\of{\alpha_{Q_0} +  w_{11} p_{I_1}^2 p_{A_1} + w_{12} p_{I_1} p_{I_2} p_{A_2} }, \nonumber \\
& k_{7_2} = \frac{1}{\alpha_8}\of{\alpha_{Q_0} +  w_{21} p_{I_1} p_{I_2} p_{A_1} + w_{22} p_{I_2}^2 p_{A_2} }. \nonumber
\end{align}
\noindent For the case of constant interaction strength, which depends only on the frequency of contact in space, $\overline{w}_{il} = w_{il}$, we have instead:

\begin{align}
& k_{1_1} = \frac{\alpha_{I_0} + \kappa_1\of{ w_{11} p_{I_1} + w_{12} p_{I_2}} }{\overline{\kappa}_3 + \kappa_2 \of{w_{11}  p_{A_1} + w_{12} p_{A_2}}},  && k_{1_2} = \frac{\alpha_{I_0} + \kappa_1\of{ w_{21} p_{I_1}  + w_{22} p_{I_2}} }{\overline{\kappa}_3 + \kappa_2 \of{w_{21} p_{A_1} + w_{22} p_{A_2}}},   \nonumber \\
& k_{4_1} = \frac{\alpha_{A_0} + \kappa_4\of{ w_{11}  q_{{}_1} + w_{12} q_{{}_2}}}{\overline{\kappa}_6 + \kappa_5\of{ w_{11} p_{I_1} + w_{12} p_{I_2}}}, &&  k_{4_2} = \frac{\alpha_{A_0} + \kappa_4\of{ w_{21} q_{{}_1} + w_{22} q_{{}_2}}}{\overline{\kappa}_6 + \kappa_5\of{ w_{21} p_{I_1} + w_{22} p_{I_2}}}, \label{eq_two_sites_full_k_const} \\
& k_{7_1} = \frac{1}{\alpha_8}\of{\alpha_{Q_0} +  w_{11}  p_{A_1} + w_{12} p_{A_2} }, && k_{7_2} = \frac{1}{\alpha_8}\of{\alpha_{Q_0} +  w_{21} p_{A_1} + w_{22} p_{A_2} }. \nonumber
\end{align}

\section{Scaling for the multi-site model}
\label{secIII}

In this section, we list the quasi-steady-state (QSS) probabilities for multinomial enzyme distributions and the corresponding mean-field equations for the multi-site model. Under the scaling introduced in Section~\ref{sec:results_scaling}, the QSS probabilities from Eq~\eqref{dist_multinomial_enzyme} become:

\begin{align}
\allowdisplaybreaks
p_{E_Z} & = \frac{1}{1+\displaystyle \sum_j k_{1_j}(\mathbf{p_I},\mathbf{p_A}) {\color{mulberry}{\frac{2}{N}}} (s_{27_0} - p_{I_j} - q_{{}_j})},  &&  \nonumber\\
p_{C_{Z_i}} &= \frac{k_{1_i}(\mathbf{p_I},\mathbf{p_A}) {\color{mulberry}{\frac{2}{N}}} (s_{27_0} - p_{I_i} - q_{{}_i})}{1+\displaystyle \sum_j k_{1_j}(\mathbf{p_I},\mathbf{p_A}) {\color{mulberry}{\frac{2}{N}}} (s_{27_0} - p_{I_j} - q_{{}_j})}, \nonumber \\[0pt]   
p_{E_M} &= \frac{1}{1+\displaystyle \sum_j k_{4_j}(\mathbf{p_I},\mathbf{q}) {\color{mulberry}{\frac{2}{N}}}(s_{4_0} - p_{A_j})},  && \nonumber \\
p_{C_{M_i}} &= \frac{k_{4_i}(\mathbf{p_I},\mathbf{q}) {\color{mulberry}{\frac{2}{N}}}(s_{4_0} - p_{A_i})}{1+\displaystyle \sum_j k_{4_j}(\mathbf{p_I},\mathbf{q}) {\color{mulberry}{\frac{2}{N}}}(s_{4_0} - p_{A_j})}, \nonumber \\[0pt]
p_{E_C} &= \frac{1}{1+\displaystyle \sum_j k_{7_j}(\mathbf{p_I},\mathbf{p_A}) {\color{mulberry}{\frac{2}{N}}}(s_{27_0} - p_{I_j} - q_{{}_j})},  &&  \label{qss_probabilities_rescaled} \\
p_{C_{C_i}} & = \frac{k_{7_i}(\mathbf{p_I},\mathbf{p_A}){\color{mulberry}{\frac{2}{N}}}(s_{27_0} - p_{I_i} - q_{{}_i})}{1+\displaystyle \sum_j k_{7_j}(\mathbf{p_I},\mathbf{p_A}) {\color{mulberry}{\frac{2}{N}}}(s_{27_0} - p_{I_j} - q_{{}_j})},  \nonumber \\[0pt]
p_{E_U} & = \frac{1}{1+\alpha_{10} \displaystyle \sum_j {\color{mulberry}{\frac{2}{N}}} \skpp p_{I_j}},  && p_{C_{U_i}} = \frac{\alpha_{10} {\color{mulberry}{\frac{2}{N}}} \skpp p_{I_i} }{1+\alpha_{10} \displaystyle \sum_j {\color{mulberry}{\frac{2}{N}}} \skpp p_{I_j} }, \nonumber \\[0pt]
p_{E_K} & = \frac{1}{1+\alpha_{13} \displaystyle \sum_j {\color{mulberry}{\frac{2}{N}}} \skpp p_{A_j}}, && p_{C_{K_i}} = \frac{\alpha_{13} {\color{mulberry}{\frac{2}{N}}} \skpp p_{A_i} }{1+\alpha_{13} \displaystyle \sum_j {\color{mulberry}{\frac{2}{N}}} \skpp p_{A_j} }, \nonumber \\[0pt]
p_{E_S} &= \frac{1}{1 +  \alpha_{16} \displaystyle \sum_j {\color{mulberry}{\frac{2}{N}}} \skpp q_{{}_j}}, && p_{C_{S_i}} = \frac{  \alpha_{16} {\color{mulberry}{\frac{2}{N}}} \skpp q_{{}_i} }{1 +  \alpha_{16} \displaystyle \sum_j {\color{mulberry}{\frac{2}{N}}} \skpp q_{{}_j} }. \nonumber
\end{align}
\noindent Here, the kinetic rates are as introduced in Eq~\eqref{eq_meanfield_constants}. The rescaling of the mean-filed equations given by Eqs~\eqref{eq_marks_meanfield_final1}-\eqref{eq_marks_meanfield_final3} yields the following:
\begin{adjustwidth}{-1.2cm}{-0.0cm}
\begin{align}
\text{H3K27me3:}~~ &\dv{p_{I_i}}{\tau} =  \frac{\kappa_3 e_{Z_0} k_{1_i}(\mathbf{p_I},\mathbf{p_A}){\color{mulberry}{\frac{2}{N}}} \of{s_{27_0} - p_{I_i} - q_{{}_i}}}{1+\displaystyle \sum_j  k_{1_j}(\mathbf{p_I},\mathbf{p_A}) {\color{mulberry}{\frac{2}{N}}} \of{s_{27_0} - p_{I_j} - q_{{}_j}}}  - \frac{\kappa_{12} e_{U_0} \alpha_{10} {\color{mulberry}{\frac{2}{N}}} \skpp p_{I_i}}{1+ \alpha_{10} {\color{mulberry}{\frac{2}{N}}} \displaystyle \sum_j p_{I_j}}, \label{eq_marks_meanfield_final1_rescaled} \\
\text{H3K4me3:} ~~~ &\dv{p_{A_i}}{\tau} =  \frac{\kappa_6 e_{M_0} k_{4_i}(\mathbf{p_I},\mathbf{q}){\color{mulberry}{\frac{2}{N}}} \of{s_{4_0} - p_{A_i}}}{1+\displaystyle \sum_j  k_{4_j}(\mathbf{p_I},\mathbf{q}) {\color{mulberry}{\frac{2}{N}}} \of{s_{4_0} - p_{A_j}}}  - \frac{\kappa_{15} e_{K_0} \alpha_{13} {\color{mulberry}{\frac{2}{N}}} \skpp p_{A_i}}{1+ \alpha_{13} {\color{mulberry}{\frac{2}{N}}} \displaystyle \sum_j p_{A_j}},  \label{eq_marks_meanfield_final2_rescaled} \\
\text{H3K27ac:} ~~ ~&\dv{q_{{}_i}}{\tau} ~ =  \frac{\kappa_9 e_{C_0} k_{7_i}(\mathbf{p_I},\mathbf{p_A}) {\color{mulberry}{\frac{2}{N}}} \of{s_{27_0} - p_{I_i} - q_{{}_i}}}{1+\displaystyle \sum_j  k_{7_j}(\mathbf{p_I},\mathbf{p_A}) {\color{mulberry}{\frac{2}{N}}} \of{s_{27_0} - p_{I_j} - q_{{}_j}}}  - \frac{\kappa_{18} e_{S_0} \alpha_{16} {\color{mulberry}{\frac{2}{N}}} \skpp q_{{}_i}}{1+ \alpha_{16} {\color{mulberry}{\frac{2}{N}}} \displaystyle \sum_j q_{{}_j}}.  \label{eq_marks_meanfield_final3_rescaled}
\end{align}
\end{adjustwidth}

\section{Mean-field equations for the two-region system}
\label{secIV}

\begin{figure}[!h]
\begin{center}
\includegraphics[width=0.6\textwidth]{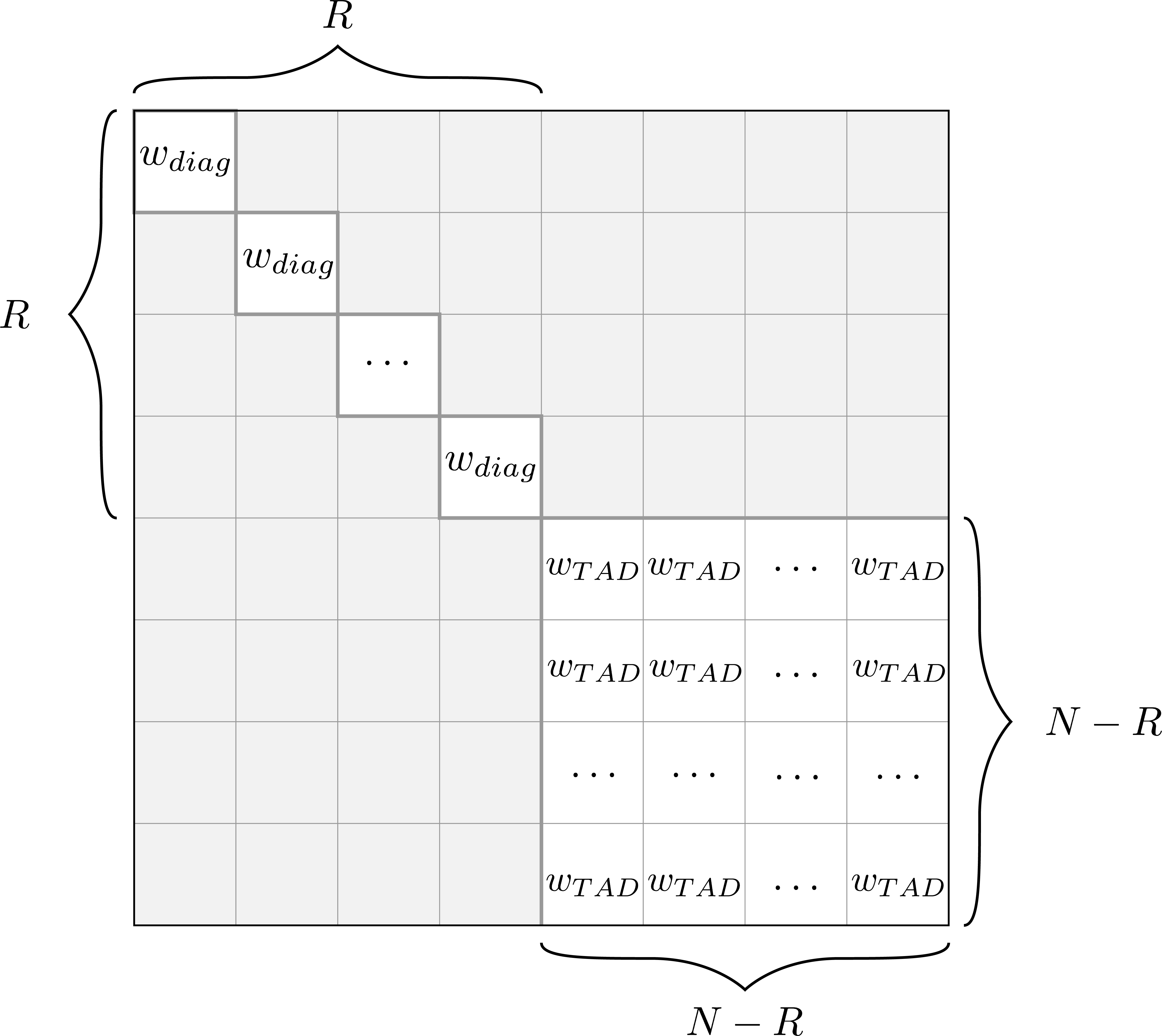}
\caption{An illustration of a contact frequency matrix, $W$, for the system with two regions. The first region of $R$ sites exhibits connections only within the same site with strength, $w_{diag}$. The second, dynamically interacting TAD-like region has a size of $(N-R)$ sites. Within the TAD region, all sites are connected among themselves with the same contact frequency of $w_{TAD}$. The shaded grey colour indicates zero interaction frequency.}
\label{SM_Figure_1}
\end{center}
\end{figure}

We now consider a chromatin architecture that comprises two regions (see Figure~\ref{SM_Figure_1}). The first block of size $R$ consists of genomic sites that only interact within the same site, with frequency, $w_{diag}$. The second region is characterised by frequent interactions among all its loci. It can be interpreted as a small TAD region. The frequency of these all-with-all interactions for the second region is given by $w_{TAD}$. The length of the TAD-like region is $(N-R)$. Given this structure of the contact frequency matrix, $W$, it is possible to simplify the governing Eqs~\eqref{eq_meanfield_constants},~\eqref{eq_marks_meanfield_final1_rescaled}-\eqref{eq_marks_meanfield_final3_rescaled}. Here, we will only consider the case of H3K27me3-dependent interaction strength. The scenario of constant interaction strength can be derived in a similar manner. Thus, for $\overline{w}_{il} = w_{il} p_{I_i} p_{I_l}$, the rate functions, $k_{1_i}$, $k_{4_i}$ and $k_{7_i}$, simplify to:
\begin{adjustwidth}{-1cm}{-0.3cm}
\begin{align}
\allowdisplaybreaks
& \text{for } i = \overline{1,R} \nonumber \\
& k_{1_i} = \frac{\alpha_{I_0} + \kappa_1 w_{diag} ~p_{I_i}^3}{\overline{\kappa}_3 + \kappa_2 w_{diag} ~p_{I_i}^2 p_{A_i}}, && k_{4_i} = \frac{\alpha_{A_0} + \kappa_4 w_{diag} ~p_{I_i}^2 q_{{}_i}}{\overline{\kappa}_6 + \kappa_5 w_{diag} ~p_{I_i}^3}, \label{eq_rates_subregion1} \\
& k_{7_i}= \frac{1}{\alpha_8}\of{\alpha_{Q_0} + w_{diag} ~p_{I_i}^2 p_{A_i}}, && \nonumber \\[10pt]
&  \text{and for }i = \overline{R+1,N} && \nonumber \\
& k_{1_i} = \frac{\alpha_{I_0} + \kappa_1 w_{TAD} ~p_{I_i} \displaystyle \sum_{l=R+1}^N p_{I_l}^2}{\overline{\kappa}_3 + \kappa_2 w_{TAD} ~p_{I_i} \displaystyle \sum_{l=R+1}^N p_{I_l} p_{A_l}}, && k_{4_i} = \frac{\alpha_{A_0} + \kappa_4 w_{TAD} ~p_{I_i} \displaystyle \sum_{l=R+1}^N p_{I_l} q_{{}_l}}{\overline{\kappa}_6 + \kappa_5 w_{TAD} ~p_{I_i} \displaystyle \sum_{l=R+1}^N p_{I_l}^2}, \label{eq_rates_subregion2}  \\
& k_{7_i}= \frac{1}{\alpha_8}\of{\alpha_{Q_0} + w_{TAD} ~p_{I_i}  \displaystyle \sum_{l=R+1}^N p_{I_l} p_{A_l}}. && \nonumber
\end{align}
\end{adjustwidth}

Since the rate functions defined in Eqs~\eqref{eq_rates_subregion1}-\eqref{eq_rates_subregion2} are identical for all genomic loci within the \textit{same} region and assuming identical initial conditions for sites from the same domain, the system dynamics can be further simplified to just two equations, one for each region. Thus, we denote by $\widetilde{p}_{I_1} = p_{I_i}$, $\widetilde{p}_{A_1} = p_{A_i}$ and $\widetilde{q}_{{}_1} = q_{{}_i}$ for $i=1,\ldots,R$ the representative variables of epigenetic profile for the first region. Similarly, we have $\widetilde{p}_{I_2} = p_{I_i}$, $\widetilde{p}_{A_2} = p_{A_i}$ and $\widetilde{q}_{{}_2} = q_{{}_i}$ for $i=R+1,\ldots,N$ corresponding to the second TAD-like region. Using this notation and the simplified rate functions from Eqs~\eqref{eq_rates_subregion1}-\eqref{eq_rates_subregion2}, we obtain the following system of equations for the two-region case:

\begin{align}
 \label{eq_two_subregions1} 
\dv{\widetilde{p}_{I_{i_s}}}{\tau} &=  \frac{\kappa_3 e_{Z_0} \widetilde{k}_{1_{i_s}} {\color{mulberry}{\frac{2}{N}}} \skpp \of{s_{27_0} - \widetilde{p}_{I_{i_s}} - \widetilde{q}_{{}_{i_s}}}}{1+ {\color{dblue}{R}} \skp \widetilde{k}_{1_1} {\color{mulberry}{\frac{2}{N}}} \skpp \of{s_{27_0} - \widetilde{p}_{I_1} - \widetilde{q}_{{}_1}} + {\color{dblue}{(N-R)}} \skp \widetilde{k}_{1_2} \skpp {\color{mulberry}{\frac{2}{N}}} \skpp \of{s_{27_0} - \widetilde{p}_{I_2} - \widetilde{q}_{{}_2}}}  \\ 
& -\frac{\kappa_{12} e_{U_0} \alpha_{10} ~ {\color{mulberry}{\frac{2}{N}}} \widetilde{p}_{I_{i_s}}}{1+ \alpha_{10} \of{ {\color{dblue}{R}} \skpp {\color{mulberry}{\frac{2}{N}}} \skpp \widetilde{p}_{I_1} + {\color{dblue}{(N-R)}} \skpp {\color{mulberry}{\frac{2}{N}}} \skpp \widetilde{p}_{I_2}}}, \nonumber \\
\label{eq_two_subregions2} 
\dv{\widetilde{p}_{A_{i_s}}}{\tau} &=  \frac{\kappa_6 e_{M_0} \widetilde{k}_{4_{i_s}} {\color{mulberry}{\frac{2}{N}}} \of{s_{4_0} - \widetilde{p}_{A_{i_s}}}}{1+ {\color{dblue}{R}} \skp \widetilde{k}_{4_1} \skpp {\color{mulberry}{\frac{2}{N}}} \skpp \of{s_{4_0} - \widetilde{p}_{A_1}} + {\color{dblue}{(N-R)}} \skp \widetilde{k}_{4_2} \skpp {\color{mulberry}{\frac{2}{N}}} \skpp \of{s_{4_0} - \widetilde{p}_{A_2}} } \\
& - \frac{\kappa_{15} e_{K_0} \alpha_{13} \skpp {\color{mulberry}{\frac{2}{N}}} \skpp \widetilde{p}_{A_{i_s}}}{1+ \alpha_{13}\of{ {\color{dblue}{R}} \skp {\color{mulberry}{\frac{2}{N}}} \skpp \widetilde{p}_{A_1} +  {\color{dblue}{(N-R)}} \skp {\color{mulberry}{\frac{2}{N}}} \skpp \widetilde{p}_{A_2} }},  \nonumber \\
 \label{eq_two_subregions3} 
\dv{\widetilde{q}_{{}_{i_s}}}{\tau} &=  \frac{\kappa_9 e_{C_0} \widetilde{k}_{7_{i_s}} {\color{mulberry}{\frac{2}{N}}} \skpp \of{s_{27_0} - \widetilde{p}_{I_{i_s}} - \widetilde{q}_{{}_i}}}{1+ {\color{dblue}{R}} \skp \widetilde{k}_{7_1} \skpp {\color{mulberry}{\frac{2}{N}}} \skpp \of{s_{27_0} - \widetilde{p}_{I_1} - \widetilde{q}_{{}_1}} + {\color{dblue}{(N-R)}} \skp \widetilde{k}_{7_2} \skpp {\color{mulberry}{\frac{2}{N}}} \skpp \of{s_{27_0} - \widetilde{p}_{I_2} - \widetilde{q}_{{}_2}}}  \\
& - \frac{\kappa_{18} e_{S_0} \alpha_{16} \skp {\color{mulberry}{\frac{2}{N}}} \skpp \widetilde{q}_{{}_{i_s}}}{1 + \alpha_{16} \of{ {\color{dblue}{R}} \skp {\color{mulberry}{\frac{2}{N}}} \skpp  \widetilde{q}_{{}_1} + {\color{dblue}{(N-R)}} \skp {\color{mulberry}{\frac{2}{N}}} \skpp \widetilde{q}_{{}_2}}}. \nonumber
\end{align}

Here, $i_s = 1,2$ denotes the index of the region and the rate functions, $\widetilde{k}_{1_{i_s}}$, $\widetilde{k}_{4_{i_s}}$, and $\widetilde{k}_{7_{i_s}}$, are given as follows:

\begin{align}
& \widetilde{k}_{1_1} = \frac{\alpha_{I_0} + \kappa_1 w_{diag} ~\of{\widetilde{p}_{I_1}}^3}{\overline{\kappa}_3 + \kappa_2 w_{diag} ~\of{\widetilde{p}_{I_1}}^2 \widetilde{p}_{A_1}}, \hspace{5pt} \nonumber \\
& k_{1_2} = \frac{\alpha_{I_0} + \kappa_1 {\color{darkgreen}{(N-R)}} \skpp w_{TAD} ~\of{\widetilde{p}_{I_2}}^3}{\overline{\kappa}_3 + \kappa_2 {\color{darkgreen}{(N-R)}} \skpp w_{TAD} ~\of{\widetilde{p}_{I_2}}^2 \widetilde{p}_{A_2}}, \nonumber \\
& \widetilde{k}_{4_1} = \frac{\alpha_{A_0} + \kappa_4 w_{diag} ~\of{\widetilde{p}_{I_1}}^2 \widetilde{q}_{{}_1}}{\overline{\kappa}_6 + \kappa_5 w_{diag} ~\of{\widetilde{p}_{I_1}}^3},  \label{eq_rates_subregions_tildek}\\
& \widetilde{k}_{4_2} = \frac{\alpha_{A_0} + \kappa_4 {\color{darkgreen}{(N-R)}} \skpp w_{TAD} ~\of{\widetilde{p}_{I_2}}^2 \widetilde{q}_{{}_2}}{\overline{\kappa}_6 + \kappa_5 {\color{darkgreen}{(N-R)}} \skpp w_{TAD} ~\of{\widetilde{p}_{I_2}}^3}, \nonumber \\
& \widetilde{k}_{7_1}= \frac{1}{\alpha_8}\of{\alpha_{Q_0} + w_{diag} ~\of{\widetilde{p}_{I_1}}^2 \widetilde{p}_{A_1}}, \nonumber \\ 
& \widetilde{k}_{7_2}= \frac{1}{\alpha_8}\of{\alpha_{Q_0} + {\color{darkgreen}{(N-R)}} \skpp w_{TAD} ~\of{\widetilde{p}_{I_2}}^2 \widetilde{p}_{A_2}} .  \nonumber
\end{align}

Comparing this system with Eqs~\eqref{eq_two_sites_full1}-\eqref{eq_two_sites_full_k} for the two-site system, we note that the equations describing dynamics of the two regions differ from the two-site system by the scaling factors. Scaling factors ${\color{dblue}{R}}$ and ${\color{dblue}{(N-R)}}$ (highlighted in blue in Eqs~\eqref{eq_two_subregions1}-\eqref{eq_two_subregions3}) represent the competition for enzyme availability for the two regions of given sizes. Moreover, the equations above also include the scaling factors corresponding to the multi-site system, ${\color{mulberry}{2/{N}}}$ (shown in magenta). From Eq~\eqref{eq_rates_subregions_tildek}, we note that the contact frequency matrix for this scenario is of dimension two, with $w_{11} = w_{diag}$, $w_{22}={\color{darkgreen}{(N-R)}} \skpp w_{TAD}$ and $w_{12} = w_{21} = 0$. Here, the factor ${\color{darkgreen}{(N-R)}}$ indicates the amplification of the interaction within the TAD-like region due to high-frequency interactions among loci within this domain. 

We emphasise that Eqs~\eqref{eq_two_subregions1}-\eqref{eq_rates_subregions_tildek} are valid only under the assumption of equal initial conditions for all loci within the same region. This assumption is made to simplify the equations for the stability analysis of the model. In simulations of our stochastic model, we use initial conditions that account for the noisy distribution of epigenetic marks among genomic sites.

\section{Supplementary figures and tables}
\label{secV}

\renewcommand{\thefigure}{\arabic{figure}}
\renewcommand{\figurename}{Supplementary Figure}
\setcounter{figure}{0}

\begin{figure} [b!]
\vspace{0pt}
\hfill
\caption{ \textit{(Figure on the next page.)} Stability analysis of the two-site system for varying cross-interaction, $w_{12}$, and the levels of catalysing enzymes: \textbf{(A)} UTX, $e_{U_0}$, \textbf{(B)} MLL, $e_{M_0}$, \textbf{(C)} KDM, $e_{K_0}$, \textbf{(D)} CBP, $e_{C_0}$, and \textbf{(E)} NuRD and SIRT, $e_{S_0}$. The left panel presents a state space diagram where the monostable regions corresponding to open and closed chromatin are shown in green and red, respectively. The classification of the state as open or closed is determined by the equilibrium value of epigenetic marks at site 1; an open (closed) state corresponds to $p_{I_1}<0.3$ ($p_{I_1}>0.7$). The bistable (tristable) parameter region is indicated in (dark) brown. The middle panel features a heatmap displaying the value of the ruggedness metric, defined by Eq~\eqref{rug_measure}. The rightmost panel shows a scatter plot of the H3K27me3 levels at sites 1 and 2 for all stable equilibria obtained by varying the considered parameters. The colour bar indicates the maximum value of the real part of the eigenvalues of the linearised system. The remaining parameters are set to their baseline levels listed in Supplementary Table~\ref{SuppTable1}.} 
\label{SuppFig1}
\end{figure}
\addtocounter{figure}{-1}
\makeatletter
\setlength{\@fptop}{0pt}
\makeatother
\begin{figure}[!t]
\vspace{-2cm}
\includegraphics[width=\textwidth]{SupplementaryFigure_1}
\captionsetup{singlelinecheck=off,justification=raggedright}
\caption{  \textit{(Caption on the previous page.)}}
\end{figure}

\pagebreak

\begin{figure}[!h]
\includegraphics[width=\textwidth]{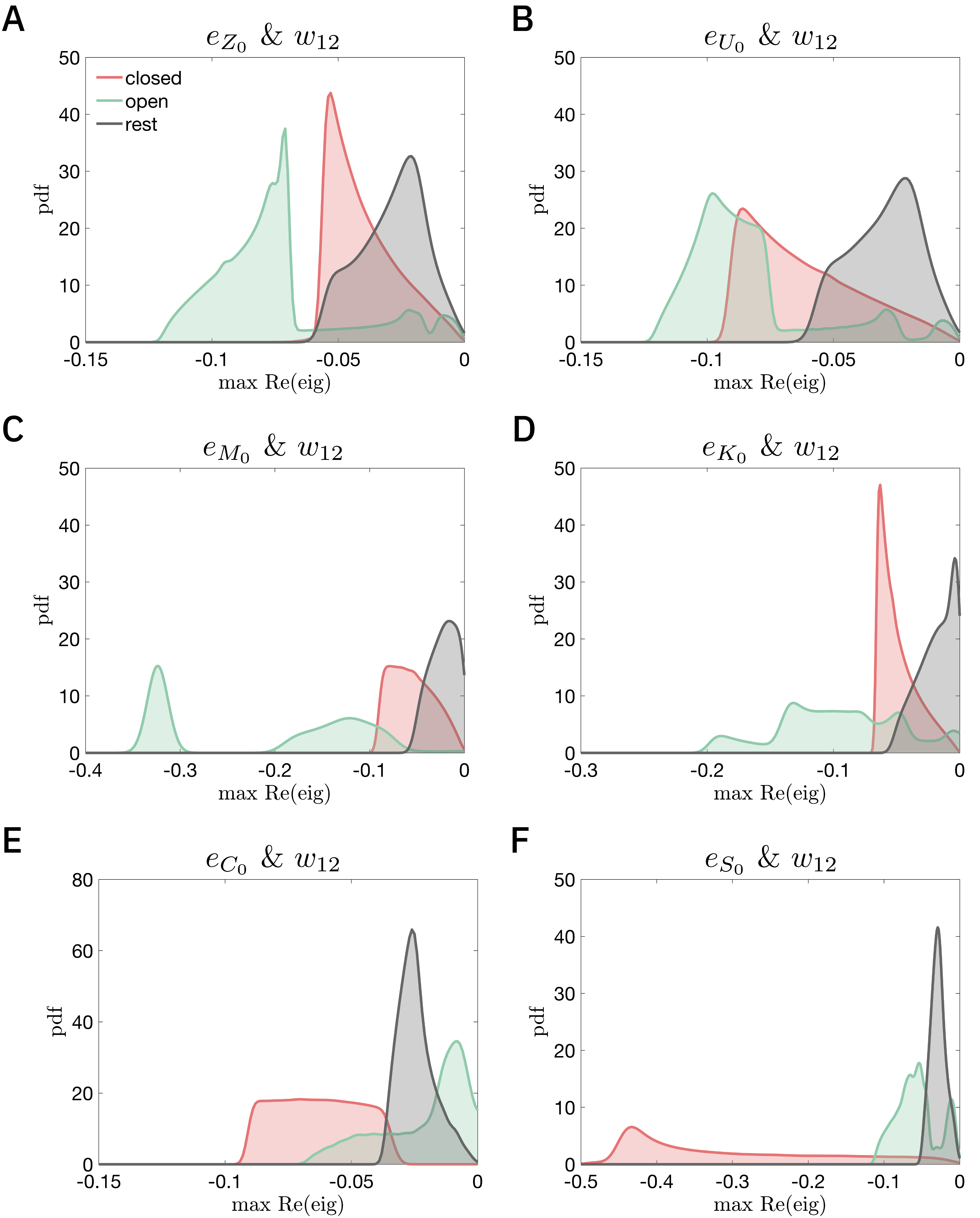}
\captionsetup{singlelinecheck=off,justification=raggedright}
\caption{  \textit{(Caption on the next page.)}}
\end{figure}
\addtocounter{figure}{-1}
\makeatletter
\setlength{\@fptop}{0pt}
\makeatother
\begin{figure} [t!]
\caption{Distributions of the maximum value of the real part of the eigenvalues of the linearised system for stable equilibria obtained by varying the strength of cross-interaction between two sites and the enzyme levels: \textbf{(A)} EZH2, $e_{Z_0}$ (corresponds to Figure~\ref{Fig4}A); \textbf{(B)} UTX, $e_{U_0}$ (corresponds to Supplementary Figure~\ref{SuppFig1}A); \textbf{(C)} MLL, $e_{M_0}$ (corresponds to Supplementary Figure~\ref{SuppFig1}B); \textbf{(D)} KDM, $e_{K_0}$ (corresponds to Supplementary Figure~\ref{SuppFig1}C); \textbf{(E)} CBP, $e_{C_0}$ (corresponds to Supplementary Figure~\ref{SuppFig1}D); \textbf{(F)} NuRD and sirtuins, $e_{S_0}$ (corresponds to Supplementary Figure~\ref{SuppFig1}E). The characterisation of the equilibrium states as open (green) and closed (red) was based on the epigenetic profile of site 1. Stable equilibria that did not fit into either the open (low H3K27me3) or closed (high H3K27me3) categories are referred to as \textit{rest} (black). These include all stable equilibria with intermediate levels of repressive and activating marks characteristic of bivalent chromatin. The remaining parameters are set to their baseline values indicated in Supplementary Table~\ref{SuppTable1}.} 
\label{SuppFig2}
\end{figure}

\begin{figure}[!h]
\vspace{-1.3cm}
 \includegraphics[width=\textwidth]{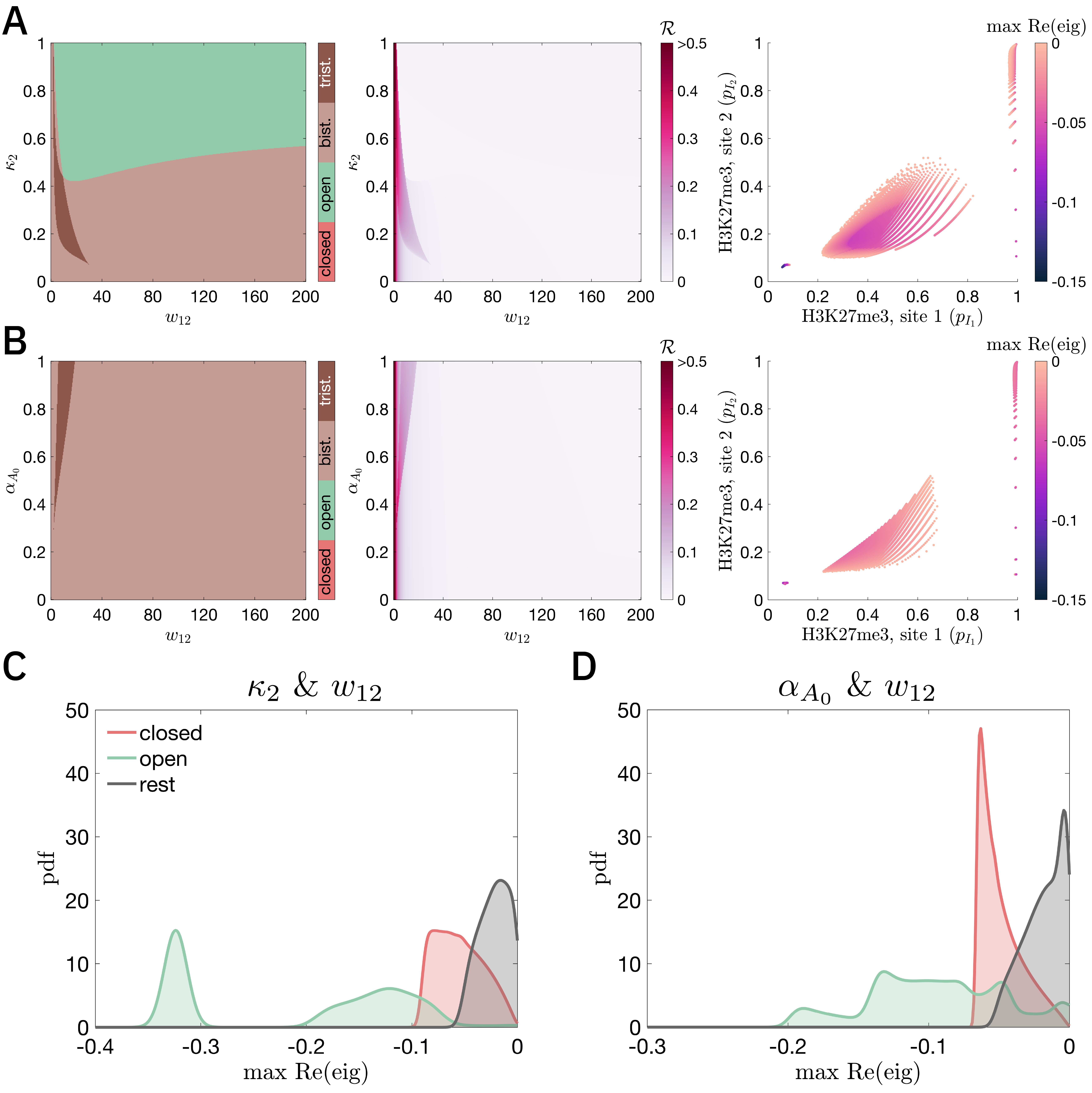}
\caption{\textbf{(A)}, \textbf{(B)} Stability analysis of the two-site system for varying cross-interaction, $w_{12}$, and \textbf{(A)} the strenght of inhibitory effects of H3K4me3 marks on production of H3K27me3, described by $\kappa_2$, \textbf{(B)} the baseline rate of H3K4 trimethylation, $\alpha_{A_0}$. The left panel presents a state space diagram where the monostable regions corresponding to open and closed chromatin are shown in green and red, respectively. The classification of the state as open or closed is determined by the equilibrium value of epigenetic marks at site 1; an open (closed) state corresponds to $p_{I_1}<0.3$ ($p_{I_1}>0.7$). The bistable (tristable) parameter region is indicated in (dark) brown. The middle panel features a heatmap displaying the value of the ruggedness metric, defined by Eq~\eqref{rug_measure}. The rightmost panel shows a scatter plot of the H3K27me3 levels at sites 1 and 2 for all stable equilibria obtained by varying the considered parameters. The colour bar indicates the maximum value of the real part of the eigenvalues of the linearised system. \textbf{(C)}, \textbf{(D)} Distributions of the maximum value of the real part of the eigenvalues of the linearised system for stable equilibria included in the analysis presented in \textbf{(A)} and \textbf{(B)}, respectively, for open, closed and bivalent states of site 1. The colour scheme is as described in Supplementary Figure~\ref{SuppFig2}. The remaining parameters are set to their baseline levels listed in Supplementary Table~\ref{SuppTable1}.} 
\label{SuppFig3}
\end{figure}

\pagebreak

\begin{figure}[!h]
\includegraphics[width=\textwidth]{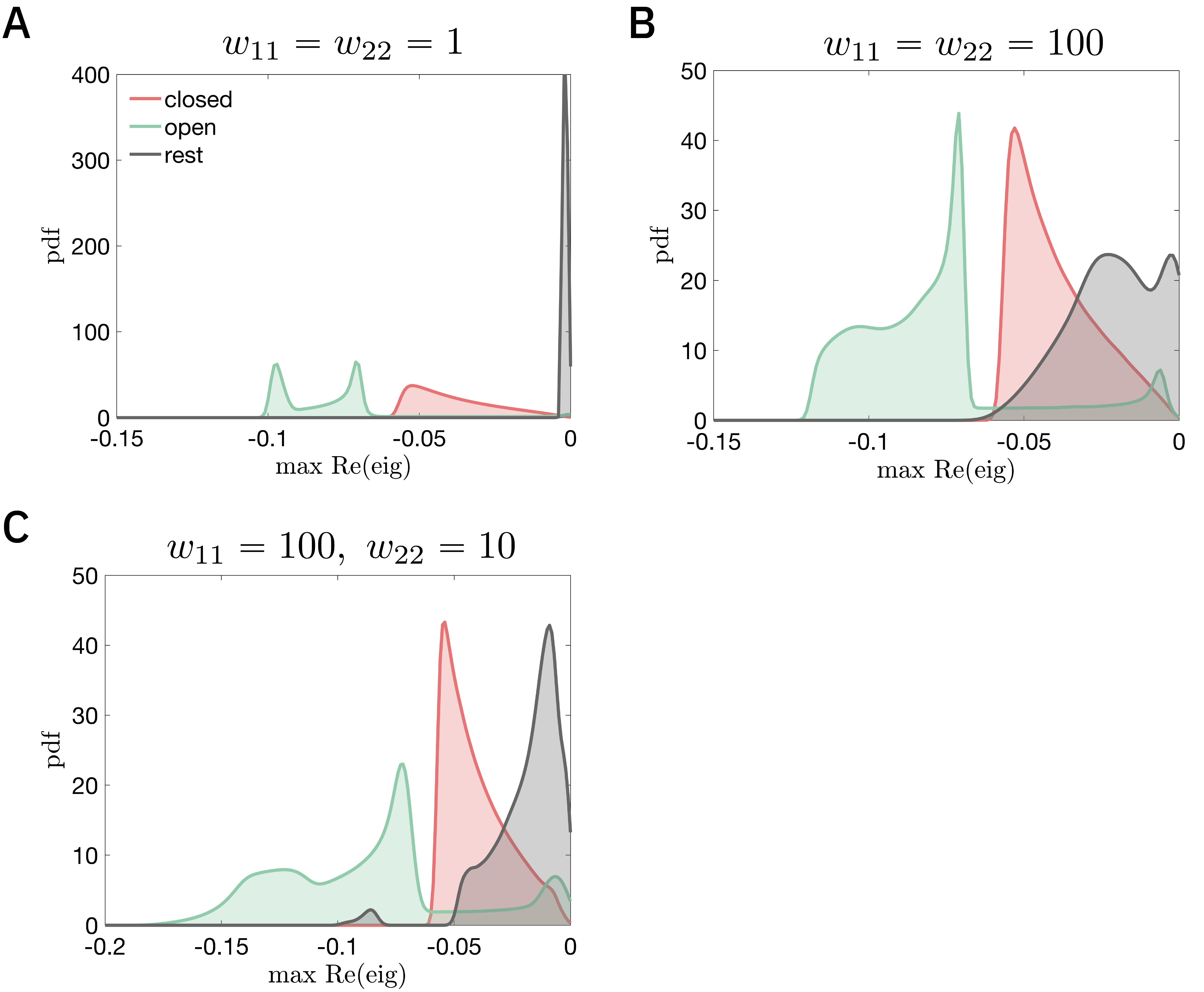}
\caption{\textbf{(A)}-\textbf{(D)} Distributions of the maximum value of the real part of the eigenvalues of the linearised system for stable equilibria included in the analysis presented in Figure~\ref{Fig5}A-D, respectively. Here, the response of the system was analysed for variations in the interactions between sites 1 and 2, $w_{12}$, and the level of EZH2, $e_{Z_0}$. Contact frequencies for interactions within the same site were set to \textbf{(A)} $w_{11}=w_{22}=1$, \textbf{(B)} $w_{11}=w_{22}=100$, \textbf{(C)} $w_{11}=100$, $w_{22}=10$. We note that the case of $w_{11}=200$, $w_{22}=0.01$ is shown in Supplementary Figure~\ref{SuppFig2}A. The classification of equilibrium states as open (green) and closed (red) was based on the epigenetic profile of site 1. Stable equilibria that did not fall into either the open (low H3K27me3) or closed (high H3K27me3) categories are labelled as \textit{rest} (black). These include all stable equilibria with intermediate levels of repressive and activating marks, characteristic of bivalent chromatin. The remaining parameters are set to their baseline levels listed in Supplementary Table~\ref{SuppTable1}.} 
\label{SuppFig4}
\end{figure}

\pagebreak

\begin{figure}[!h]
\vspace{-2cm}
\includegraphics[width=\textwidth]{SupplementaryFigure_5}
\captionsetup{singlelinecheck=off,justification=raggedright}
\caption{  \textit{(Caption on the next page.)}}
\end{figure}
\addtocounter{figure}{-1}
\makeatletter
\setlength{\@fptop}{0pt}
\makeatother
\begin{figure} [t!]
\caption{Stability analysis of the two-site system for varying cross-interaction, $w_{12}$, and contact frequency for interaction within site 2, $w_{22}$.
Interaction frequency within site 1 were set to \textbf{(A)} $w_{11}=0.01$, \textbf{(B)} $w_{11}=1$, \textbf{(C)} $w_{11}=10$, \textbf{(C)} $w_{11}=100$, and \textbf{(D)} $w_{11}=200$. Left panels: state space diagrams where the monostable regions corresponding to open and closed chromatin are shown in green and red, respectively. The classification of the state as open or closed is determined by the equilibrium value of epigenetic marks at site 1; an open (closed) state corresponds to $p_{I_1}<0.3$ ($p_{I_1}>0.7$). The bistable (tristable) parameter region is indicated in (dark) brown. In panels \textbf{(D)}-\textbf{(E)}, we observe the emergence of regions where four stable equilibria coexist (light blue). In panel \textbf{(E)}, a small region with five stable equilibria is highlighted in dark grey. Middle panels: heatmaps displaying the value of the ruggedness metric, defined by Eq~\eqref{rug_measure}. Right panels: scatter plots of the H3K27me3 levels at sites 1 and 2 for all stable equilibria obtained by varying the considered parameters. The colour bar indicates the maximum value of the real part of the eigenvalues of the linearised system. The remaining parameters are set to their baseline levels listed in Supplementary Table~\ref{SuppTable1}.} 
\label{SuppFig5}
\end{figure}

\begin{figure}[!b]
\captionsetup{singlelinecheck=off,justification=raggedright}
\caption{ \textit{(Figure on the next page.)} Analysis and simulations of the two-region system for the constant interaction matrix, $\overline{W}$, where we assume $\overline{w}_{il} = w_{il}$. We consider chromatin geometry in which region 1 exhibits only connections within the same site with frequency, $w_{diag}$, while TAD-like region 2 is characterised by all-to-all interactions with the uniform strength of $w_{TAD}$. \textbf{(A)} Stability analysis for variations in $w_{diag}$ and $w_{TAD}$. The full modelling equations are presented in Section~\ref{secIV}. We note that this analysis is valid only for a particular case when the initial conditions within each region are the same. The colour schemes of the leftmost and rightmost panels are as in Figure~\ref{Fig5}B. The middle panel shows a heatmap describing which rugged equilibrium patterns exist for the corresponding values of parameters. Here, rugged pattern 1 (rugged pattern 2) shown in magenta (pastel purple) is characterised by region 1 enriched with activating (repressive) marks, whereas TAD-like region 2 displays high levels of repressive (activating) modifications. The parameter space where both rugged patterns coexist is shown in dark red. \textbf{(B)}, \textbf{(C)} Stochastic simulations of the multi-site system for the two-region scenario with constant interaction matrix, $\overline{W}$, and a fixed value of $w_{TAD}=10$. The value of $w_{diag}= 1$ in \textbf{(B)} and $w_{diag}= 3$ in \textbf{(C)}. Final configurations are marked as rugged pattern 1 and rugged pattern 2 as described for panel \textbf{(A)}. Here, we used the total number of sites, $N=10$, the size of the region 1, $R=5$, and the final simulation time, $T=5000$. All other parameters are set to their baseline values listed in Supplementary Table~\ref{SuppTable1}.}
\label{SuppFig6}
\end{figure}
\addtocounter{figure}{-1}
\makeatletter
\setlength{\@fptop}{0pt}
\makeatother
\begin{figure}[!t]
\begin{center}
\includegraphics[width=\textwidth]{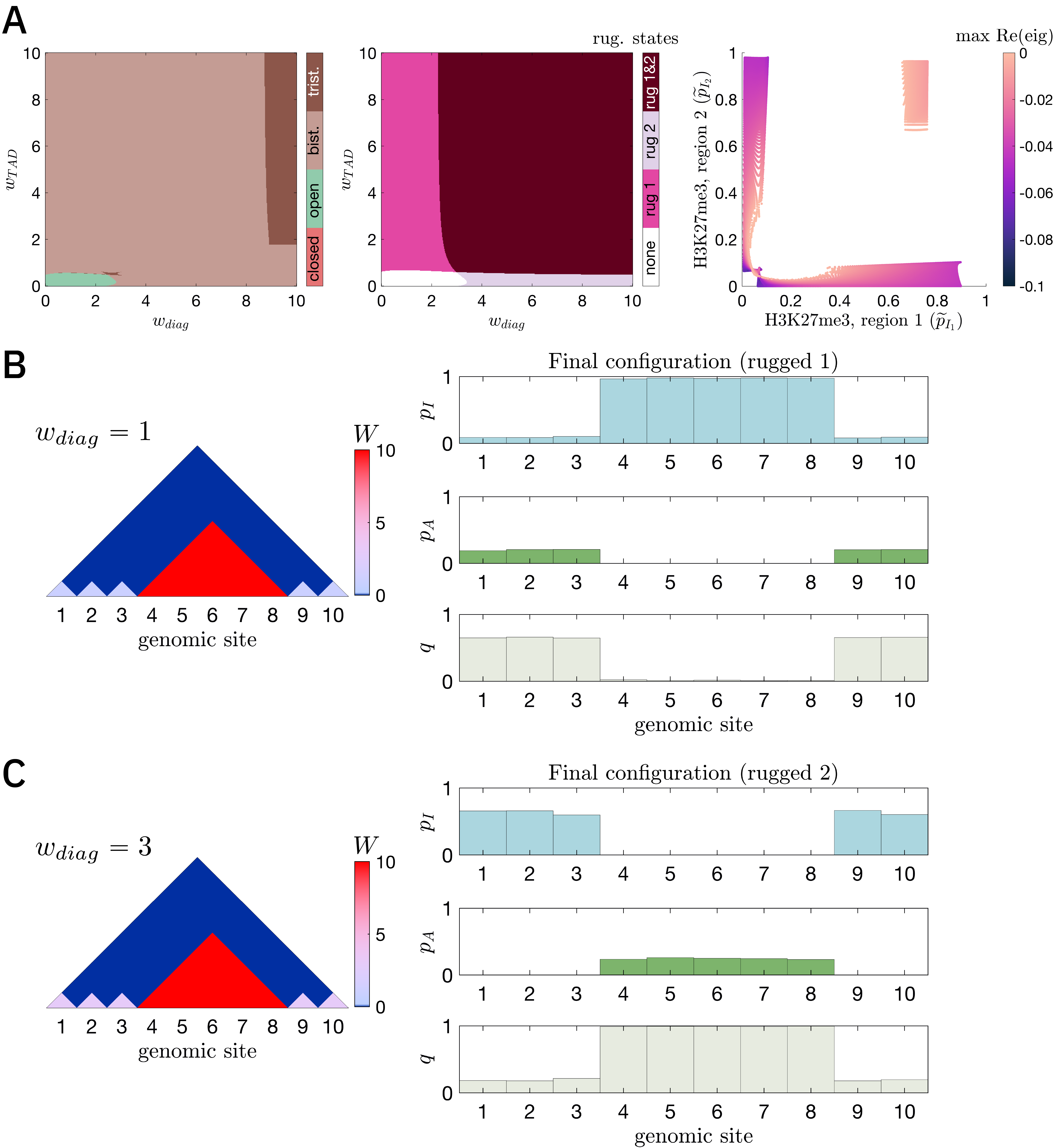}
\captionsetup{singlelinecheck=off,justification=raggedright}
\caption{\textit{(Caption on the previous page.)} }
\end{center}
\end{figure}

\clearpage

\renewcommand{\thetable}{\arabic{table}}
\renewcommand{\tablename}{Supplementary Table}
\setcounter{table}{0}

\renewcommand{\arraystretch}{1.4}
\setlength{\doublerulesep}{2\arrayrulewidth}
\begin{longtable}{| p{1cm} | p{1.7cm} | p{10.0cm} |}
\caption{Descriptions of parameters included in our model and their baseline values used in this work.} \label{SuppTable1} \\
\endfirsthead
\caption{\textit{Continued.}} \\
\endhead
\hline
\textbf{Par.} & \textbf{Value} & \textbf{Description} \\ \hline \hline
$e_{Z_0}$ & $0.75$ & Dimensionless parameter describing levels of methyltransferase EZH2 (Enhancer of Zeste 2) catalysing deposition of trimethyl to H3K27. \\
$e_{U_0}$ & $0.5$ & Dimensionless parameter describing levels of demethylase UTX (Ubiquitously Transcribed Tetratricopeptide Repeat on chromosome X) mediating the removal of H3K27me3 marks. \\
$e_{M_0}$ & $0.1$ & Dimensionless parameter describing levels of methyltransferase MLL (Mixed-Lineage Leukemia) catalysing deposition of trimethyl to H3K4. \\
$e_{K}$ & $0.5$ & Dimensionless parameter describing levels of histone lysine demethylase (KDM) mediating the removal of H3K4me3 marks. \\
$e_{C_0}$ & $1.0$ & Dimensionless parameter describing levels of acetyltransferase CBP (CREB-binding protein) catalysing deposition of acetyl to H3K27. \\
$e_{S_0}$ & $0.1$ & Dimensionless parameter describing levels of deacetylases NuRD (Nucleosome Remodeling and Deacetylase) and sirtuins. \\ \hline
$\kappa_1$ & $1.0$ & Dimensionless parameter describing the strength of the self-reinforcing feedback mechanisms on the production of H3K27me3 marks. \\
$\kappa_2$ & $0.2$ & Dimensionless parameter describing the strength of the antagonistic feedback mechanisms of H3K4me3 on the production of H3K27me3 marks. \\
$\kappa_4$ & $1.0$ & Dimensionless parameter describing the strength of the reinforcing feedback mechanisms of H3K27ac on the production of H3K27me3 marks. \\
$\kappa_5$ & $1.0$ & Dimensionless parameter describing the strength of the antagonistic feedback mechanisms of H3K27me3 on the production of H3K4me3 marks. \\ \hline
$s_{{27}_0}$ & $1.0$ & Dimensionless parameter describing the number of lysine 27 residues of histone 3 in one genomic site (of a given length in kb). \\
$s_{{4}_0}$ & $1.0$ & Dimensionless parameter describing the number of lysine 4 residues of histone 3 in one genomic site (of a given length in kb). \\ \hline
$\kappa_3$ & $1.0$ & Dimensionless rate of trimethylation of H3K27 by EZH2. \\
$\kappa_{12}$ & $1.0$ & Dimensionless rate of demethylation of H3K27me3 by UTX. \\
$\kappa_6$ & $1.0$ & Dimensionless rate of trimethylation of H3K4 by MLL. \\
$\kappa_{15}$ & $1.0$ & Dimensionless rate of demethylation of H3K4me3 by KDM. \\
$\kappa_9$ & $1.0$ & Dimensionless rate of acetylation of H3K27 by CBP. \\
$\kappa_{18}$ & $1.0$ & Dimensionless rate of deacetylation of H3K27ac by NuRD and sirtuins. \\ \hline
$\kappa_2 \kappa_{2_0}$ & $0.0$ & Dimensionless parameter describing the baseline rate of disassociation of a complex between EZH2 and an unmodified H3K27 residue. \\
$\kappa_5 \kappa_{5_0}$ & $0.0$ & Dimensionless parameter describing the baseline rate of disassociation of a complex between MLL and an unmodified H3K4 residue. \\
$\kappa_8$ & $1.0$ & Dimensionless parameter describing the baseline rate of disassociation of a complex between CBP and an unmodified H3K27 residue. \\ 
$\overline{\kappa}_3$ & $\kappa_3 + \kappa_2 \kappa_{2_0}$ & Dimensionless parameter describing the rate of decrease of levels of EZH2-H3K27 complex due to either its disassociation or release of product reactant (H3K27me3 mark). \\
$\overline{\kappa}_6$ & $\kappa_6 + \kappa_5 \kappa_{5_0}$ & Dimensionless parameter describing the rate of decrease of levels of MLL-H3K4 complex due to either its disassociation or release of product reactant (H3K4me3 mark). \\
$\alpha_8$ & $\kappa_8 + \kappa_9$ & Dimensionless parameter describing the rate of decrease of levels of CBP-H3K27 complex due to either its disassociation or release of product reactant (H3K27ac mark). \\ \hline
$\alpha_{10}$ & $1.0$ & The inverse of the Michaelis (or half-saturating) constant \citep{ingalls2013mathematical} for the enzymatic reaction of removal of H3K27me3 marks. \\
$\alpha_{13}$ & $1.0$ & The inverse of the Michaelis (or half-saturating) constant \citep{ingalls2013mathematical} for the enzymatic reaction of removal of H3K4me3 marks. \\
$\alpha_{16}$ & $0.(90)$ & The inverse of the Michaelis (or half-saturating) constant \citep{ingalls2013mathematical} for the enzymatic reaction of removal of H3K27ac marks. \\ \hline
$\alpha_{I_0}$ & $0.1$ & Dimensionless parameter describing the baseline rate of formation of a complex between EZH2 and an unmodified H3K27 residue. \\
$\alpha_{A_0}$ & $1.0$ & Dimensionless parameter describing the baseline rate of formation of a complex between MLL and an unmodified H3K4 residue. \\
$\alpha_{Q_0}$ & $0.1$ & Dimensionless parameter describing the baseline rate of formation of a complex between CBP and an unmodified H3K27 residue. \\ \hline
\end{longtable}


\bibliographystyle{elsarticle-harv} 




\makeatletter\@input{xxsupp.tex}\makeatother